\documentclass[journal, twoside]{IEEEtran}

\ifx\pdfoutput\undefined
\usepackage{graphicx}
\graphicspath{{eps_figs/}}
\else
\usepackage[pdftex]{graphicx}
\graphicspath{{pdf_figs/}}
\fi

\usepackage{latexsym,amssymb,amsmath,ifthen,afterpage,cite,amsfonts}

\usepackage{lipsum}
\usepackage{algorithm}
\usepackage{algorithmic}
\usepackage{enumerate}
\usepackage{subfig}
\usepackage{url}
\usepackage{tikz} % To generate the plot from csv
\usepackage{pgfplots}
\usepackage{pgfplotstable}
\usepackage{filecontents}
\usepgfplotslibrary{statistics}

\newcommand{\Fig}[1]{Fig.~\ref{#1}}
\newcommand{\eqdef}{\stackrel{\scriptscriptstyle\bigtriangleup}{=} }
\newcommand{\smpl}[2]{#1^{(#2)}}
\newcommand{\Z}{\mathbb{Z}}

\newcommand{\N}{\mathcal{N}}
\newcommand{\G}{\mathcal{G}}
\newcommand{\EE}{\mathcal{E}}
\newcommand{\VV}{\mathcal{V}}

\newcommand{\overbar}[1]{\mkern 1.5mu\overline{\mkern-1.5mu#1\mkern-1.5mu}\mkern 1.5mu}

\newcommand{\calX}{\mathcal{X}}

\newcommand{\calU}{\mathcal{U}}

\newcommand{\calW}{\mathcal{W}}
\newcommand{\calS}{\mathcal{S}}

\newcommand{\X}{{\bf X}}
\newcommand{\x}{{\bf x}}
\newcommand{\Y}{{\bf Y}}
\newcommand{\y}{{\bf y}}
\newcommand{\z}{{\bf z}}
\newcommand{\ZZ}{{\bf Z}}

\newcommand{\F}{{\overbar{\mathcal{T}}}}
\newcommand{\T}{\mathcal{T}}

\newcommand{\E}{\operatorname{E}}
\newcommand{\Var}{\operatorname{Var}}

\newcommand{\I}{\operatorname{I}}

\newcounter{examplecntr}
{\begin{trivlist}\small\item[]\refstepcounter{examplecntr}%
 {\bfseries Example~\theexamplecntr%
  \ifthenelse{\equal{#1}{}}{}{ (#1)}.
}}%
{\end{trivlist}}

\newcounter{propositioncntr}
\newenvironment{proposition}[1][]%
{\begin{trivlist}\item[]\refstepcounter{propositioncntr}%
{\bfseries Proposition~\thepropositioncntr%
  \ifthenelse{\equal{#1}{}}{}{ (#1)}.
}}%
{\hfill$\Box$\end{trivlist}}

\newcounter{theoremcntr}
{\begin{trivlist}\item[]\refstepcounter{theoremcntr}%
{\bfseries Theorem~\thetheoremcntr%
  \ifthenelse{\equal{#1}{}}{}{ (#1)}.
}}%
{\hfill$\Box$\end{trivlist}}

\newenvironment{proofof}[1]{\begin{trivlist}\item[]{\bfseries Proof\ifthenelse{\equal{#1}{}}{}{ #1}:}
}{\hfill$\Box$\end{trivlist}}

\newcommand{\eproofnegspace}{\\[-1.5\baselineskip]\rule{0em}{0ex}}

\newcommand{\pos}[2]{\makebox(0,0)[#1]{#2}}

\begin{document}
\DeclareGraphicsExtensions{.pdf}

\title{Monte Carlo Methods for the Ferromagnetic\\ Potts Model Using Factor Graph Duality}

\author{Mehdi Molkaraie and Vicen\c{c} G\'{o}mez
\thanks{M.~Molkaraie is with the 
Department of Information Technology and Electrical Engineering, ETH Zurich, 
CH-8092 Z\"urich, Switzerland (email: mehdi.molkaraie@alumni.ethz.ch). 
V.~G\'{o}mez is with the Artificial Intelligence and Machine Learning group at the Universitat 
Pompeu Fabra, 08018 Barcelona, Spain (email: vicen.gomez@upf.edu).}
\thanks{Parts of this work were presented in \cite{MoLo:ISIT2013,MeMo:2014a,Mo:IZS2016}.}
}

\maketitle 
 
\begin{abstract}
Normal factor graph duality offers new possibilities for Monte Carlo algorithms 
in graphical models. 
Specifically, we consider the problem of estimating the partition function 
of the ferromagnetic Ising and Potts models by Monte Carlo methods, 
which are known to work well at high temperatures,
but to fail at low temperatures. 
We propose Monte Carlo methods (uniform sampling and importance 
sampling) in the dual normal factor graph, and demonstrate that they
behave differently: they work particularly well at 
low temperatures. %and their convergence improves as temperature decreases. 
By comparing the relative error in estimating the partition 
function, we show that the proposed importance sampling algorithm significantly outperforms 
the state-of-the-art deterministic and Monte Carlo methods. For the ferromagnetic Ising model in an 
external field, we show 
the equivalence between the valid configurations in the dual normal factor graph and the terms that appear in the 
high-temperature series expansion of the partition function. Following this result, we discuss connections with 
Jerrum--Sinclair's polynomial randomized approximation scheme (the subgraphs-world process)
for evaluating the partition function of ferromagnetic 
Ising models.
\end{abstract} 


\section{Introduction}

%Let $f(x_1,\ldots,x_n)$ be a nonnegative real function 
%of finite-alphabet variables $x_1,\ldots,x_n$. 

Many quantities of interest in statistical physics, combinatorics, 
information theory, and machine learning can be expressed as a \emph{partition function} 
%(German: \emph{Zustandssumme})
\begin{IEEEeqnarray}{c} 
\label{eqn:defZf}
Z \eqdef \sum_{x_1,\ldots,x_N} f(x_1,\ldots,x_N),
\end{IEEEeqnarray} 
where $f(x_1,\ldots,x_N)$ is a nonnegative real function 
of finite-valued variables $x_1,\ldots,x_N$ 
and where the sum runs over all possible values of 
these variables.
For example, if $f$ takes values in $\{0,1\}$, then 
(\ref{eqn:defZf}) counts the number of configurations $x_1,\ldots,x_N$ for which 
$f(x_1,\ldots,x_N)\neq 0$. 
%In Bayesian model selection, the marginal likelihood (the evidence)~\cite[Chapter 5]{murphy:2012}
%is a partition function, 
%and estimates of this quantity are also required for
%maximum-likelihood training of deep belief networks \cite{SM:2008}.
In statistical physics, $Z$ is usually considered as a function of temperature 
(cf.\ Section~\ref{sec:Ising}) and hence is called the partition function.
%Note that, 
For large $N$, we are usually interested in the free energy per site
%(good-enough approximations of) 
$\frac{1}{N}\ln Z$ rather than in $Z$ itself. 

Naive computation of (\ref{eqn:defZf}) is exponential in $N$ 
and practically possible only for small $N$. 
If the function $f$ in (\ref{eqn:defZf}) has 
a cycle-free factor graph 
(and no state variables with excessively many states), 
%(and the variables do not have excessively many states),
then the partition function can be computed exactly by 
sum-product message passing \cite{KFL:01, Lg:ifg2004, AM:00} 
(with complexity typically linear in $N$), 
%which includes as a special case the transfer function method from statistical physics~\cite[Chapter 5]{Yeo:92}.
which, in this context, coincides with the transfer matrix method from statistical 
physics~\cite[Chapter 2]{Baxter07},~\cite[Chapter 5]{Yeo:92}.
%(which may be viewed as a special case of sum

In general, however, the exact computation of the partition function is intractable for large $N$, 
%Unfortunately, partition sums are often very difficult 
%(i.e., practically impossible) to compute numerically,
and even good approximations can be hard to obtain.
Empirically good approximations 
%of the partition sum 
are often achieved with deterministic methods 
(including the belief propagation (BP), the generalized belief propagation (GBP)~\cite{Yedidia:2005}, 
and the tree expectation propagation (TreeEP)~\cite{Minka:04} algorithms),
but the accuracy of such approximations is often difficult to assess theoretically.

In this paper, we pursue the idea (proposed in \cite{MoLo:ISIT2013, MeMo:2014a, Mo:IZS2016}) 
that normal factor graph duality offers new opportunities for Monte Carlo algorithms. 
We develop and test this idea for two-dimensional (2D) nearest-neighbor Ising and Potts models. 
%mehdi
%(The generalization to more general Potts models is straightforward.)
%mehdi
Both the Ising model~\cite{Ising:1925,Cipra:87} and the more general 
Potts model~\cite{Potts:52,WU:82} play an important role
in many areas, including statistical physics~\cite{Baxter07}, image processing~\cite{boykov2001}, 
spatial statistics~\cite{besag1993}, and graph theory~\cite{wu1988potts,Sokal:2001,Sokal:2005}.

Exact computation of the partition function of the Potts model 
is possible only in some special cases (e.g., in the one-dimensional (1D) case). 
%(??? what about infinite grid ???)
For the planar Ising model without an external field,
%the problem can be reduced to evaluating  
the problem is tractable and can be reduced to evaluating 
a certain determinant~\cite{Kasteleyn:1963, Fisher:66},
and this approach can also be used to obtain accurate approximations 
in more general settings~\cite{Gomez:10}. 
Also there is a polynomial randomized approximation scheme~\cite[Chapter 28]{Vaz:2013}
for the partition function of general ferromagnetic Ising models in an external field due to Jerrum 
and Sinclair \cite{JS:93}. Connections among the dual normal factor 
graph representation 
of the Ising model, the approximation scheme of Jerrum and Sinclair, and the high-temperature series expansions of the partition function 
will be discussed in detail in this paper. 
However, under reasonable complexity assumptions, there is no polynomial randomized approximation scheme
for the partition function of the Potts model. Indeed, for ferromagnetic 
Potts models, approximating the partition function is as hard as approximating the number of independent sets in 
a bipartite graph, which is among the presumably intractable 
problems~\cite{goldberg2012,Gold:12,Galan:16}.

Known Monte Carlo algorithms for the partition function work very well for 
the Potts model at high temperature (i.e., when local correlations decay quickly)~\cite{HH:64,Neal:proinf1993r,BH:10,LoMo:IT2013}. At low temperatures, however, 
%known 
Monte Carlo methods 
%(including the celebrated Swendsen-Wang algorithm~\cite{SW:87}) 
suffer from slow and erratic convergence \cite{Potam:97, BH:10}.  
More advanced Monte Carlo methods (e.g., nested 
sampling~\cite{MMGS:2005} and the Swendsen-Wang algorithm~\cite{SW:87}) require sampling 
from a large sequence of constraints 
or intermediate distributions at different temperatures 
to estimate the partition function. The main challenge is, therefore, to design Monte 
Carlo methods   
that achieve sufficiently fast convergence in the low-temperature regime. 
%and nested sampling methods~\cite{MMGS:2005} require an excessive number 
%of auxiliary distributions at intermediate temperatures.

The approach of this paper is based on the notions of the
dual normal realization as introduced by Forney \cite{Forney:01} and the dual 
normal factor 
graph~\cite{AY:2011, Forney:11, FV:2011}. According to the normal factor graph duality theorem, 
the partition function of the dual normal factor graph 
equals that of the primal normal factor graph up to a known scale factor 
\cite{AY:2011}. 
The relation of normal factor graph duality to Kramers--Wannier duality~\cite{KW:41}, \cite[Section~II]{WU:82} 
in statistical physics has been worked out in \cite{AlVo:ISIT2015, AlVo:2016, Forney:16}.

Using Monte Carlo methods in the dual normal factor graph has been investigated
in \cite{MoLo:ISIT2013},~\cite{AY:2014},~\cite{MeMo:2014a},~\cite{Mo:IZS2016}.
It was demonstrated (by simulations) in~\cite{MoLo:ISIT2013} that for the 2D Ising model without an external field, 
baseline Monte Carlo methods converge faster at low temperature in the dual normal factor graph than in the 
primal normal factor graph. 
%For one specific example (the nearest-neighbor two-dimensional (2D) Ising
%model without an external field), it was demonstrated that 
%both uniform sampling and Gibbs sampling in the dual 
%NFG excel at low temperatures, where these methods fail 
%in the primal NFG. 
Some pertinent analytical and numerical results regarding the variance of Monte Carlo methods in 
the two domains were given in~\cite{AY:2014}.
%The good low-temperature performance of MC methods in the dual NFG
%were confirmed by some analytical results in~\cite{AY:2014}. 
%Algorithms and simulation results for more general Ising models 
%were given in \cite{MeMo:2014a}. 
%The methods of~\cite{MoLo:ISIT2013} were further 
%generalized in~\cite{MeMo:2014a} to design an importance sampling algorithm for the 2D Ising 
%model in a strong external field.
A suitable partitioning of variables, which allows drawing independent samples according to an
auxiliary distribution, was introduced in~\cite{MeMo:2014a} to propose an importance sampling algorithm 
to estimate the partition function of the Ising model in a 
strong external field. The methods of~\cite{MeMo:2014a} were
further generalized in~\cite{Mo:IZS2016} to models with a mixture of 
strong and weak couplings. % in an external field.

In this paper, 
%we develop these ideas more clearly and more generally; 
%we corroborate the basic idea of \cite{MoLo:ISIT2013} and further explore 
we further explore the use of Monte Carlo methods in the dual normal factor graph. 
Specifically, we propose Monte Carlo methods
for estimating the partition function of %the nearest-neighbor 2D 
ferromagnetic $q$-state Potts models, with or without an 
external field. We consider uniform sampling and importance sampling algorithms, 
both of which are shown to work very well for strong couplings or, equivalently, at low temperature. 
%
%For the importance sampling algorithm, our analytical results
%prove that very accurate estimates of
%the partition function can be obtained, when coupling parameters are strong on
%only a subset of the edges (which forms a spanning tree in the dual NFG). 
%Based on this result, we 
%propose a heuristic strategy to partition models with a mixture of strong and weak couplings and with 
%arbitrary topology. 

Our experimental results show that, in various settings, 
the importance sampling algorithm significantly improves 
upon the state-of-the-art Monte Carlo and deterministic methods. Indeed, in contrast to Monte Carlo methods in the primal domain, the dual-domain
Monte Carlo algorithms of this paper excel at low temperatures.
%The good low-temperature performance of the
%proposed algorithms are confirmed by some analytical results. 

The paper is organized as follows. 
In Section~\ref{sec:Ising}, we review the Potts model. 
The primal and the dual normal factor graphs of the model will be presented in 
Section~\ref{sec:NFGD} and Section~\ref{sec:DualFGPotts}, respectively.
Specific Monte Carlo algorithms that use the dual normal factor graph
are proposed in Section~\ref{sec:IS}, 
and pertinent experimental results 
(including comparisons with standard deterministic and Monte Carlo methods) 
are presented in Section~\ref{sec:Num}.
Extensions of our Monte Carlo methods to the Potts model in an external field
are discussed in Section~\ref{sec:PottsExt}.
In Section~\ref{sec:HightTSExt}, we establish the connection among the valid configurations in the 
dual normal factor graph representation of the Ising model in an external 
field, high temperature series expansions of the partition function, and the randomized 
approximation scheme of Jerrum and Sinclair.
Appendix~\ref{appsec:VarIsing} compares the variance of Monte Carlo methods in the 
primal and in the dual normal factor graphs of the 2D Ising model to demonstrate their opposite behavior.

\section{The 2D Potts Model without External Field}
\label{sec:Ising}

Let $X_1, X_2, \ldots, X_N$ be a collection of $N$ random variables
that take values in the set $\calX$, which in this context is identical to $\Z/q\Z$, the 
ring of integers modulo $q$ for some 
fixed integer $q\geq 2$. 
%The special case $q=2$ is the Ising model. 
(In the special case where $q=2$, we obtain the Ising model.)
%For $q = 2$, the Potts model is equivalent to the Ising model.
%(Later on, we will identify $\calX$ with the integers modulo $q$.)
Let $x_i$ represent a
possible realization of $X_i$ and $\X$ stand for 
$(X_1, X_2, \ldots, X_N)$.
The vectors $\x \in \calX^N$ will be called configurations.

The variables $X_1, X_2, \ldots, X_N$ are associated with the vertices of a simple and connected graph 
$\G = (\VV, \EE)$ that has $N = |\VV|$ vertices and $|\EE|$ edges. In the Potts model, each variable 
represents the $q$ possible states of a particle, and two variables interact if their 
corresponding vertices are connected by an edge in $\EE$.

For illustrative purposes, 
we will take $\G$ to be a grid with size $N=M\times M$, % (as illustrated in Fig.~\ref{fig:2DGrid}), 
and assume periodic boundary conditions, so that each variable 
has exactly four neighbors. However, the methods of this 
paper are easily adapted to Potts models with arbitrary topology. Indeed, we will consider 
2D Potts models with free boundary conditions in our numerical experiments.
%and each variable interacts only with its nearest neighbors
%(as illustrated in Fig.~\ref{fig:2DGrid}).

Conventionally $\EE$ is defined as the set of all the unordered pairs 
$(k,\ell) \in \{1,\ldots,N \} \times \{1,\ldots,N \}$ 
such that $X_k$ and $X_\ell$ are nearest neighbors. 
Thus
\begin{IEEEeqnarray}{c} 
\label{eqn:cardB2DTorus}
|\EE|=2N
\end{IEEEeqnarray}
for periodic boundary conditions.
A real coupling parameter 
$J_{k,\ell}$ is associated with each pair $(k,\ell) \in \EE$. 
The energy 
of a configuration $\x \in \calX^N$ is given by the Hamiltonian
\begin{IEEEeqnarray}{c}
\label{eqn:HamiltonianI}
\mathcal{H}(\x) = - \!\!\sum_{(k,\ell)\in \EE} J_{k, \ell} \cdot \delta(x_k - x_{\ell}),
%\mathcal{H}(\x) = - \!\!\sum_{(k,\ell)\in \EE} J_{k, \ell} \delta_{=}(x_k, x_{\ell}),
\end{IEEEeqnarray}
where $\delta(\cdot)$ is the Kronecker delta, which evaluates to one if its argument is zero, and to 
zero otherwise.

%%%%%%%%%%%%%%%%%%%%%%%%%%%%%%%%%%%%%%%%%%%%%%%%%%%%%%%%%%%%%
\begin{figure}[t]
\setlength{\unitlength}{0.87mm}
%\begin{minipage}{0.5\linewidth}
\centering
\begin{picture}(70,27)(0,-5)
%\put(0,-5){\dashbox(70,24){}}
\put(0,2.5){\line(1,0){12.5}}    \put(5,3.5){\pos{cb}{$X_1$}}
\put(12.5,0){\framebox(5,5){}}   \put(15,-1.25){\pos{ct}{$f_1$}}
 \put(15,15){\line(0,-1){10}}
 \put(15,15){\line(1,0){40}}     \put(35,16){\pos{cb}{$X_5$}}
 \put(55,15){\line(0,-1){10}}
\put(17.5,2.5){\line(1,0){15}}   \put(25,3.5){\pos{cb}{$X_2$}}
\put(32.5,0){\framebox(5,5){}}   \put(35,-1.25){\pos{ct}{$f_2$}}
\put(37.5,2.5){\line(1,0){15}}   \put(45,3.5){\pos{cb}{$X_3$}}
\put(52.5,0){\framebox(5,5){}}   \put(55,-1.25){\pos{ct}{$f_3$}}
\put(57.5,2.5){\line(1,0){12.5}}  \put(65,3.5){\pos{cb}{$X_4$}}
\end{picture}
\vspace{2\unitlength}
\caption{\label{fig:FGEx1}%
Normal factor graph of (\ref{eqn:FactEx1}).}
%\end{minipage}
%\end{figure}
\vspace{3\unitlength}
%\begin{figure}
%\begin{minipage}{0.5\linewidth}
\centering
\begin{picture}(70,25)(0,-4)
%\put(0,-4){\dashbox(70,25){}}
\put(0,2.5){\line(1,0){12.5}}      \put(5,3.5){\pos{cb}{$X_2$}}
\put(12.5,0){\framebox(5,5){}}     \put(15,-1.5){\pos{ct}{$g_1$}}
\put(17.5,2.5){\line(1,0){15}}     \put(25,3.5){\pos{cb}{$X_1$}}
\put(32.5,0){\framebox(5,5){$=$}}  \put(35,-1.25){\pos{ct}{$\I_=$}}
 \put(20,17.5){\line(1,0){12.5}}   \put(25,18.5){\pos{cb}{$X_3$}}
 \put(32.5,15){\framebox(5,5){}}   \put(38.75,17.5){\pos{cl}{$g_2$}}
 \put(35,15){\line(0,-1){10}}      \put(36,10){\pos{cl}{$X_1'$}}
\put(37.5,2.5){\line(1,0){15}}     \put(45,3.5){\pos{cb}{$X_1''$}}
\put(52.5,0){\framebox(5,5){}}     \put(55,-1.5){\pos{ct}{$g_3$}}
\put(57.5,2.5){\line(1,0){12.5}}   \put(65,3.5){\pos{cb}{$X_4$}}

\end{picture}
\vspace{2\unitlength}
\caption{\label{fig:FGEx2}%
Normal factor graph of (\ref{eqn:FactEx2}).}
%\end{minipage}
\end{figure}

%%%%%%%%%%%%%%%%%%%%%%%%%%%%%%%%%%%%%%%%%%%%%%%%%%%%%%%%%%

In this paper, we focus on ferromagnetic models, which are 
characterized by the condition $J_{k, \ell} \ge 0$ for all $(k, \ell) \in \EE$, 
i.e., configurations in which adjacent 
variables take on the same value have lower energy. 
The probability of a configuration $\x \in \calX^N$ is
given by the Boltzmann distribution
\begin{IEEEeqnarray}{c}
\label{eqn:Prob}
p_{\text{B}}(\x) = \frac{e^{-\beta\mathcal{H}(\x)}}{Z}.
%, \, \quad \forall \x \in \calX^{N}
%p_{\text{B}}(\x) = 
%\frac{1}{Z}\text{exp}\big(-\mathcal{H}(\x)/T\big) 
\end{IEEEeqnarray}

%Here $\beta = 1/k_\text{B}T$ denotes 
Here, $\beta$ denotes 
the inverse temperature and the normalization constant $Z$ 
is the partition function given by
\begin{IEEEeqnarray}{c}
\label{eqn:PartitionfunctionDef}
Z = \sum_{\x \in \calX^N} e^{-\beta\mathcal{H}(\x)},
\end{IEEEeqnarray}
where the sum runs over all the configurations~\cite{Baxter07}.

We will find it convenient to omit the parameter $\beta$ 
(i.e., we set $\beta = 1$) and to work with varying values of the coupling 
parameters $J_{k,\ell}$. 
%Hence, large values of $|J_{k,\ell}|$ correspond to 
In this set-up, large values of $J_{k,\ell}$ correspond to 
low temperature and small values correspond to high temperature.
In particular, the special case where $J_{k,\ell}=0$ for all $(k,\ell)\in\EE$ 
corresponds to infinite temperature.

We now let 
\begin{IEEEeqnarray}{rCl}
f(\x) & = & e^{-\mathcal{H}(\x)}\\
  & = & \prod_{(k,\ell)\in \EE} \kappa_{k, \ell}(x_k, x_{\ell}) \label{eqn:factorF}
\end{IEEEeqnarray}
with
\begin{equation} \label{eqn:IsingA}
\kappa_{k,\ell}(x_k, x_\ell) = \left\{ \begin{array}{ll}
       e^{J_{k,\ell}}, & \text{if $x_k = x_\ell$} \\
       1, & \text{otherwise.}
      \end{array}\right.   
\end{equation}

Thus (\ref{eqn:PartitionfunctionDef}) becomes 
%From (\ref{eqn:HamiltonianI})--(\ref{eqn:factorF}), $Z$ in (\ref{eqn:PartitionfunctionDef})
%can also be expressed as 
\begin{IEEEeqnarray}{c} 
\label{eqn:PartFunction}
Z = \sum_{\x \in \calX^N} f(\x),
\end{IEEEeqnarray} 
%which is of the form (\ref{eqn:defZf}).
in agreement with (\ref{eqn:defZf}).
%mehdi
%We will sometimes denote the partition function by $Z(J)$ to emphasize that it is a function of the 
%coupling parameters.
%mehdi

The factorization (\ref{eqn:factorF}) will be used in the next section.

\section{Primal Normal Factor Graph\\ of the Potts Model}
\label{sec:NFGD}

We use normal factor graphs as in~\cite{Forney:01},\cite{Lg:ifg2004},\cite{AY:2011, Forney:11}
(also called Forney factor graphs),
where variables are represented by edges and factors are represented by nodes/boxes.
%(rather than by variable nodes as in \cite{KFL:01}). 
(By contrast, factor graphs as in \cite{KFL:01} represent both factors and variables by nodes.)
For example, the factorization 
\begin{multline}
\label{eqn:FactEx1}
f(x_1,x_2,x_3,x_4,x_5) \\ = f_1(x_1,x_2,x_5) f_2(x_2,x_3) f_3(x_3,x_4,x_5)
\end{multline}
is represented by the normal factor graph in \Fig{fig:FGEx1}.
We say that a configuration $\x$ is \emph{valid} iff $f(\x)\neq 0$,
and we note that only valid configurations contribute to the partition function (\ref{eqn:defZf}).

As observed in~\cite{Forney:01}, in order to represent the variables by edges, each variable must be involved in only
one or two factors.
If some variable appears in more than two factors as, e.g., in 
\begin{IEEEeqnarray}{c} 
\label{eqn:FactEx2}
g(x_1,x_2,x_3,x_4) = g_1(x_1,x_2) g_2(x_1,x_3) g_3(x_1,x_4),
\end{IEEEeqnarray}
we introduce auxiliary variables $x_1'$ and $x_1''$, and define an additional equality indicator 
function
\begin{IEEEeqnarray}{c}
\label{eqn:DefEqFact}
%\Phi_=(x_1,x_1',x_1'') \eqdef [x_1=x_1'=x_1'']
\I_=(x_1,x_1',x_1'') \eqdef \delta(x_1 - x_1')\cdot\delta(x_1-x_1'')
\end{IEEEeqnarray}
(as shown in \Fig{fig:FGEx2})
such that $x_1=x_1'=x_1''$ in all valid configurations. The partition function
is not affected by such replications.

We next note that the Hamiltonian~(\ref{eqn:HamiltonianI}) and
factors~(\ref{eqn:IsingA}) can equivalently be written as a function 
of \mbox{$y_{k,\ell} = x_{k}-x_{\ell}$}. 
To simplify notation, we will henceforth denote the elements of $\EE$ by a single 
index variable $e \in \EE$ rather than by a vertex index pair $(k,\ell) \in \VV^2$, with adjacencies 
continuing to be determined by the graph $\G  = (\VV, \EE)$.

Hence, each factor~(\ref{eqn:IsingA}) can be written as
\begin{equation} 
\label{eqn:IsingKernelDualMod2}
\kappa_{e}(y_e) = \left\{ \begin{array}{ll}
      e^{J_e}, & \text{if $y_{e} = 0$} \\
      1, & \text{otherwise.}
  \end{array} \right.
\end{equation}

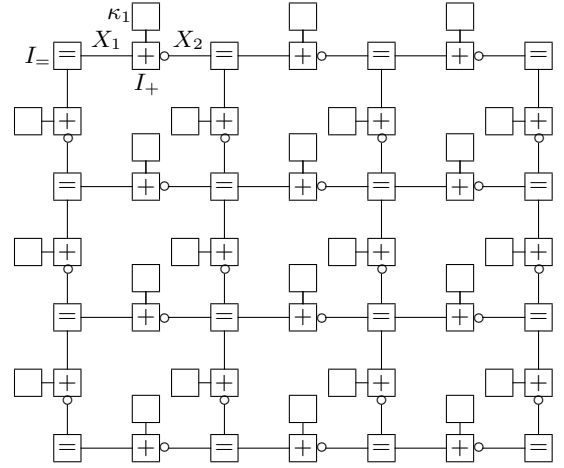
\begin{figure}[t!]
\setlength{\unitlength}{0.87mm}
\centering
\begin{picture}(82,73)(-6,0)
%\put(-6,0){\dashbox(82,73.5){}}
\small
\put(0,60){\framebox(4,4){$=$}} \put(-2.5,60.5){\pos{bc}{$I_=$}} 
\put(4,62){\line(1,0){8}}          \put(8,63){\pos{bc}{$X_{1}$}}
\put(12,60){\framebox(4,4){$+$}} \put(14.2,56){\pos{bc}{$I_+$}} 
%%%%%%%%%%%%%%%%%%%
\put(15.95,61){$\circ$}					 \put(20.5,63){\pos{bc}{$X_{2}$}}
\put(17.38,62){\line(1,0){6.6}}    %\put(20,63){\pos{bc}{$\tilde X_{1,2}'$}}
%%%%%%%%%%%%%%%%%%
\put(24,60){\framebox(4,4){$=$}}
\put(28,62){\line(1,0){8}}         %\put(32,63){\pos{bc}{$\tilde X_{2,3}$}}
\put(36,60){\framebox(4,4){$+$}}
%%%%%%%%%%%%%%%%%%
\put(39.95,61){$\circ$}
\put(41.38,62){\line(1,0){6.6}}    %\put(44,63){\pos{bc}{$\tilde X_{2,3}'$}}
%%%%%%%%%%%%%%%%%
\put(48,60){\framebox(4,4){$=$}}
\put(52,62){\line(1,0){8}}      
\put(60,60){\framebox(4,4){$+$}}
%%%%%%%%%%%%%%%%%
\put(63.95,61){$\circ$}
\put(65.38,62){\line(1,0){6.6}}
%%%%%%%%%%%%%%%%
\put(72,60){\framebox(4,4){$=$}}
%%%%%%%%%%%%%%%%%%%
%%%%%%%%%%%%%%%%%%%
\put(0,50){\framebox(4,4){$+$}}
%%%%%%%%%%%%%
\put(1.2,48.45){$\circ$}
%%%%%%%%%%%%%
\put(24,50){\framebox(4,4){$+$}}
%%%%%%%%%%%%%
\put(25.2,48.45){$\circ$}
%%%%%%%%%%%%%
\put(48,50){\framebox(4,4){$+$}}
%%%%%%%%%%%%%
\put(49.2,48.45){$\circ$}
%%%%%%%%%%%%%
\put(72,50){\framebox(4,4){$+$}}
%%%%%%%%%%%%%
\put(73.2,48.45){$\circ$}
%%%%%%%%%%%%%%%%%%
%%%%%%%%%%%%%%%%%%
\put(0,40){\framebox(4,4){$=$}}
\put(4,42){\line(1,0){8}}
\put(12,40){\framebox(4,4){$+$}}
%%%%%%%%%%%%%%%%%
\put(15.95,41.2){$\circ$}
\put(17.4,42){\line(1,0){6.6}}
%%%%%%%%%%%%%%%%%
\put(24,40){\framebox(4,4){$=$}}
\put(28,42){\line(1,0){8}}
\put(36,40){\framebox(4,4){$+$}}
%%%%%%%%%%%%%%%%%
\put(39.95,41.2){$\circ$}
\put(41.4,42){\line(1,0){6.6}}
%%%%%%%%%%%%%%%%%
\put(48,40){\framebox(4,4){$=$}}
\put(52,42){\line(1,0){8}}
\put(60,40){\framebox(4,4){$+$}}
%%%%%%%%%%%%%%%%%
\put(63.98,41.2){$\circ$}
\put(65.38,42){\line(1,0){6.6}}
%%%%%%%%%%%%%%%%%
\put(72,40){\framebox(4,4){$=$}}
%%%%%%%%%%%%%%%%%%%%%
%%%%%%%%%%%%%%%%%%%%%
\put(0,30){\framebox(4,4){$+$}}
%%%%%%%%%%%%%
\put(1.2,28.45){$\circ$}
%%%%%%%%%%%%%
\put(24,30){\framebox(4,4){$+$}}
%%%%%%%%%%%%%
\put(25.2,28.45){$\circ$}
%%%%%%%%%%%%%
\put(48,30){\framebox(4,4){$+$}}
%%%%%%%%%%%%%
\put(49.2,28.45){$\circ$}
%%%%%%%%%%%%
\put(72,30){\framebox(4,4){$+$}}
%%%%%%%%%%%%%
\put(73.2,28.45){$\circ$}
%%%%%%%%%%%%%%%%%%%%
%%%%%%%%%%%%%%%%%%%%
\put(0,20){\framebox(4,4){$=$}}
\put(4,22){\line(1,0){8}}
\put(12,20){\framebox(4,4){$+$}}
%%%%%%%%%%%%%%%%
\put(15.95,21){$\circ$}
\put(17.38,22){\line(1,0){6.6}}
%%%%%%%%%%%%%%%%
\put(24,20){\framebox(4,4){$=$}}
\put(28,22){\line(1,0){8}}
\put(36,20){\framebox(4,4){$+$}}
%%%%%%%%%%%%%%%
\put(39.95,21){$\circ$}
\put(41.38,22){\line(1,0){6.6}}
%%%%%%%%%%%%%%%
\put(48,20){\framebox(4,4){$=$}}
\put(52,22){\line(1,0){8}}
\put(60,20){\framebox(4,4){$+$}}
%%%%%%%%%%%%%%%
\put(63.98,21){$\circ$}
\put(65.38,22){\line(1,0){6.6}}
%%%%%%%%%%%%%%%
\put(72,20){\framebox(4,4){$=$}}
%%%%%%%%%%%%%%%%%%%%%
%%%%%%%%%%%%%%%%%%%%%
\put(0,10){\framebox(4,4){$+$}}
%%%%%%%%%%%%%
\put(1.1,8.45){$\circ$}
%%%%%%%%%%%%%
\put(24,10){\framebox(4,4){$+$}}
%%%%%%%%%%%%%
\put(25.1,8.45){$\circ$}
%%%%%%%%%%%%%
\put(48,10){\framebox(4,4){$+$}}
%%%%%%%%%%%%%
\put(49.1,8.45){$\circ$}
%%%%%%%%%%%%%
\put(72,10){\framebox(4,4){$+$}}
%%%%%%%%%%%%%%
\put(73.1,8.45){$\circ$}
%%%%%%%%%%%%%%%%%%%%
\put(0,0){\framebox(4,4){$=$}}
\put(12,0){\framebox(4,4){$+$}}
%%%%%%%%%%%%%
\put(15.95,1.2){$\circ$}
%%%%%%%%%%%%%
\put(24,0){\framebox(4,4){$=$}}
\put(36,0){\framebox(4,4){$+$}}
%%%%%%%%%%%%%
\put(39.96,1.2){$\circ$}
%%%%%%%%%%%%%
\put(48,0){\framebox(4,4){$=$}}
\put(60,0){\framebox(4,4){$+$}}
%%%%%%%%%%%%%
\put(63.96,1.2){$\circ$}
%%%%%%%%%%%%%
\put(72,0){\framebox(4,4){$=$}}
%%%%%%%%%%%%%%%%%%%%%
%%%
\put(14,64){\line(0,1){2}}
\put(38,64){\line(0,1){2}}
\put(62,64){\line(0,1){2}}
\put(12,66){\framebox(4,4){}}    \put(10,67.2){\pos{cb}{$\kappa_{1}$}}
\put(36,66){\framebox(4,4){}}    %\put(38,71){\pos{cb}{$\gamma_{2,3}$}}
\put(60,66){\framebox(4,4){}}
\put(14,44){\line(0,1){2}}
\put(38,44){\line(0,1){2}}
\put(62,44){\line(0,1){2}}
\put(12,46){\framebox(4,4){$$}}
\put(36,46){\framebox(4,4){$$}}
\put(60,46){\framebox(4,4){$$}}
\put(14,24){\line(0,1){2}}
\put(38,24){\line(0,1){2}}
\put(62,24){\line(0,1){2}}
\put(12,26){\framebox(4,4){$$}}
\put(36,26){\framebox(4,4){$$}}
\put(60,26){\framebox(4,4){$$}}
\put(14,4){\line(0,1){2}}
\put(38,4){\line(0,1){2}}
\put(62,4){\line(0,1){2}}
\put(12,6){\framebox(4,4){$$}}
\put(36,6){\framebox(4,4){$$}}
\put(60,6){\framebox(4,4){$$}}
\put(0,52){\line(-1,0){2}}
\put(24,52){\line(-1,0){2}}
\put(48,52){\line(-1,0){2}}
\put(72,52){\line(-1,0){2}}
\put(-6,50){\framebox(4,4){$$}}
\put(18,50){\framebox(4,4){$$}}
\put(42,50){\framebox(4,4){$$}}
\put(66,50){\framebox(4,4){$$}}
\put(0,32){\line(-1,0){2}}
\put(24,32){\line(-1,0){2}}
\put(48,32){\line(-1,0){2}}
\put(72,32){\line(-1,0){2}}
\put(-6,30){\framebox(4,4){$$}}
\put(18,30){\framebox(4,4){$$}}
\put(42,30){\framebox(4,4){$$}}
\put(66,30){\framebox(4,4){$$}}
\put(0,12){\line(-1,0){2}}
\put(24,12){\line(-1,0){2}}
\put(48,12){\line(-1,0){2}}
\put(72,12){\line(-1,0){2}}
\put(-6,10){\framebox(4,4){$$}}
\put(18,10){\framebox(4,4){$$}}
\put(42,10){\framebox(4,4){$$}}
\put(66,10){\framebox(4,4){$$}}

%\put(4,62){\line(1,0){8}}        
%\put(28,62){\line(1,0){8}}       
%\put(52,62){\line(1,0){8}}      
%\put(4,42){\line(1,0){8}}
%\put(28,42){\line(1,0){8}}
%\put(52,42){\line(1,0){8}}
%\put(4,22){\line(1,0){8}}
%\put(28,22){\line(1,0){8}}
%\put(52,22){\line(1,0){8}}

%
\put(14,64){\line(0,1){2}}
\put(38,64){\line(0,1){2}}
\put(62,64){\line(0,1){2}}
\put(14,44){\line(0,1){2}}
\put(38,44){\line(0,1){2}}
\put(62,44){\line(0,1){2}}
\put(14,24){\line(0,1){2}}
\put(38,24){\line(0,1){2}}
\put(62,24){\line(0,1){2}}

%%%%%%%%%%%%%%%%%%%%%%%%%%%%%%%%%
\put(2,54){\line(0,1){6}}
\put(2,48.7){\line(0,-1){4.6}}
\put(26,54){\line(0,1){6}}
\put(26,48.7){\line(0,-1){4.6}}
\put(50,54){\line(0,1){6}}
\put(50,48.7){\line(0,-1){4.6}}
\put(74,54){\line(0,1){6}}
\put(74,48.7){\line(0,-1){4.6}}
\put(2,34){\line(0,1){6}}
\put(2.0,28.7){\line(0,-1){4.6}}
\put(26,34){\line(0,1){6}}
\put(26.0,28.7){\line(0,-1){4.6}}
\put(50,34){\line(0,1){6}}
\put(50,28.7){\line(0,-1){4.6}}
\put(74,34){\line(0,1){6}}
\put(74,28.7){\line(0,-1){4.6}}
\put(2,14){\line(0,1){6}}
\put(2.0,8.7){\line(0,-1){4.6}}
\put(26,14){\line(0,1){6}}
\put(26.0,8.7){\line(0,-1){4.6}}
\put(50,14){\line(0,1){6}}
\put(50.0,8.7){\line(0,-1){4.6}}
\put(74,14){\line(0,1){6}}
\put(74,8.7){\line(0,-1){4.6}}
\put(4,2){\line(1,0){8}}
\put(17.38,2){\line(1,0){6.6}}
\put(28,2){\line(1,0){8}}
\put(41.38,2){\line(1,0){6.6}}
\put(52,2){\line(1,0){8}}
\put(65.38,2){\line(1,0){6.6}}
\end{picture}
\vspace{2ex}
\caption{\label{fig:2DGridM}%
Primal normal factor graph of the 2D Potts model. 
The empty boxes represent the factors~(\ref{eqn:IsingA}),
the boxes labeled ``$=$'' are equality indicator functions given by~(\ref{eqn:DefEqFact}),
the boxes labeled ``$+$'' 
are zero-sum indicator functions given by~(\ref{eqn:DefModqsumFact}), and 
the symbol ``$\circ$'' indicates a sign inversion.
The periodic boundary conditions are not shown.
}
\end{figure}

Applying factors~(\ref{eqn:IsingKernelDualMod2}) in~(\ref{eqn:factorF}), we can construct the 
primal normal factor graph of the Potts model as shown in~\Fig{fig:2DGridM}, in which the empty boxes 
represent~(\ref{eqn:IsingKernelDualMod2}), the boxes labeled ``$=$'' are instances of equality indicator functions as in~(\ref{eqn:DefEqFact}), the boxes labeled ``$+$'' 
are instances of the zero-sum indicator functions defined as
\begin{IEEEeqnarray}{c}
\label{eqn:DefModqsumFact}
\I_+(y_{e}, x_k, x_\ell) \eqdef \delta(y_{e}+x_k+ x_\ell), 
\end{IEEEeqnarray}
and in analogy with the logic NAND gate the symbol ``$\circ$'' is used to 
indicate a sign inversion. 
%See~\cite{Forney:01} for a similar 
%approach in the context of the dual of non-binary linear codes.
(Recall that all arithmetic manipulations are done modulo $q$.)

%%%%%%%%%%%%%%%%%%%%%%%%%%%%%%%%%%%%%%%%
%%%%%%%%%%%%%%%%%%%%%%%%%%%%%%%%%%%%%%%%%%%%%%%

For $q=2$, the Potts model is equivalent to the Ising model. However, the standard convention is 
to define the Hamiltonian of the model as
\begin{IEEEeqnarray}{c} 
\label{eqn:HamiltonianIsing}
%\kappa_{k, \ell}(x_k, x_{\ell}) \eqdef e^{J_{k, \ell}\cdot\delta(x_k - x_{\ell})}.
\mathcal{H}(\x) = - \!\!\sum_{(k,\ell)\in \EE} J_{k, \ell}\cdot\big(2\delta(x_k - x_{\ell}) - 1\big).
\end{IEEEeqnarray}
%\begin{equation} \label{eqn:IsingPrimal}
%%\kappa_{k, \ell}(x_k, x_{\ell}) \eqdef e^{J_{k, \ell}\cdot\delta(x_k - x_{\ell})}.
%\kappa_{k, \ell}(x_k, x_{\ell}) = e^{J_{k, \ell}\cdot([x_k = x_{\ell}] - [x_k \ne x_{\ell}])}.
%\end{equation}
Following a similar approach, we can obtain the primal normal factor graph of the 
Ising model (also shown in~\Fig{fig:2DGridM}), where the empty boxes represent factors given by
\begin{equation} 
\label{eqn:IsingPrimal}
\kappa_{e}(y_{e}) = \left\{ \begin{array}{ll}
      e^{J_{e}}, & \text{if $y_{e} = 0$} \\
      e^{-J_{e}}, & \text{if $y_{e} = 1$.}
  \end{array} \right.
\end{equation}
Note that, for the Ising model the ``$\circ$'' symbols are immaterial and can be removed 
from~\Fig{fig:2DGridM}.

In the special case of the 2D Ising model
with constant couplings $J_e = J$ and without an external field, the partition function is analytically 
available in the thermodynamic limit (i.e., as $N \to \infty$) from Onsager's solution~\cite{Onsager:44}. 
In Appendix~\ref{appsec:VarIsing}, we will use the analytical solution of the partition function 
to analyze the variance of Monte Carlo methods of this paper. For the (nonbinary) 2D Potts model, no such 
analytical solution for the partition 
function is yet available.

Next, we will describe the corresponding dual normal factor graphs of the models in this 
section.

\section{Dual Normal Factor Graph\\ of the Potts Model}
\label{sec:DualFGPotts}

The dual normal factor graph of some given (primal) normal factor graph
has the same topology as the primal normal factor graph, but all 
factors are replaced by their Fourier transforms (which includes replacing 
equality indicator functions by zero-sum indicator functions, and vice versa). 
See~\cite{Forney:01},\cite{Lg:ifg2004, AY:2011,FV:2011, AY:2014} for
more details. 

In the dual normal factor graph, all variables are replaced by their corresponding dual (frequency)
variables, which take values in the same alphabet
as the primal variables. We will use the tilde 
symbol to denote variables in the dual domain.
The dual normal factor graph has the same partition function 
as the primal normal factor graph, up to 
some known scale factor~\cite[Theorem 2]{AY:2011}.
%For more details, see~\cite[Theorem 2]{AY:2011},~\cite{Forney:11}.
We denote the partition function of the dual normal factor graph by $Z_\text{d}$.

%%%%%%%%%%%%%%%%%%%%%%%%%%%%%%%%%%%%%%%%%%%%%%%%%%%%%%%%%%%%%%%%%%%

\begin{figure}[t!!!!]
\setlength{\unitlength}{0.87mm}
\centering
\begin{picture}(82,73)(-6,0)
%\put(-6,0){\dashbox(82,73.5){}}
\small
\put(0,60){\framebox(4,4){$+$}} \put(-2.5,60.5){\pos{bc}{$I_+$}} 
\put(4,62){\line(1,0){8}}          \put(8,63){\pos{bc}{$\tilde Y_{1}$}}
\put(12,60){\framebox(4,4){$=$}}
%%%%%%%%%%%%%%%%%%%
\put(15.95,61){$\circ$}
\put(17.38,62){\line(1,0){6.6}}    \put(20.2,63){\pos{bc}{${-}\!\tilde Y_{1}$}}
%%%%%%%%%%%%%%%%%%
\put(24,60){\framebox(4,4){$+$}}
\put(28,62){\line(1,0){8}}         \put(14.2,56){\pos{bc}{$I_=$}} 
\put(36,60){\framebox(4,4){$=$}}
%%%%%%%%%%%%%%%%%%
\put(39.95,61){$\circ$}
\put(41.38,62){\line(1,0){6.6}}   % \put(44,63){\pos{bc}{$\tilde X_{2,3}'$}}
%%%%%%%%%%%%%%%%%
\put(48,60){\framebox(4,4){$+$}}
\put(52,62){\line(1,0){8}}      
\put(60,60){\framebox(4,4){$=$}}
%%%%%%%%%%%%%%%%%
\put(63.95,61){$\circ$}
\put(65.38,62){\line(1,0){6.6}}
%%%%%%%%%%%%%%%%
\put(72,60){\framebox(4,4){$+$}}
%%%%%%%%%%%%%%%%%%%
%%%%%%%%%%%%%%%%%%%
\put(0,50){\framebox(4,4){$=$}}
%%%%%%%%%%%%%
\put(1.2,48.45){$\circ$}
%%%%%%%%%%%%%
\put(24,50){\framebox(4,4){$=$}}
%%%%%%%%%%%%%
\put(25.2,48.45){$\circ$}
%%%%%%%%%%%%%
\put(48,50){\framebox(4,4){$=$}}
%%%%%%%%%%%%%
\put(49.2,48.45){$\circ$}
%%%%%%%%%%%%%
\put(72,50){\framebox(4,4){$=$}}
%%%%%%%%%%%%%
\put(73.2,48.45){$\circ$}
%%%%%%%%%%%%%%%%%%
%%%%%%%%%%%%%%%%%%
\put(0,40){\framebox(4,4){$+$}}
\put(4,42){\line(1,0){8}}
\put(12,40){\framebox(4,4){$=$}}
%%%%%%%%%%%%%%%%%
\put(15.95,41.2){$\circ$}
\put(17.4,42){\line(1,0){6.6}}
%%%%%%%%%%%%%%%%%
\put(24,40){\framebox(4,4){$+$}}
\put(28,42){\line(1,0){8}}
\put(36,40){\framebox(4,4){$=$}}
%%%%%%%%%%%%%%%%%
\put(39.95,41.2){$\circ$}
\put(41.4,42){\line(1,0){6.6}}
%%%%%%%%%%%%%%%%%
\put(48,40){\framebox(4,4){$+$}}
\put(52,42){\line(1,0){8}}
\put(60,40){\framebox(4,4){$=$}}
%%%%%%%%%%%%%%%%%
\put(63.98,41.2){$\circ$}
\put(65.38,42){\line(1,0){6.6}}
%%%%%%%%%%%%%%%%%
\put(72,40){\framebox(4,4){$+$}}
%%%%%%%%%%%%%%%%%%%%%
%%%%%%%%%%%%%%%%%%%%%
\put(0,30){\framebox(4,4){$=$}}
%%%%%%%%%%%%%
\put(1.2,28.45){$\circ$}
%%%%%%%%%%%%%
\put(24,30){\framebox(4,4){$=$}}
%%%%%%%%%%%%%
\put(25.2,28.45){$\circ$}
%%%%%%%%%%%%%
\put(48,30){\framebox(4,4){$=$}}
%%%%%%%%%%%%%
\put(49.2,28.45){$\circ$}
%%%%%%%%%%%%
\put(72,30){\framebox(4,4){$=$}}
%%%%%%%%%%%%%
\put(73.2,28.45){$\circ$}
%%%%%%%%%%%%%%%%%%%%
%%%%%%%%%%%%%%%%%%%%
\put(0,20){\framebox(4,4){$+$}}
\put(4,22){\line(1,0){8}}
\put(12,20){\framebox(4,4){$=$}}
%%%%%%%%%%%%%%%%
\put(15.95,21){$\circ$}
\put(17.38,22){\line(1,0){6.6}}
%%%%%%%%%%%%%%%%
\put(24,20){\framebox(4,4){$+$}}
\put(28,22){\line(1,0){8}}
\put(36,20){\framebox(4,4){$=$}}
%%%%%%%%%%%%%%%
\put(39.95,21){$\circ$}
\put(41.38,22){\line(1,0){6.6}}
%%%%%%%%%%%%%%%
\put(48,20){\framebox(4,4){$+$}}
\put(52,22){\line(1,0){8}}
\put(60,20){\framebox(4,4){$=$}}
%%%%%%%%%%%%%%%
\put(63.98,21){$\circ$}
\put(65.38,22){\line(1,0){6.6}}
%%%%%%%%%%%%%%%
\put(72,20){\framebox(4,4){$+$}}
%%%%%%%%%%%%%%%%%%%%%
%%%%%%%%%%%%%%%%%%%%%
\put(0,10){\framebox(4,4){$=$}}
%%%%%%%%%%%%%
\put(1.1,8.45){$\circ$}
%%%%%%%%%%%%%
\put(24,10){\framebox(4,4){$=$}}
%%%%%%%%%%%%%
\put(25.1,8.45){$\circ$}
%%%%%%%%%%%%%
\put(48,10){\framebox(4,4){$=$}}
%%%%%%%%%%%%%
\put(49.1,8.45){$\circ$}
%%%%%%%%%%%%%
\put(72,10){\framebox(4,4){$=$}}
%%%%%%%%%%%%%%
\put(73.1,8.45){$\circ$}
%%%%%%%%%%%%%%%%%%%%
\put(0,0){\framebox(4,4){$+$}}
\put(12,0){\framebox(4,4){$=$}}
%%%%%%%%%%%%%
\put(15.95,1.2){$\circ$}
%%%%%%%%%%%%%
\put(24,0){\framebox(4,4){$+$}}
\put(36,0){\framebox(4,4){$=$}}
%%%%%%%%%%%%%
\put(39.96,1.2){$\circ$}
%%%%%%%%%%%%%
\put(48,0){\framebox(4,4){$+$}}
\put(60,0){\framebox(4,4){$=$}}
%%%%%%%%%%%%%
\put(63.96,1.2){$\circ$}
%%%%%%%%%%%%%
\put(72,0){\framebox(4,4){$+$}}
%%%%%%%%%%%%%%%%%%%%%
%%%
\put(14,64){\line(0,1){2}}
\put(38,64){\line(0,1){2}}
\put(62,64){\line(0,1){2}}
\put(12,66){\framebox(4,4){}}    \put(10,67.2){\pos{cb}{$\gamma_{1}$}}
\put(36,66){\framebox(4,4){}}    %\put(38,71){\pos{cb}{$\gamma_{2,3}$}}
\put(60,66){\framebox(4,4){}}
\put(14,44){\line(0,1){2}}
\put(38,44){\line(0,1){2}}
\put(62,44){\line(0,1){2}}
\put(12,46){\framebox(4,4){$$}}
\put(36,46){\framebox(4,4){$$}}
\put(60,46){\framebox(4,4){$$}}
\put(14,24){\line(0,1){2}}
\put(38,24){\line(0,1){2}}
\put(62,24){\line(0,1){2}}
\put(12,26){\framebox(4,4){$$}}
\put(36,26){\framebox(4,4){$$}}
\put(60,26){\framebox(4,4){$$}}
\put(14,4){\line(0,1){2}}
\put(38,4){\line(0,1){2}}
\put(62,4){\line(0,1){2}}
\put(12,6){\framebox(4,4){$$}}
\put(36,6){\framebox(4,4){$$}}
\put(60,6){\framebox(4,4){$$}}
\put(0,52){\line(-1,0){2}}
\put(24,52){\line(-1,0){2}}
\put(48,52){\line(-1,0){2}}
\put(72,52){\line(-1,0){2}}
\put(-6,50){\framebox(4,4){$$}}
\put(18,50){\framebox(4,4){$$}}
\put(42,50){\framebox(4,4){$$}}
\put(66,50){\framebox(4,4){$$}}
\put(0,32){\line(-1,0){2}}
\put(24,32){\line(-1,0){2}}
\put(48,32){\line(-1,0){2}}
\put(72,32){\line(-1,0){2}}
\put(-6,30){\framebox(4,4){$$}}
\put(18,30){\framebox(4,4){$$}}
\put(42,30){\framebox(4,4){$$}}
\put(66,30){\framebox(4,4){$$}}
\put(0,12){\line(-1,0){2}}
\put(24,12){\line(-1,0){2}}
\put(48,12){\line(-1,0){2}}
\put(72,12){\line(-1,0){2}}
\put(-6,10){\framebox(4,4){$$}}
\put(18,10){\framebox(4,4){$$}}
\put(42,10){\framebox(4,4){$$}}
\put(66,10){\framebox(4,4){$$}}

%\put(4,62){\line(1,0){8}}        
%\put(28,62){\line(1,0){8}}       
%\put(52,62){\line(1,0){8}}      
%\put(4,42){\line(1,0){8}}
%\put(28,42){\line(1,0){8}}
%\put(52,42){\line(1,0){8}}
%\put(4,22){\line(1,0){8}}
%\put(28,22){\line(1,0){8}}
%\put(52,22){\line(1,0){8}}

%
\put(14,64){\line(0,1){2}}
\put(38,64){\line(0,1){2}}
\put(62,64){\line(0,1){2}}
\put(14,44){\line(0,1){2}}
\put(38,44){\line(0,1){2}}
\put(62,44){\line(0,1){2}}
\put(14,24){\line(0,1){2}}
\put(38,24){\line(0,1){2}}
\put(62,24){\line(0,1){2}}

%%%%%%%%%%%%%%%%%%%%%%%%%%%%%%%%%
\put(2,54){\line(0,1){6}}
\put(2,48.7){\line(0,-1){4.6}}
\put(26,54){\line(0,1){6}}
\put(26,48.7){\line(0,-1){4.6}}
\put(50,54){\line(0,1){6}}
\put(50,48.7){\line(0,-1){4.6}}
\put(74,54){\line(0,1){6}}
\put(74,48.7){\line(0,-1){4.6}}
\put(2,34){\line(0,1){6}}
\put(2.0,28.7){\line(0,-1){4.6}}
\put(26,34){\line(0,1){6}}
\put(26.0,28.7){\line(0,-1){4.6}}
\put(50,34){\line(0,1){6}}
\put(50,28.7){\line(0,-1){4.6}}
\put(74,34){\line(0,1){6}}
\put(74,28.7){\line(0,-1){4.6}}
\put(2,14){\line(0,1){6}}
\put(2.0,8.7){\line(0,-1){4.6}}
\put(26,14){\line(0,1){6}}
\put(26.0,8.7){\line(0,-1){4.6}}
\put(50,14){\line(0,1){6}}
\put(50.0,8.7){\line(0,-1){4.6}}
\put(74,14){\line(0,1){6}}
\put(74,8.7){\line(0,-1){4.6}}
\put(4,2){\line(1,0){8}}
\put(17.38,2){\line(1,0){6.6}}
\put(28,2){\line(1,0){8}}
\put(41.38,2){\line(1,0){6.6}}
\put(52,2){\line(1,0){8}}
\put(65.38,2){\line(1,0){6.6}}
\end{picture}
\vspace{2ex}
\caption{\label{fig:2DGridDM}%
Dual normal factor graph of the 2D Potts model. 
%The boxes labeled ``$=$'' are equality constraints as in (\ref{eqn:DefEqFact}) 
%and the added little circle denotes a sign inversion (mod $q$). 
The empty boxes represent the factors~(\ref{eqn:IsingKernelDual}),
the boxes labeled ``$=$'' are equality indicator functions given by~(\ref{eqn:DefEqFact}), the boxes 
labeled ``$+$'' are zero-sum indicator functions given by~(\ref{eqn:DefModqsumFact}), and 
the symbol ``$\circ$'' indicates a sign inversion. 
%Note that $\tilde X_e' = - \tilde X_e$ in every valid configuration.
The periodic boundary conditions are not shown.
}
\end{figure}
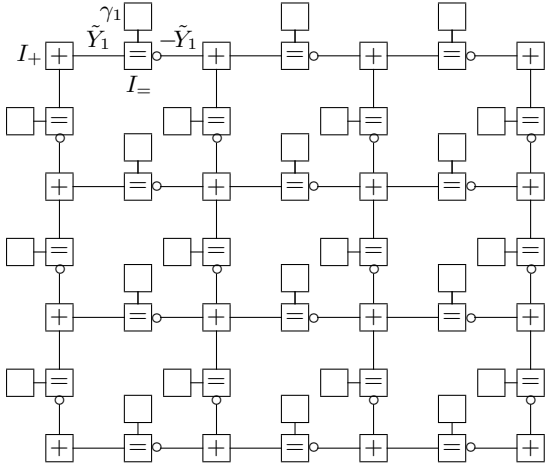

%%%%%%%%%%%%%%%%%%%%%%%%%%%%%%%%%%%%

Following~\cite{AY:2014}, we can obtain the dual normal factor graph of the 
2D Potts model %\Fig{fig:2DGridM} is
as shown in \Fig{fig:2DGridDM}, in which the empty boxes represent factors 
that are the 1D Fourier transforms %~\cite{Brace:1999}
%represent the Fourier transform 
of~(\ref{eqn:IsingKernelDualMod2}) given by 
\begin{IEEEeqnarray}{c}
\label{eqn:FourierEdgeFactor}
\gamma_{e}(\tilde{y}_{e}) = \sum_{y_e \in \calX}\kappa_{e}(y_{e}) e^{-\mathrm{i}2\pi y_e \tilde y_e /q}.
\end{IEEEeqnarray}

Thus
\begin{equation} 
\label{eqn:IsingKernelDual}
\gamma_{e}(\tilde{y}_{e}) = \left\{ \begin{array}{ll}
      e^{J_{e}} -1 + q, & \text{if $\tilde{y}_{e} = 0$} \\
     e^{J_{e}} -1, & \text{otherwise,}
  \end{array} \right.
\end{equation}
%where $\mathrm{i}$ denotes the unit imaginary number. 
which is nonnegative 
due to the ferromagnetic 
assumption (i.e.,  \mbox{$J_{e} \ge 0$}).

Similarly, we can obtain the dual normal factor graph of the 2D
Ising model (shown in~\Fig{fig:2DGridDM}), where the empty 
boxes represent factors as in
%represent the Fourier transform 
\begin{equation} 
\label{eqn:IsingDualAdding}
\gamma_{e}(\tilde{y}_{e}) = \left\{ \begin{array}{ll}
      2\cosh(J_{e}), & \text{if $\tilde{y}_{e} = 0$} \\
      2\sinh(J_{e}), & \text{if $\tilde{y}_{e} = 1,$}
  \end{array} \right.
\end{equation}
which is the 1D Fourier transform of~(\ref{eqn:IsingPrimal}). Notice that~(\ref{eqn:IsingDualAdding}) is 
also nonnegative due to ferromagnetic assumption.
Again, for the Ising model the ``$\circ$'' symbols can be safely removed from the dual normal factor graph.

The partition function of the dual normal factor graph is thus
\begin{IEEEeqnarray}{c} 
\label{eqn:DualPartitionSum}
Z_\text{d} = \sum_{\text{valid $\tilde\y$}}\, \prod_{e\in\EE} \gamma_e(\tilde{y}_e),
\end{IEEEeqnarray}
where the sum runs over all valid configurations in the dual normal factor graph.

%%%%%%%%%%%%%%%%%%%%%%%%%%%%%%%%%%%%%%%%%%%%%%%%%%%%%%%
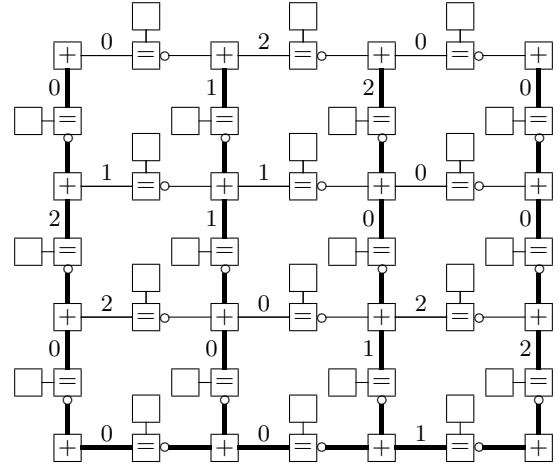
\begin{figure}[t]
\setlength{\unitlength}{0.87mm}
\centering
\begin{picture}(73,72.5)(0,0)
\small
\put(8,63){\pos{bc}{\bf {$0$}}}
\put(32,63){\pos{bc}{\bf {$2$}}}
\put(56,63){\pos{bc}{\bf {$0$}}}
\put(8,43){\pos{bc}{\bf {$1$}}}
\put(32,43){\pos{bc}{\bf {$1$}}}
\put(56,43){\pos{bc}{\bf {$0$}}}
\put(8,23){\pos{bc}{$2$}}
\put(32,23){\pos{bc}{$0$}}
\put(56,23){\pos{bc}{$2$}}
\put(0,56){\pos{bc}{$0$}}
\put(24,56){\pos{bc}{$1$}}
\put(48,56){\pos{bc}{$2$}}
\put(72,56){\pos{bc}{$0$}}
\put(0,36){\pos{bc}{$2$}}
\put(24,36){\pos{bc}{$1$}}
\put(48,36){\pos{bc}{$0$}}
\put(72,36){\pos{bc}{$0$}}
\put(0,16){\pos{bc}{$0$}}
\put(24,16){\pos{bc}{$0$}}
\put(48,16){\pos{bc}{$1$}}
\put(72,16){\pos{bc}{$2$}}
\put(8,3){\pos{bc}{\bf {$0$}}}
\put(32,3){\pos{bc}{\bf {$0$}}}
\put(56,3){\pos{bc}{\bf {$1$}}}

\put(0,60){\framebox(4,4){$+$}}
\put(4,62){\line(1,0){8}}        
\put(12,60){\framebox(4,4){$=$}}
%%%%%%%%%%%%%%%%%%%
\put(15.95,61){$\circ$}
\put(17.38,62){\line(1,0){6.6}}
%%%%%%%%%%%%%%%%%%
\put(24,60){\framebox(4,4){$+$}}
\put(28,62){\line(1,0){8}}       
\put(36,60){\framebox(4,4){$=$}}
%%%%%%%%%%%%%%%%%%
\put(39.95,61){$\circ$}
\put(41.38,62){\line(1,0){6.6}}
%%%%%%%%%%%%%%%%%
\put(48,60){\framebox(4,4){$+$}}
\put(52,62){\line(1,0){8}}      
\put(60,60){\framebox(4,4){$=$}}
%%%%%%%%%%%%%%%%%
\put(63.95,61){$\circ$}
\put(65.38,62){\line(1,0){6.6}}
%%%%%%%%%%%%%%%%
\put(72,60){\framebox(4,4){$+$}}
%%%%%%%%%%%%%%%%%%%
%%%%%%%%%%%%%%%%%%%
\put(0,50){\framebox(4,4){$=$}}
%%%%%%%%%%%%%
\put(1.2,48.45){$\circ$}
%%%%%%%%%%%%%
\put(24,50){\framebox(4,4){$=$}}
%%%%%%%%%%%%%
\put(25.2,48.45){$\circ$}
%%%%%%%%%%%%%
\put(48,50){\framebox(4,4){$=$}}
%%%%%%%%%%%%%
\put(49.2,48.45){$\circ$}
%%%%%%%%%%%%%
\put(72,50){\framebox(4,4){$=$}}
%%%%%%%%%%%%%
\put(73.2,48.45){$\circ$}
%%%%%%%%%%%%%%%%%%
%%%%%%%%%%%%%%%%%%
\put(0,40){\framebox(4,4){$+$}}
\put(4,42){\line(1,0){8}}
\put(12,40){\framebox(4,4){$=$}}
%%%%%%%%%%%%%%%%%
\put(15.95,41.2){$\circ$}
\put(17.4,42){\line(1,0){6.6}}
%%%%%%%%%%%%%%%%%
\put(24,40){\framebox(4,4){$+$}}
\put(28,42){\line(1,0){8}}
\put(36,40){\framebox(4,4){$=$}}
%%%%%%%%%%%%%%%%%
\put(39.95,41.2){$\circ$}
\put(41.4,42){\line(1,0){6.6}}
%%%%%%%%%%%%%%%%%
\put(48,40){\framebox(4,4){$+$}}
\put(52,42){\line(1,0){8}}
\put(60,40){\framebox(4,4){$=$}}
%%%%%%%%%%%%%%%%%
\put(63.98,41.2){$\circ$}
\put(65.38,42){\line(1,0){6.6}}
%%%%%%%%%%%%%%%%%
\put(72,40){\framebox(4,4){$+$}}
%%%%%%%%%%%%%%%%%%%%%
%%%%%%%%%%%%%%%%%%%%%
\put(0,30){\framebox(4,4){$=$}}
%%%%%%%%%%%%%
\put(1.2,28.45){$\circ$}
%%%%%%%%%%%%%
\put(24,30){\framebox(4,4){$=$}}
%%%%%%%%%%%%%
\put(25.2,28.45){$\circ$}
%%%%%%%%%%%%%
\put(48,30){\framebox(4,4){$=$}}
%%%%%%%%%%%%%
\put(49.2,28.45){$\circ$}
%%%%%%%%%%%%
\put(72,30){\framebox(4,4){$=$}}
%%%%%%%%%%%%%
\put(73.2,28.45){$\circ$}
%%%%%%%%%%%%%%%%%%%%
%%%%%%%%%%%%%%%%%%%%
\put(0,20){\framebox(4,4){$+$}}
\put(4,22){\line(1,0){8}}
\put(12,20){\framebox(4,4){$=$}}
%%%%%%%%%%%%%%%%
\put(15.95,21){$\circ$}
\put(17.38,22){\line(1,0){6.6}}
%%%%%%%%%%%%%%%%
\put(24,20){\framebox(4,4){$+$}}
\put(28,22){\line(1,0){8}}
\put(36,20){\framebox(4,4){$=$}}
%%%%%%%%%%%%%%%
\put(39.95,21){$\circ$}
\put(41.38,22){\line(1,0){6.6}}
%%%%%%%%%%%%%%%
\put(48,20){\framebox(4,4){$+$}}
\put(52,22){\line(1,0){8}}
\put(60,20){\framebox(4,4){$=$}}
%%%%%%%%%%%%%%%
\put(63.98,21){$\circ$}
\put(65.38,22){\line(1,0){6.6}}
%%%%%%%%%%%%%%%
\put(72,20){\framebox(4,4){$+$}}
%%%%%%%%%%%%%%%%%%%%%
%%%%%%%%%%%%%%%%%%%%%
\put(0,10){\framebox(4,4){$=$}}
%%%%%%%%%%%%%
\put(1.1,8.45){$\circ$}
%%%%%%%%%%%%%
\put(24,10){\framebox(4,4){$=$}}
%%%%%%%%%%%%%
\put(25.1,8.45){$\circ$}
%%%%%%%%%%%%%
\put(48,10){\framebox(4,4){$=$}}
%%%%%%%%%%%%%
\put(49.1,8.45){$\circ$}
%%%%%%%%%%%%%
\put(72,10){\framebox(4,4){$=$}}
%%%%%%%%%%%%%%
\put(73.1,8.45){$\circ$}
%%%%%%%%%%%%%%%%%%%%
\put(0,0){\framebox(4,4){$+$}}
\put(12,0){\framebox(4,4){$=$}}
%%%%%%%%%%%%%
\put(15.95,1.2){$\circ$}
%%%%%%%%%%%%%
\put(24,0){\framebox(4,4){$+$}}
\put(36,0){\framebox(4,4){$=$}}
%%%%%%%%%%%%%
\put(39.96,1.2){$\circ$}
%%%%%%%%%%%%%
\put(48,0){\framebox(4,4){$+$}}
\put(60,0){\framebox(4,4){$=$}}
%%%%%%%%%%%%%
\put(63.96,1.2){$\circ$}
%%%%%%%%%%%%%
\put(72,0){\framebox(4,4){$+$}}
%%%%%%%%%%%%%%%%%%%%%
%%%
\put(14,64){\line(0,1){2}}
\put(38,64){\line(0,1){2}}
\put(62,64){\line(0,1){2}}
\put(12,66){\framebox(4,4){$$}}
\put(36,66){\framebox(4,4){$$}}
\put(60,66){\framebox(4,4){$$}}
\put(14,44){\line(0,1){2}}
\put(38,44){\line(0,1){2}}
\put(62,44){\line(0,1){2}}
\put(12,46){\framebox(4,4){$$}}
\put(36,46){\framebox(4,4){$$}}
\put(60,46){\framebox(4,4){$$}}
\put(14,24){\line(0,1){2}}
\put(38,24){\line(0,1){2}}
\put(62,24){\line(0,1){2}}
\put(12,26){\framebox(4,4){$$}}
\put(36,26){\framebox(4,4){$$}}
\put(60,26){\framebox(4,4){$$}}
\put(14,4){\line(0,1){2}}
\put(38,4){\line(0,1){2}}
\put(62,4){\line(0,1){2}}
\put(12,6){\framebox(4,4){$$}}
\put(36,6){\framebox(4,4){$$}}
\put(60,6){\framebox(4,4){$$}}
\put(0,52){\line(-1,0){2}}
\put(24,52){\line(-1,0){2}}
\put(48,52){\line(-1,0){2}}
\put(72,52){\line(-1,0){2}}
\put(-6,50){\framebox(4,4){$$}}
\put(18,50){\framebox(4,4){$$}}
\put(42,50){\framebox(4,4){$$}}
\put(66,50){\framebox(4,4){$$}}
\put(0,32){\line(-1,0){2}}
\put(24,32){\line(-1,0){2}}
\put(48,32){\line(-1,0){2}}
\put(72,32){\line(-1,0){2}}
\put(-6,30){\framebox(4,4){$$}}
\put(18,30){\framebox(4,4){$$}}
\put(42,30){\framebox(4,4){$$}}
\put(66,30){\framebox(4,4){$$}}
\put(0,12){\line(-1,0){2}}
\put(24,12){\line(-1,0){2}}
\put(48,12){\line(-1,0){2}}
\put(72,12){\line(-1,0){2}}
\put(-6,10){\framebox(4,4){$$}}
\put(18,10){\framebox(4,4){$$}}
\put(42,10){\framebox(4,4){$$}}
\put(66,10){\framebox(4,4){$$}}

\put(4,62){\line(1,0){8}}        

\put(28,62){\line(1,0){8}}       

\put(52,62){\line(1,0){8}}      

\put(4,42){\line(1,0){8}}

\put(28,42){\line(1,0){8}}

\put(52,42){\line(1,0){8}}

\put(4,22){\line(1,0){8}}

\put(28,22){\line(1,0){8}}

\put(52,22){\line(1,0){8}}

\put(14,64){\line(0,1){2}}
\put(38,64){\line(0,1){2}}
\put(62,64){\line(0,1){2}}
\put(14,44){\line(0,1){2}}
\put(38,44){\line(0,1){2}}
\put(62,44){\line(0,1){2}}
\put(14,24){\line(0,1){2}}
\put(38,24){\line(0,1){2}}
\put(62,24){\line(0,1){2}}

%%%%%%%%%%%%%%%%%%%%%%%%%%%%%%%%%%%%%%%%%

%%%%%%%%%%%%%%%%%%
\linethickness{0.66mm}
%%%%%%%%%%%%%%%%%%
\put(2,54){\line(0,1){6}}
\put(2,48.7){\line(0,-1){4.6}}
\put(26,54){\line(0,1){6}}
\put(26,48.7){\line(0,-1){4.6}}
\put(50,54){\line(0,1){6}}
\put(50,48.7){\line(0,-1){4.6}}
\put(74,54){\line(0,1){6}}
\put(74,48.7){\line(0,-1){4.6}}
\put(2,34){\line(0,1){6}}
\put(2.0,28.7){\line(0,-1){4.6}}
\put(26,34){\line(0,1){6}}
\put(26.0,28.7){\line(0,-1){4.6}}
\put(50,34){\line(0,1){6}}
\put(50,28.7){\line(0,-1){4.6}}
\put(74,34){\line(0,1){6}}
\put(74,28.7){\line(0,-1){4.6}}
\put(2,14){\line(0,1){6}}
\put(2.0,8.7){\line(0,-1){4.6}}
\put(26,14){\line(0,1){6}}
\put(26.0,8.7){\line(0,-1){4.6}}
\put(50,14){\line(0,1){6}}
\put(50.0,8.7){\line(0,-1){4.6}}
\put(74,14){\line(0,1){6}}
\put(74,8.7){\line(0,-1){4.6}}
\put(4,2){\line(1,0){8}}
\put(17.38,2){\line(1,0){6.6}}
\put(28,2){\line(1,0){8}}
\put(41.38,2){\line(1,0){6.6}}
\put(52,2){\line(1,0){8}}
\put(65.38,2){\line(1,0){6.6}}
\end{picture}
\vspace{2ex}
\caption{\label{fig:2DGridDPart}%
A partitioning of the variables $\tilde \Y = \{ \tilde Y_e: e\in\EE \}$
in Fig.~\ref{fig:2DGridDM}
as in Section~\ref{sec:SpanningTree}. %(but assuming no boundary conditions). 
The edges in $\T$ (drawn with thick edges) 
%represent the variables $\tilde\X_B$, which are linearly 
represent the variables $\tilde\Y_\T$, which are linearly 
%dependent of the remaining variables $\tilde\X_A$. 
dependent on the remaining variables $\tilde\Y_\F$. 
This spanning tree works both with and without periodic boundary conditions.
Also shown is an example of a valid configuration for $q=3$  
(assuming no boundary conditions).
%
%The thin edges represent the variables $\tilde \X_A$ 
%and thick edges represent $\tilde \X_B$. 
%Here, $\tilde \X_B$ is a 
%linear combination of $\tilde \X_A$. A numerical example of 
%a valid configuration in a \mbox{3-state} Potts model is shown on the edges. 
}
\end{figure}

%%%%%%%%%%%%%%%%%%%%%%%%%%%%%%%%%%%%%%%%%%%%%%%%%%%%%%%%%%%%%%%%%

For the dual 2D Potts and Ising models with factors as in~(\ref{eqn:IsingKernelDual})
and~(\ref{eqn:IsingDualAdding}), 
and with periodic boundary conditions, it holds that\footnote{In general, the scale factor $\alpha(\G) = Z_{\text{d}}/ Z$
depends on the topology of $\G$ and on the local scale factors used in the Fourier transforms. In our setup, the scale factor is given by $\alpha(\G) = q^{|\EE|-|\VV|}$. 
For example, in a 2D torus $|\EE| = 2N$, and therefore $\alpha(\G) = q^{N}$ as in~(\ref{eqn:NDual}); in a 1D model with periodic boundary conditions 
$|\EE| = |\VV|$, and thus $\alpha(\G) = 1$. For more details, see~\cite{Mo:2017Arxiv},\cite[Section 3.3]{Forney:16}.}
\begin{equation}\label{eqn:NDual}
Z_\text{d} = q^{N}Z
\end{equation}
see~\cite{AY:2011}.

\Fig{fig:2DGridDM} is the basis of the Monte Carlo algorithms of this paper to estimate $Z_\text{d}$.
The estimates are then used to compute an estimate of $Z$ via~(\ref{eqn:NDual}).

\subsection{Independent Variables and Spanning Trees in the Dual Normal Factor Graph}
\label{sec:SpanningTree}

In our Monte Carlo methods, we will use partitions of $\G$ into two disjoint subsets 
$\G = \T \cup \F$  such that 
%$\B_\T$ has no cycles and reaches every zero-sum node/factor, 
$\T$ is a spanning tree (that reaches every zero-sum factor) and $\F$ is the corresponding cospanning tree, 
as illustrated in \Fig{fig:2DGridDPart}. 
 The edges of $\T$ are called the \emph{branches} 
and the edges of $\F$ are called the \emph{chords} of $\G$ with respect to $\T$.
%, or simply the chords of $\T$. 
Although $\T$ is always without cycles, $\F$ need not 
be cycle-free.
Any such partition induces a corresponding partition 
of the variables $\tilde \Y = \{ \tilde Y_e: e\in\EE \}$
into $\tilde\Y_\F$ and $\tilde\Y_\T$.

\begin{proposition} \label{prop:cutset}
Consider a valid configuration in the dual normal factor graph of the Potts model. Suppose
variables $\tilde Y_1, \tilde Y_2, \ldots$ form a cutset in the dual normal factor graph, then it holds that
\begin{IEEEeqnarray}{c}
\label{eqn:cut}
\sum_\text{$e \in$ Cutset} \tilde y_{e} = 0
%\bigoplus_{m=1}^c Y_{m} = 0
\end{IEEEeqnarray}
%attached to the unlabeled
%boxes and to the XOR factors in the cycle, 
%, it holds that
%, we have
%\begin{lemma}
\eproofnegspace
\end{proposition}

\begin{trivlist}
\item \emph{Proof. } 
Removing all the edges that represent variables $\tilde Y_1, \tilde Y_2, \ldots$ 
%(and their corresponding edges that represent variables $-\tilde Y_1, -\tilde Y_2, \ldots$) 
partitions $\G$ into $\G_1\cup \,\G_2$.
Suppose in $\G_1$ we 
write down the equations associated with all the zero-sum indicator factors. 
Since each variable, say $\tilde y_t$ for $t \in \EE_1$ appears twice in the 
summation, once as $\tilde y_t$ and once as $-\tilde y_t$ (see~\Fig{fig:2DGridDM}), the sum over all 
these equations is equal to zero. Furthermore,
the same sum in $\G$ is 
equal to $\sum_\text{$e \in$ Cutset} \tilde y_{e}$. This completes the proof. 
\hfill$\blacksquare$
%\end{IEEEproof}
\end{trivlist}
The proof follows along the same lines in $\G_2$. 
For more details, see~\cite{Mo:2017Arxiv},\cite[Section 2.5]{Forney:16}.

Removing a branch $b \in \T$ partitions $\T=\T_1 \cup \T_2$. The edges that connect $\T_1$ and 
$\T_2$ form a unique cutset in $\G$, called the fundamental cutset belonging to $b$. Each fundamental 
cutset has exactly one branch of $\T$ that does not appear 
in any other fundamental cutset, along with edges (chords) that belong to $\F$. Indeed, each spanning tree defines a set of $|\T|$ fundamental cutsets: 
one for each branch of the spanning tree~\cite[Chapter 2]{Bolob}. 
According to Proposition~\ref{prop:cutset}, for each $b \in \T$ we can compute $\tilde Y_b$ as a 
linear combination of 
$\tilde \Y_\F$ by applying~(\ref{eqn:cut}) on the fundamental cutset belonging to $b$. %This is due to the fact that.
We conclude that the variables in $\tilde\Y_\F$ are linearly independent
and the variables in $\tilde\Y_\T$ are fully determined by $\tilde\Y_\F$ via a linear transformation.

It follows that the number of valid configurations in the dual normal factor graph of the Potts model 
is $q^{|\F|}$. In any such partitioning
the number of variables in $\tilde\Y_\T$ is 
\begin{IEEEeqnarray}{c} 
\label{eqn:CardBF}
|\T| = N-1
\end{IEEEeqnarray}
and the number of variables in $\tilde\Y_\F$ is
$|\F| = |\EE| - |\T|$. 
For the 2D torus, 
%periodic boundary conditions, 
we thus have
\begin{IEEEeqnarray}{c} 
\label{eqn:CardBFPeriodic}
|\F| =  N+1
\end{IEEEeqnarray}
from (\ref{eqn:cardB2DTorus}).

\section{Monte Carlo Methods for the Partition Function of the Dual Normal Factor Graph}
\label{sec:IS}

We propose two basic
Monte Carlo algorithms for 
estimating the partition function. 
Both algorithms use partitions of $\EE$ and $\Y$ as in 
Section~\ref{sec:SpanningTree} and \Fig{fig:2DGridDPart}.

In both Monte Carlo algorithms, 
we draw \emph{independent} samples $\smpl{\tilde\y_\F}{1},\ldots,\smpl{\tilde\y_\F}{L} \in \calX^{|\F|}$ 
according to some auxiliary probability distribution,
and each of these samples $\smpl{\tilde\y_\F}{\ell}$ 
is completed 
(by computing the corresponding $\smpl{\tilde\y_\T}{\ell} \in \calX^{|\T|}$)
to a valid configuration 
$\smpl{\tilde\y}{\ell} = (\smpl{\tilde\y_\F}{\ell},\smpl{\tilde\y_\T}{\ell}) \in \calX^{|\EE|}$.
Computing $\smpl{\tilde\y_\T}{\ell}$ from $\smpl{\tilde\y_\F}{\ell}$ 
is easy and linear in $|\T|$. 

We will also use the quantities 
\begin{IEEEeqnarray}{rCl} 
\Gamma_\F(\tilde\y_\F) & = & \prod_{e\in\F} \gamma_e(\tilde y_e),  \label{eqn:GammaF} \\
\Gamma_\T(\tilde\y_\T) & = & \prod_{e\in\T} \gamma_e(\tilde y_e), \label{eqn:GammaT}
\end{IEEEeqnarray}
and 
\begin{IEEEeqnarray}{c} 
\Gamma(\tilde\y) =  \Gamma_\F(\tilde\y_\F) \Gamma_\T(\tilde\y_\T)
= \prod_{e\in\EE} \gamma_e(\tilde y_e),
\end{IEEEeqnarray}
%with which the partition sum (\ref{eqn:DualPartitionSum}) 
therefore (\ref{eqn:DualPartitionSum}) becomes
\begin{IEEEeqnarray}{c} 
\label{eqn:DualPartitionSumGamma}
%Z_\text{d} = \sum_{\text{valid $\tilde\x$}} \Gamma_\F(\tilde\x_\F)\, \Gamma_\T(\tilde\x_\T).
Z_\text{d} = \sum_{\text{valid $\tilde\y$}} \Gamma(\tilde\y).
\end{IEEEeqnarray}

We propose uniform sampling and importance sampling algorithms to 
estimate $Z_\text{d}$. Given the partitioning, the computational 
complexity of our algorithms 
is $O(|\EE|)$ per sample and $O(L |\EE|)$ in total. The variance of both methods is derived 
in Section~\ref{sec:Analysis}.

%%%%%%%%%%%%%%%%%%%%%%%%%%%%%%%%%%%%%%%%%%%%%%%%%%%%%%

\subsection{Uniform Sampling}
\label{sec:USampling}

As a %naive 
baseline algorithm (used in \cite{MoLo:ISIT2013} and \cite{AY:2014}), 
we use independent samples $\smpl{\tilde\y_\F}{1},\ldots,\smpl{\tilde\y_\F}{L}$ 
drawn uniformly over $\calX^{|\F|}$,
which are completed to valid configurations 
$\smpl{\tilde\y}{\ell} \in\calX^{|\EE|}$ as described above.
We then use the estimate 
\begin{IEEEeqnarray}{c} \label{eqn:EstU}
%\hat{Z}_\text{d} 
%\hat{Z}_\text{d}^\text{uni}
%\hat{Z}_\text{d,uni}
\hat{Z}^{\text{Uni}}_\text{d}
    = \frac{q^{|\F|}}{L} \sum_{\ell=1}^L \Gamma(\smpl{\tilde\y}{\ell}).
\end{IEEEeqnarray}
It is easily verified that
%$\E[ \hat Z_\text{d} ] = Z_\text{d}$,
$\E[ \hat{Z}^{\text{Uni}}_\text{d}] = Z_\text{d}$,
i.e., the estimator is unbiased:
\begin{IEEEeqnarray}{rCl}
\E\!\left[ \frac{q^{|\F|}}{L} \sum_{\ell=1}^L \Gamma(\smpl{\tilde\Y}{\ell}) \right] 
 & = & q^{|\F|} \E\!\left[ \Gamma(\smpl{\tilde\Y}{1}) \right] \\
 & = & q^{|\F|} \sum_{\text{valid $\tilde\y$}} \frac{1}{q^{|\F|}} \Gamma(\tilde\y) \\
 & = & Z_\text{d},
\end{IEEEeqnarray}
where the last step follows from (\ref{eqn:DualPartitionSumGamma}).

The accuracy of~(\ref{eqn:EstU}) depends on the fluctuations of 
$\Gamma(\smpl{\tilde\y}{\ell})$.
In the low-temperature limit (i.e., for $e^{J_e}\gg q$), these fluctuations disappear,
%(since $\gamma_b \approx e^{J_b}$, cf.\ Section~\ref{sec:LowTempLim}). 
because $\gamma_e(\tilde y_e) \approx e^{J_e}$ becomes constant. 
The estimator (\ref{eqn:EstU}) can therefore be expected to work well at sufficiently low temperatures.
%The variance of the estimator is given in (\ref{eqn:VarUnifChi2div}).

\subsection{Importance Sampling}
\label{sec:ImpSampling}

%A better estimate is obtained by using independent samples 
An importance sampling estimator (proposed in \cite{MeMo:2014a, Mo:IZS2016}) is obtained by 
drawing independent samples 
$\smpl{\tilde\y_\F}{1},\ldots,\smpl{\tilde\y_\F}{L} 
\in \calX^{|\F|}$ according to the auxiliary probability distribution %given by
%\in \calX^{|\B_\F|}$ drawn according to the auxiliary probability distribution 
%\begin{equation}  \label{eqn:AuxDist}
%p_\F(\tilde \x_\F) = \prod_{b\in\B_\F} \frac{\gamma_b(\tilde{x}_b)}{\sum_{\xi=0}^{q-1}\gamma_b(\xi)}
%\end{equation}
\begin{IEEEeqnarray}{rCl}
p_\F(\tilde \y_\F) & = & \frac{\Gamma_\F(\tilde\y_\F)}{Z_\F}\\
 & = &  \prod_{e\in\F} \frac{\gamma_e(\tilde{y}_e)}{\sum_{\xi=0}^{q-1}\gamma_e(\xi)},
        \label{eqn:ISPropProb}
\end{IEEEeqnarray}
where $Z_\F$ is available in closed-form as
\begin{IEEEeqnarray}{rCl}
Z_\F & = & \prod_{e\in\F} \sum_{\xi=0}^{q-1} \gamma_e(\xi) \\
  & = & \prod_{e\in\F} q e^{J_e} \\
  %& = & q^{N+1} \exp\!\Big( \sum_{b\in\B_\F} J_b \Big).
  & = & q^{|\F|} \exp\!\Big( \sum_{e\in\F} J_e \Big).
      \label{eqn:ZFis}
\end{IEEEeqnarray}
%
%It is obvious from (\ref{eqn:ISPropProb}) that $p_\F(\tilde \x_\F)$
%%This 
%%is an i.i.d.\ distribution and 
%is easy to sample from:

%%%%%%%%%%%%%%%%%%%%%%%%%%%%%%%%%%%%%%%%%%%%%%%%%%%%%%%%%%%%%%%%
\begin{figure}[t]
\centering
\begin{tikzpicture}
\begin{axis}[
			legend style={at = {(0.99,0.85)} ,font=\tiny},		
			height = 42.0ex,
			width = 48.0ex,
			grid = major,
			tick pos=left, 
			ymode=log,
			xminorticks = false,	
		    yminorticks = false,	
		    y tick label style={
        /pgf/number format/.cd,
            fixed,
        /tikz/.cd
    		}, 				
			ytick={1e-7, 1e-6, 1e-5, 1e-4, 1e-3, 1e-2, 1e-1, 1},
			xtick={0.1, 0.5, 0.9, 1.3, 1.7, 2.1, 2.5},
		xlabel= $J$ ={font=\normalsize},
			xmin = 0.1,
			xmax = 2.5,
			ymin = 0.0000001,
			ymax = 1,
			ylabel = Relative Error = {font=\tiny},
			yticklabel style = {font=\tiny,yshift=0.5ex},
            xticklabel style = {font=\tiny,xshift=0.0ex}			
			]
\pgfplotstableread{./1/constant_BP.txt}\mydataone
\pgfplotstableread{./1/constant_HAK.txt}\mydatatwo
\pgfplotstableread{./1/constant_TREEEP.txt}\mydatathree
\pgfplotstableread{./1/constant_MEHDI_UNI.txt}\mydatafour
\pgfplotstableread{./1/constant_MEHDI_IMP.txt}\mydatafive

		\addplot [
		 color = black,
		 mark = o,
		]		
		 table[y = Z] from \mydataone;
		 
 		 \addplot [
 		 color = black,
 		 mark = triangle,
 		 ]
 		  table[y = Z] from \mydatatwo;	 
 
  		 \addplot [
 		 color = black,
 		 mark = star,
 		 ]
 		  table[y = Z] from \mydatathree;	
 		  
 		 \addplot [
 		 color = blue,
 		 mark = x,
 		 ]
 		  table[y = Z] from \mydatafour;	
 		  
 		 \addplot [
 		 color = blue,
 		 mark = diamond,
 		 ]
 		  table[y = Z] from \mydatafive;

 		 \legend{BP, GBP, TreeEP, Uni, Imp};	  	

%		\legend{BP, GBPLoop, TREEEP, Unif. Sampling, Imp. 
%       Sampling};
		%\addlegendentry{Uniform Sampling};

%table [x index=0, y index=1]{\mytable};
\end{axis}
\end{tikzpicture}
%%%%%%%%%%%%%%%%%%%%%%%%%%%%%%%%%%%%%%%%%%%%%%%%%%%%%%%%%
\caption{\label{fig:GBP}%
Comparison with deterministic algorithms (BP, GBP, and TreeEP): 
experimental results for a
Potts model with $q=3$, $N = 8\times 8$, periodic boundary conditions, and constant couplings $J$. 
The plot shows the relative error~(\ref{eqn:RelLogError})
%$|\ln Z  - \ln \hat Z|/\ln Z$ 
as a function of $J$. 
(Recall that large $J$ corresponds to the low-temperature regime.)}
\end{figure}
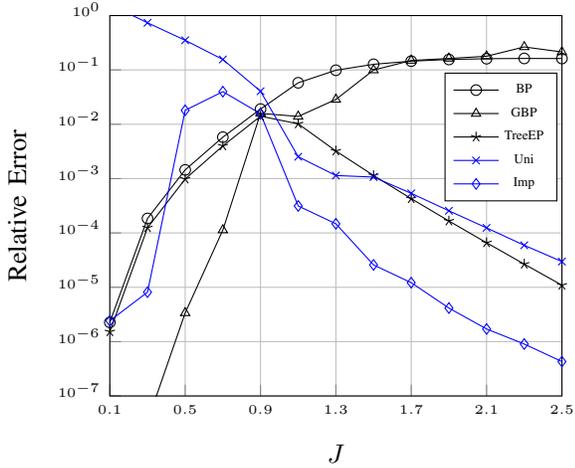
%%%%%%%%%%%%%%%%%%%%%%%%%%%%%%%%%%%%%%%%%%%%%%%%%%%%%%%%%%%
%

The product form in~(\ref{eqn:ISPropProb}) indicates that to draw 
samples according to $p_\F(\tilde \y_\F)$,
we can draw each component $\smpl{\tilde{y}_e}{\ell}$ of $\smpl{\tilde\y_\F}{\ell}$
\emph{independently} with probability
\begin{multline} 
\label{eqn:SampleProbIS}
P\big(\smpl{\tilde{y}_e}{\ell} = \xi\big) \\[1.0ex] =
 {%
 \renewcommand{\arraystretch}{1.5}
 \left\{\begin{array}{ll}
   \dfrac{1+(q-1)e^{-{J_e}}}{q}, & \text{if $\xi=0$} \\ [1.5ex]
   \dfrac{1-e^{-J_e}}{q},  & \text{for $\xi = 1,2,\ldots, q-1$.}
 \end{array}\right.
 }
\end{multline}
Again, the samples $\smpl{\tilde\y_\F}{1},\ldots,\smpl{\tilde\y_\F}{L}$ 
are completed to valid configurations $\smpl{\tilde\y}{1},\ldots,\smpl{\tilde\y}{L} \in \calX^{|\EE|}$.
We then use the estimate
\begin{equation} \label{eqn:EstIS}
%\hat Z_{\text{d}} = 
%\hat Z_\text{d}^\text{imp} \eqdef
\hat Z^{\text{Imp}}_\text{d} =
     \frac{Z_\F}{L} \sum_{\ell = 1}^{L} 
     \Gamma_\T(\smpl{\tilde\y_\T}{\ell}),
\end{equation}
%which is easily verified to be unbiased:
which is unbiased:
\begin{IEEEeqnarray}{rCl}
\E\!\left[ \, \frac{Z_\F}{L} \sum_{\ell=1}^L \Gamma_\T(\smpl{\tilde\Y_\T}{\ell}) \right] 
 & = & Z_\F \E\!\left[ \Gamma_\T(\smpl{\tilde\Y_\T}{1}) \right] \\
 & = & Z_\F \sum_{\text{valid $\tilde\y$}} 
       p_\F(\tilde\y_{\F}) \Gamma_\T(\tilde\y_\T) \\
 & = & \sum_{\text{valid $\tilde\y$}} \Gamma_\F(\tilde\y_\F) \Gamma_\T(\tilde\y_\T) \\
 & = & Z_\text{d}. \label{eqn:hatZisUnbiasedProved}
\end{IEEEeqnarray}

%The accuracy of the estimate (\ref{eqn:EstIS}) mainly depends on the fluctuations of 
The accuracy of (\ref{eqn:EstIS}) mainly depends on the fluctuations of 
$\Gamma_\T(\smpl{\tilde\y_\T}{\ell})$. 
The estimator can therefore be expected to work well at sufficiently low 
temperatures where these fluctuations disappear.
%The variance of the estimator is given in (\ref{eqn:VarISChi2div}).% below.

%%%%%%%%%%%%%%%%%%%%%%%%%%%%%%%%%%%%%%%%%%%%%%%%%%%%%%%%%%%%%%%%
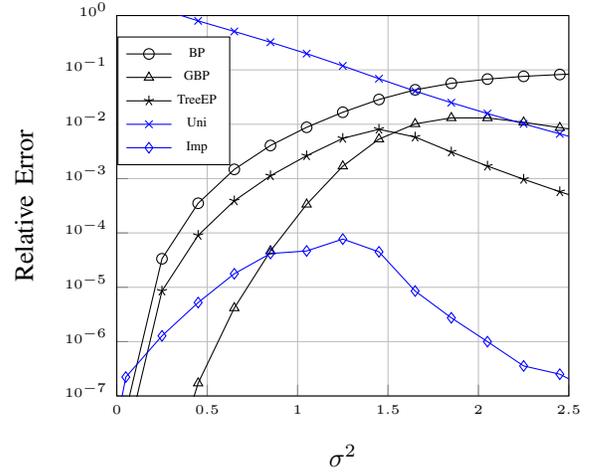
\begin{figure}[t]
\centering
\begin{tikzpicture}
\begin{axis}[
			legend style={at = {(0.251,0.945)} ,font=\tiny},		
			height = 42.0ex,
			width = 48.0ex,
			grid = major,
			tick pos=left, 
			ymode=log,
			xminorticks = false,	
		    yminorticks = false,	
		    y tick label style={
        /pgf/number format/.cd,
            fixed,
        /tikz/.cd
    		}, 				
			ytick={1e-7, 1e-6, 1e-5, 1e-4, 1e-3, 1e-2, 1e-1,1e0},
			xtick={0, 0.5, 1, 1.5, 2, 2.5},
		xlabel= $\sigma^2$={font=\normalsize},
			xmin = 0,
			xmax = 2.5,
			ymin = 0.0000001,
			ymax = 1e0,
			ylabel = Relative Error = {font=\tiny},
			yticklabel style = {font=\tiny,yshift=0.5ex},
            xticklabel style = {font=\tiny,xshift=0.0ex}			
			]
\pgfplotstableread{./2/BP.txt}\mydataone
\pgfplotstableread{./2/GBPLoop4.txt}\mydatatwo
\pgfplotstableread{./2/Tree.txt}\mydatathree
\pgfplotstableread{./2/Unif.txt}\mydatafour
\pgfplotstableread{./2/Imp.txt}\mydatafive

		\addplot [
		 color = black,
		 mark = o,
		]		
		 table[y = Z] from \mydataone;
		 
 		 \addplot [
 		 color = black,
 		 mark = triangle,
 		 ]
 		  table[y = Z] from \mydatatwo;	 
 
  		 \addplot [
 		 color = black,
 		 mark = star,
 		 ]
 		  table[y = Z] from \mydatathree;	
 		  
		 \addplot [
		 color = blue,
 		 mark = x,
 		 ]
 		  table[y = Z] from \mydatafour;	
 		  
 		 \addplot [
 		 color = blue,
 		 mark = diamond,
 		 ]
 		  table[y = Z] from \mydatafive;

 		 \legend{BP, GBP, TreeEP, Uni, Imp};	  	

%		\legend{BP, GBPLoop, TREEEP, Unif. Sampling, Imp. 
%       Sampling};
		%\addlegendentry{Uniform Sampling};

%table [x index=0, y index=1]{\mytable};
\end{axis}
\end{tikzpicture}
  %%%%%%%%%%%%%%%%%%%%%%%%%%%%%%%%%%%%%%%%%%%%%%%%%%%%%%%%%
\caption{\label{fig:Gauss}%
Comparison with deterministic algorithms (BP, GBP, and TreeEP): 
experimental results for a Potts model with \mbox{$q=3$}, $N = 8\times 8$, periodic boundary 
conditions, and coupling parameters $J_e$: $J_e = |J'_e|$ with $J'_e~\overset{\text{i.i.d.}}{\sim} \N(0,\sigma^2)$.
The plot shows the relative error (\ref{eqn:RelLogError})
%$|\ln Z  - \ln \hat Z|/\ln Z$ 
as a function of $\sigma^2$. 
%The plot shows the relative error $|\ln Z  - \ln \hat Z|/\ln Z$ 
%vs.\ $\sigma^2$. 
%The relative error $|\ln Z  - \ln \hat Z|/\ln Z$ as a function of the variance $\sigma^2$ 
%for an $8\times 8$ \mbox{3-state} Potts model with 
%couplings $J_k$, where $J_k = |J'_k|$ and $J'_k~\sim \N(0,\sigma^2)$.
}
\end{figure}
%%%%%%%%%%%%%%%%%%%%%%%%%%%%%%%%%%%%%%%%%%%%%%%%%%%%%%%%%%%

%\section{Analysis of the Variance}
\subsection{Variance of the Estimates}
\label{sec:Analysis}

The variance of the importance sampling estimator (\ref{eqn:EstIS}) is 
\begin{IEEEeqnarray}{rCl}
\Var[\hat Z^{\text{Imp}}_\text{d}] 
 & = & \E\big[\big(\hat Z^{\text{Imp}}_\text{d}\big)^2\big] - \E\big[\hat Z^{\text{Imp}}_\text{d}\big]^2 \\
%mehdi 20 fev 
% & = & 
%       \frac{Z_\F^2}{L} 
%       \E\left[ \Big( \Gamma_\T(\smpl{\tilde\x_\T}{1}) \Big)^2 \right]
%       - Z_\text{d}^2 \\
%mehdi 20 fev
 & = & \frac{Z_\F^2}{L} \left( \sum_{\text{valid $\tilde \y$}} p_\F(\tilde \y_\F)
        \Gamma_\T(\tilde\y_\T)^2 \right)
        - \frac{Z_\text{d}^2}{L} \\
 & = & \frac{1}{L} \left( \sum_{\text{valid $\tilde \y$}} 
        \frac{\Gamma_\F(\tilde\y_\F)^2}{p_\F(\tilde \y_\F)}
        \Gamma_\T(\tilde\y_\T)^2 \right)
        - \frac{Z_\text{d}^2}{L} \\
% & = & \frac{1}{L}  \sum_{\text{valid $\tilde \x$}} 
%        \frac{ \Gamma(\tilde\x)^2}{p_\F(\tilde \x_\F)} 
%        - Z_\text{d}^2 \\
 & = & \frac{Z_\text{d}^2}{L}  \left(
        \sum_{\text{valid $\tilde \y$}} 
        \frac{ p_\text{d}(\tilde\y)^2}{p_\F(\tilde \y)} 
        - 1  \right),
\end{IEEEeqnarray}
where both $p_\text{d}(\tilde\y) = \Gamma(\tilde\y)/Z_\text{d}$
and $p_\F(\tilde\y_\F) = \Gamma_\F(\tilde\y_\F)/Z_\F$ 
are probability mass functions defined on the valid configurations
of the dual normal factor graph.

We thus have
\begin{IEEEeqnarray}{c} 
\label{eqn:VarISChi2div}
%\Var[\hat Z_\text{d}] = \frac{Z_\text{d}^2}{L} \chi^2\big( p_\text{d}, p_\F \big),
\Var[\hat Z^{\text{Imp}}_\text{d}] \frac{L}{Z_\text{d}^2} = \chi^2\big( p_\text{d}, p_\F \big),
\end{IEEEeqnarray}
where $\chi^2(\cdot, \cdot)$ denotes the chi-squared divergence, 
which is always nonnegative 
%with equality to zero 
and is equal to zero 
if and only if its two arguments are equal~\cite[Chapter 4]{Csiszar:04}.
% (see also~\citep{AS:66}).

An analogous derivation for the uniform sampling estimator (\ref{eqn:EstU}) yields 
\begin{IEEEeqnarray}{c}
\label{eqn:VarUnifChi2div}
\Var[\hat Z^{\text{Uni}}_\text{d}] \frac{L}{Z_\text{d}^2} = \chi^2\big( p_\text{d}, p_\text{u} \big),
\end{IEEEeqnarray}
where $p_\text{u}(\tilde\y)$ is the uniform distribution over the valid configurations.

%In particular, 
In the low-temperature limit 
%with constant couplings $J_b=J\gg q$,
with $e^{J_e} \gg q$ for all $e\in\EE$,
both $p_\text{d}$ and $p_\F$ become uniform over the valid configurations
and both (\ref{eqn:VarISChi2div}) and (\ref{eqn:VarUnifChi2div}) vanish. 
More importantly, the variance of the importance sampling estimator~(\ref{eqn:VarISChi2div}) 
%(but not (\ref{eqn:VarUnifChi2div})) 
vanishes under the weaker condition (weaker for nonconstant couplings) 
\begin{IEEEeqnarray}{c}
\label{eqn:CondJbLargeOnTree}
\text{$e^{J_e} \gg q$ for $e\in \T$,}
\end{IEEEeqnarray}
since in this case $p_\text{d}(\tilde\y) \propto \Gamma_\F(\tilde\y_\F) \Gamma_\T(\tilde\y_\T)$ 
converges to $p_\F(\tilde\y) \propto \Gamma_\F(\tilde\y_\F)$ 
if $\Gamma_\T(\tilde\y_\T)$ becomes constant.

%mehdi
For the Ising model on a 2D torus, a more detailed analysis of the variance of 
our proposed Monte Carlo methods is given in Appendix~\ref{appsec:VarIsing}.

%mehdu

\subsection{Choosing the Partitioning}
\label{sec:PartitionIS}

The choice of $\T$ and $\F$ does not affect the performance 
of the uniform sampling estimator (\ref{eqn:EstU}), 
but it can affect the performance (i.e., the convergence) 
of the importance sampling estimator (\ref{eqn:EstIS})
for nonconstant couplings. 
Recall that the normalized variance (\ref{eqn:VarISChi2div}) 
vanishes if (\ref{eqn:CondJbLargeOnTree}) holds.
This result suggests 
to include into $\T$ only edges $e$ with stronger couplings (i.e., with large $J_e$). 
%if possible. 
%From (\ref{eqn:VarISChi2div}), it seems a good idea (why ???)
%to include edges $b$ with stronger couplings (i.e., with large $J_b$) 
%in $\B_\T$, if possible. 
To this end, the following heuristic strategy can be used: 
choose $\T$ to be a spanning tree that maximizes $\sum_{e\in\T} J_e$. 
This is a maximum-spanning tree problem, that can be solved efficiently 
with complexity linear in $N$ and $|\EE|$ (see~\cite[Chapter.~VI]{Thomas:2009}).
We will use this heuristic strategy in our numerical experiments 
in Section~\ref{sec:Num}.

\section{Numerical experiments}
\label{sec:Num}

In this section, 
we demonstrate the methods of Section~\ref{sec:IS} 
with some numerical experiments. 
In Sections~\ref{sec:Num1} to \ref{sec:ErrorAnal}, we work with tractable 
models where the partition function can 
be computed exactly via the junction tree algorithm~\cite{murphy:2012}; 
larger grids are considered in Section~\ref{sec:Num2}.

\subsection{Comparison with Deterministic Algorithms}
\label{sec:Num1}

We first consider the Potts model with $q=3$ on an $8\times 8$ grid 
with periodic boundary conditions. For this size of grid, we were able to compute the exact value of
the partition function.
%For such a small grid, 
%it is possible to compute the partition sum exactly
%(essentially by brute force, using a suitable junction tree). 

In Figs.~\ref{fig:GBP} and \ref{fig:Gauss},
%we compare the proposed methods 
we compare the accuracy of the proposed methods 
with three standard deterministic algorithms: 
%(implemented in~\cite{Mooij:2010})
BP, GBP~\cite{yedidia:00, Heskes:02,Yedidia:2005},
and TreeEP~\cite{Minka:04}. 
These three algorithms turned out to perform best, in our setting, 
among all deterministic 
%state-of-the-art 
methods implemented in~\cite{Mooij:2010}.
%Among the state-of-the-art deterministic
%the algorithms implemented in~\cite{Mooij:2010}, 
%these three algorithms are the state-of-the-art deterministic methods 
%that performed best in our set-up. 
(Among the different versions of GBP in \cite{Mooij:2010}, we selected the one with 
the best performance.)
%For GBP, we used versions that allow for loops of different sizes as clusters, 
%and one cluster for each maximal factor ???).

%Note that 
The accuracy of the proposed Monte Carlo methods depends on the number of samples, 
but the result is exact (with probability one) in the limit of infinitely many samples.
By contrast, the deterministic algorithms (BP, GBP, and TreeEP) yield 
approximations whose accuracy is not improved beyond convergence. 
However, it should be emphasized that deterministic algorithms 
converge much faster than our Monte Carlo methods.

Figs.~\ref{fig:GBP} and \ref{fig:Gauss}
show the relative error 
%$|\ln Z - \ln \hat Z|/\ln Z$ 
\begin{equation} \label{eqn:RelLogError}
%\frac{| \ln\hat{Z} - \ln Z |}{\ln Z}
\frac{| \log\hat{Z} - \log Z |}{\log Z}
\end{equation}
for the different estimates $\hat Z$. 
The labels ``Uni'' and ``Imp'' 
refer to uniform sampling as in Section~\ref{sec:USampling} 
and importance sampling as in Section~\ref{sec:ImpSampling}, respectively.

In \Fig{fig:GBP}, the couplings $J_e=J$ are constant. 
For the proposed Monte Carlo methods, $\log \hat{Z}$ in (\ref{eqn:RelLogError}) 
is averaged over 50 trials, 
each with $L=10^8$ samples (taking about two minutes on a 2GHz Intel Xeon CPU).

%In \Fig{fig:Gauss}, the couplings are random with 
In \Fig{fig:Gauss}, the couplings are chosen randomly according to  
a half-normal distribution: $J_e = |J'_e|$
with \mbox{$J'_e \overset{\text{i.i.d.}}{\sim} \N(0,\sigma^2)$} for all $e\in\EE$.
The spanning tree was chosen according to the heuristic strategy 
proposed in Section~\ref{sec:PartitionIS}. 
The plot shows the relative error (\ref{eqn:RelLogError}),
where $\log Z$ is averaged over 25 independent realizations of the couplings
and for every realization, $\log \hat{Z}$ is averaged over $L=10^8$ samples.

Both Figs.~\ref{fig:GBP} and \ref{fig:Gauss} clearly 
show that, for sufficiently strong couplings (i.e., at low temperature), 
the importance sampling algorithm~(\ref{eqn:EstIS})
%(with a practical number of samples)
yields much better estimates of the partition function than the other algorithms.
In particular, from Fig.~\ref{fig:Gauss}, we observe that importance 
sampling outperforms the second best approach (TreeEP) by more than 
two orders of magnitude for $\sigma^2 > 1.5$. 
The accuracy of the proposed methods 
can still be improved, of course, by increasing the number of samples.

%%%%%%%%%%%%%%%%%%%%%%%%%%%%%%%%%%%%%%%%%%%%%%%%%%%%%%%%%%%%%%%%
\begin{figure}[t]
\centering
\begin{tikzpicture}
\begin{axis}[
			%legend pos=south east,
			legend style={legend pos = north west,font=\tiny},
			height = 42.0ex,
			width = 48.0ex,
			grid = major,
			tick pos=left, 
		    y tick label style={
        /pgf/number format/.cd,
            fixed,
            fixed zerofill,
            precision = 1,
        /tikz/.cd
    		}, 
			xminorticks = false,			
			ytick={0.5, 1.0, 1.5, 2.0, 2.5, 3.0, 
			3.5},
			xtick={0.0, 0.25, 0.5, 0.75, 1.0, 1.25, 1.5},
		xlabel= $J$, style={font=\normalsize},
			xmin = 0,
			xmax = 1.5,
			ymax = 3.5,
			ymin = 0,
			%ylabel = Estimated $\ln(Z)/N$,
			ylabel = $\ln(\hat{Z})/N$,
			yticklabel style = {font=\tiny,xshift=0.2ex},
            xticklabel style = {font=\tiny,yshift=0.5ex}			
			]
\pgfplotstableread{./3/Exact.txt}\mydataone
\pgfplotstableread{./3/SW.txt}\mydatatwo
\pgfplotstableread{./3/UniP.txt}\mydatathree
\pgfplotstableread{./3/Imp.txt}\mydatafour
\pgfplotstableread{./3/UniD.txt}\mydatafive

		\addplot [
		only marks,
		 color = black,
		 mark = star,
		]		
		 table[y = Z] from \mydataone;

		\addplot [
		 color = red,
		 dashdotted,
		 line width = 0.2mm,
		]		
		 table[y = Z] from \mydatatwo;		
		 
		\addplot [
		 color = red,
		 densely dotted,
		 line width = 0.2mm,
		]		
		 table[y = Z] from \mydatathree;

		\addplot [
		 color = blue,
		]		
		 table[y = Z] from \mydatafour;	 
		 
		 		\addplot [
		 color = blue,
		 dashed,
		 line width = 0.2mm,
		]		
		 table[y = Z] from \mydatafive;	 

		\legend{Exact $\ln(Z)/N$, Swendsen-Wang, Uni (primal), Imp, Uni (dual)};
		%\addlegendentry{Uniform Sampling};

%table [x index=0, y index=1]{\mytable};
\end{axis}
\end{tikzpicture}
%%%%%%%%%%%%%%%%%%%%%%%%%%%%%%%%%%%%%%%%%%%%%%%%
\caption{\label{fig:SmallGridSWAbs}%
Comparison with 
%standard Monte Carlo algorithms (using uniform sampling or the Swendsen-Wang algorithm): 
uniform sampling and the Swendsen-Wang algorithm: 
experimental results for a Potts model with $q=3$, $N=8\times 8$, periodic boundary conditions, and 
constant couplings $J$. 
The plot shows $\ln(\hat Z)/N$ 
as a function of $J$.
%the Swendsen-Wang algorithm (dashdotted red line) and uniform
%sampling (dotted red line) in the primal NFG, and importance sampling (solid blue line) and uniform sampling 
%(dashed blue line) in the dual NFG. The exact value of 
%$\ln(Z)/N$ is marked by ``$*$".}
}
\end{figure}
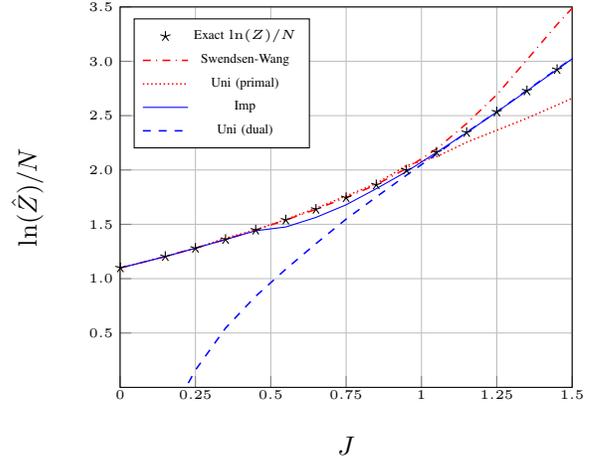
%%%%%%%%%%%%%%%%%%%%%%%%%%%%%%%%%%%%%%%%%%%%%%%

\subsection{Comparison with Standard Monte Carlo Methods}
\label{sec:CompStandMC}

We again consider the Potts model with $q=3$ on an $8\times 8$ grid with periodic 
boundary conditions.
In \Fig{fig:SmallGridSWAbs}, we compare 
the proposed methods with two Monte Carlo methods that operate  
in the primal %original 
Potts model. 
The first of these (standard) Monte Carlo algorithms 
(labeled ``Uni (primal)'' in \Fig{fig:SmallGridSWAbs})
is a %naive 
baseline algorithm:
we use uniform samples $\smpl{\x}{1},\ldots,\smpl{\x}{L} \in \calX^N$ 
and form the (unbiased) estimate 
\begin{IEEEeqnarray}{c} 
\label{eqn:UnifPrimal}
\hat{Z}^\text{Uni} = \frac{q^N}{L} \sum_{\ell=1}^L f(\smpl{\x}{\ell})
\end{IEEEeqnarray}
%\begin{equation}
%\hat{Z}_\text{u} \eqdef \frac{q^N}{L} \sum_{\ell=1}^L f(\smpl{\x}{\ell})
%\end{equation}
with $f$ as in (\ref{eqn:factorF}).

The second standard Monte Carlo algorithm 
uses the Swendsen-Wang algorithm~\cite{SW:87}
to obtain %(dependent) 
samples \sloppy\mbox{$\smpl{\x}{1},\ldots,\smpl{\x}{L} \in \calX^N$}
according to the Boltzmann distribution (\ref{eqn:Prob}). From these samples,
we form the Ogata-Tanemura estimate
\begin{IEEEeqnarray}{c}
%\hat{Z}^{-1} = \frac{1}{L q^N} \sum_{\ell=1}^L \frac{1}{f(\smpl{\x}{\ell})}
\hat{Z}_\text{OT} =
   \left( \frac{1}{L q^N} \sum_{\ell=1}^L \frac{1}{f(\smpl{\x}{\ell})} \right)^{\!-1},
\end{IEEEeqnarray}
which satisfies $\E[\hat{Z}_\text{OT}^{-1}] = Z^{-1}$. For more details on 
the Ogata-Tanemura estimator, see~\cite{OgTa:eip1981}, \cite{Potam:97}.
%,~\cite[Section~III]{LoMo:IT2013}.
%\begin{equation}
%%\hat{Z}^{-1} = \frac{1}{L q^N} \sum_{\ell=1}^L \frac{1}{f(\smpl{\x}{\ell})}
%\hat{Z}_\text{OT} \eqdef 
%   \left( \frac{1}{L q^N} \sum_{\ell=1}^L \frac{1}{f(\smpl{\x}{\ell})} \right)^{\!-1},
%\end{equation}
%which satisfies $\E[\hat{Z}_\text{OT}^{-1}] = Z^{-1}$ \cite{Potam:97}.

\Fig{fig:SmallGridSWAbs} shows the estimated $\ln(Z)/N$, where the results were obtained 
by averaging over 50 trials, 
each with $L=10^8$ samples.
%It is obvious from \Fig{fig:SmallGridSWAbs} that 
It is clear from \Fig{fig:SmallGridSWAbs} that 
importance sampling as in Section~\ref{sec:ImpSampling} 
works well for large couplings (low temperatures), 
where both standard Monte Carlo algorithms fail.

We do not here compare the proposed methods with annealed (i.e., multi-temperature) 
Monte Carlo methods~\cite{NealIS:2001}: 
since, in principle, annealing can also be used in the dual normal factor graph; 
%(see Appendix~\ref{appsec:AnnealedIS}); 
the advantage of the dual graph over the primal graph
at low temperatures extends also to Monte Carlo methods with annealing.

%%%%%%%%%%%%%%%%%%%%%%%%%%%%%%%%%%%%%%%%%%%%%%%%%%%%%%%%%%%%%%%%
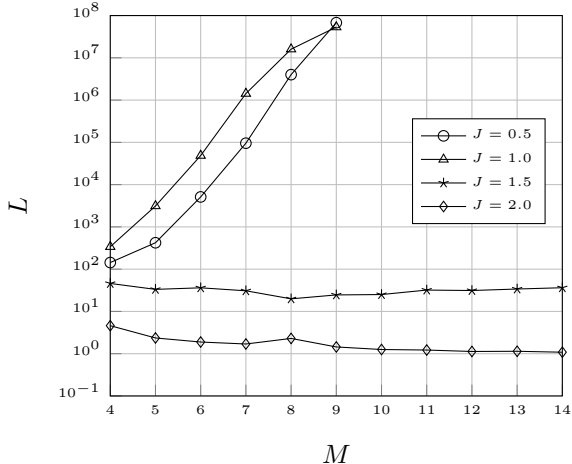
\begin{figure}[t]
\centering
\begin{tikzpicture}
\begin{axis}[
			legend style={at = {(0.963,0.73)} ,font=\tiny},		
			height = 42.0ex,
			width = 48.0ex,
			grid = major,
			tick pos=left, 
			ymode=log,
			xminorticks = false,	
		    yminorticks = false,	
		    y tick label style={
        /pgf/number format/.cd,
            fixed,
        /tikz/.cd
    		}, 				
			ytick={1e-1,1,1e1,1e2,1e3,1e4,1e5,1e6,1e7,1e8},
			xtick={4, 5, 6, 7, 8, 9,10,11,12,13,14},
		xlabel= $M$={font=\normalsize},
			xmin = 4,
			xmax = 14,
			ymin = 0.1,
			ymax = 1e8,
			ylabel = $L$ = {font=\normalsize},
			yticklabel style = {font=\tiny,yshift=0.5ex},
            xticklabel style = {font=\tiny,xshift=0.0ex}			
			]
\pgfplotstableread{./4/minL_beta1_thresh0.01.txt}\mydataone
\pgfplotstableread{./4/minL_beta2_thresh0.01.txt}\mydatatwo
\pgfplotstableread{./4/minL_beta3_thresh0.01.txt}\mydatathree
\pgfplotstableread{./4/minL_beta4_thresh0.01.txt}\mydatafive

		\addplot [
		 color = black,
		 mark = o,
		]		
		 table[y = Z] from \mydataone;
		 
 		 \addplot [
 		 color = black,
 		 mark = triangle,
 		 ]
 		  table[y = Z] from \mydatatwo;	 
 
  		 \addplot [
 		 color = black,
 		 mark = star,
 		 ]
 		  table[y = Z] from \mydatathree;	
 		  
%		 \addplot [
%		 color = blue,
% 		 mark = x,
% 		 ]
% 		  table[y = Z] from \mydatafour;	
 		  
 		 \addplot [
 		 color = black,
 		 mark = diamond,
 		 ]
 		  table[y = Z] from \mydatafive;

 		 \legend{$J = 0.5$, $J = 1.0$, $J = 1.5$, $J = 2.0$};	  	

%		\legend{BP, GBPLoop, TREEEP, Unif. Sampling, Imp. 
%       Sampling};
		%\addlegendentry{Uniform Sampling};

%table [x index=0, y index=1]{\mytable};
\end{axis}
\end{tikzpicture}
  %%%%%%%%%%%%%%%%%%%%%%%%%%%%%%%%%%%%%%%%%%%%%%%%%%%%%%%%%
\caption{\label{fig:scaling1}%
Number of required samples as a function of the width of the grid $M$ to achieve a relative error (\ref{eqn:RelLogError}) of $10^{-2}$, 
for a 2D Potts model with $q=3$,  free boundary conditions, and constant couplings $J$.
%$|\ln Z  - \ln \hat Z|/\ln Z$ 
%The plot shows the relative error $|\ln Z  - \ln \hat Z|/\ln Z$ 
%vs.\ $\sigma^2$. 
%The relative error $|\ln Z  - \ln \hat Z|/\ln Z$ as a function of the variance $\sigma^2$ 
%for an $8\times 8$ \mbox{3-state} Potts model with 
%couplings $J_k$, where $J_k = |J'_k|$ and $J'_k~\sim \N(0,\sigma^2)$.
}
\end{figure}
%%%%%%%%%%%%%%%%%%%%%%%%%%%%%
%\vspace{3mm}

\subsection{Scaling Behavior of the Importance Sampling Algorithm}
\label{sec:ErrorAnal}

We analyze the performance of the importance sampling algorithm in the dual normal factor graph 
in terms of the required number of samples $L$ to achieve a given relative 
error as a function of the width of the gird $M$. If the desired relative 
error was not achieved after $L=10^8$ samples, we stopped the simulations.

We consider a 3-state Potts model with constant couplings $J_e=J$, with free boundary 
conditions, and on an $M\times M$ grid, where up to $M = 14$ we were able to compute the exact value of the partition function.

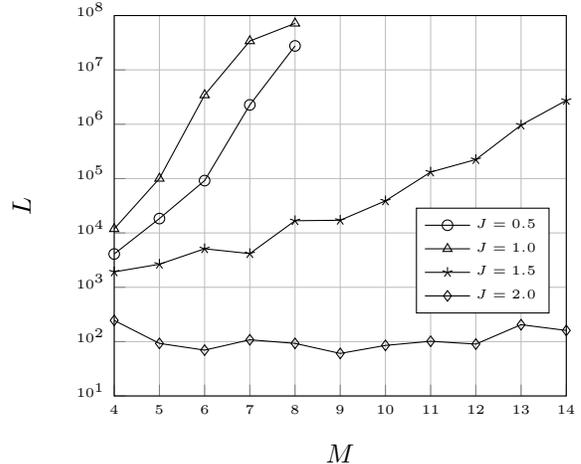
\begin{figure}[t]
\centering
\begin{tikzpicture}
\begin{axis}[
			legend style={at = {(0.963,0.495)} ,font=\tiny},		
			height = 42.0ex,
			width = 48.0ex,
			grid = major,
			tick pos=left, 
			ymode=log,
			xminorticks = false,	
		    yminorticks = false,	
		    y tick label style={
        /pgf/number format/.cd,
            fixed,
        /tikz/.cd
    		}, 				
			ytick={1e1,1e2,1e3,1e4,1e5,1e6,1e7,1e8,1e9},
			xtick={4, 5, 6, 7, 8, 9,10,11,12,13,14},
		xlabel= $M$={font=\normalsize},
			xmin = 4,
			xmax = 14,
			ymin = 1e1,
			ymax = 1e8,
			ylabel = $L$ = {font=\normalsize},
			yticklabel style = {font=\tiny,yshift=0.5ex},
            xticklabel style = {font=\tiny,xshift=0.0ex}			
			]
\pgfplotstableread{./4/minL_beta1_thresh0.001.txt}\mydataone
\pgfplotstableread{./4/minL_beta2_thresh0.001.txt}\mydatatwo
\pgfplotstableread{./4/minL_beta3_thresh0.001.txt}\mydatathree
\pgfplotstableread{./4/minL_beta4_thresh0.001.txt}\mydatafive

		\addplot [
		 color = black,
		 mark = o,
		]		
		 table[y = Z] from \mydataone;
		 
 		 \addplot [
 		 color = black,
 		 mark = triangle,
 		 ]
 		  table[y = Z] from \mydatatwo;	 
 
  		 \addplot [
 		 color = black,
 		 mark = star,
 		 ]
 		  table[y = Z] from \mydatathree;	
 		  
%		 \addplot [
%		 color = blue,
% 		 mark = x,
% 		 ]
% 		  table[y = Z] from \mydatafour;	
 		  
 		 \addplot [
 		 color = black,
 		 mark = diamond,
 		 ]
 		  table[y = Z] from \mydatafive;

 		 \legend{$J = 0.5$, $J = 1.0$, $J = 1.5$, $J = 2.0$};	  	

%		\legend{BP, GBPLoop, TREEEP, Unif. Sampling, Imp. 
%       Sampling};
		%\addlegendentry{Uniform Sampling};

%table [x index=0, y index=1]{\mytable};
\end{axis}
\end{tikzpicture}
  %%%%%%%%%%%%%%%%%%%%%%%%%%%%%%%%%%%%%%%%%%%%%%%%%%%%%%%%%
\caption{\label{fig:scaling2}%
Number of required samples as a function of the width of the grid $M$ to achieve a relative error (\ref{eqn:RelLogError}) of $10^{-3}$, 
for a 2D Potts model with $q=3$,  free boundary conditions, and constant couplings $J$.
}
\end{figure}

Figs.~\ref{fig:scaling1} and~\ref{fig:scaling2} show experimental results to achieve a relative 
error of $10^{-2}$ and $10^{-3}$, respectively.
For $J=2$ (i.e., when the temperature is low enough), the number of required samples is almost independent of $M$.
For $J=1.5$,  to achieve a relative error of $10^{-3}$, the number of required samples increases with $M$;
but it remains almost constant to achieve a relative error of $10^{-2}$. We take these results as evidence that the importance sampling algorithm is robust at low temperature. On the other hand, 
for weaker couplings, $L$ grows quickly as a function of $M$.

%%%%%%%%%%%%%%%%%%%%%%%%%%%%%%%%%%%%%%%%%%%%%%%%%%%%%%%%%%%%%%%%%
%%%%%%%%%%%%%%%%%%%%%%%%%%%%%%%%%%%%%%%%%%%%%%%%%%%%%%%%%%%

\subsection{Larger Grids}
\label{sec:Num2}

\begin{figure}[t]
\centering
%\begin{subfigure}{0.48\textwidth}
\begin{tikzpicture}
\begin{axis}[
			%legend pos=south east,
			legend style={at = {(0.95,0.38)} ,font=\tiny},
			height = 42.0ex,
			width = 48.0ex,
			grid = major,
			tick pos=left, 
			xmode=log,
		    y tick label style={
        /pgf/number format/.cd,
            fixed,
            fixed zerofill,
            precision=3,
        /tikz/.cd
    		}, 
			xminorticks = false,			
			ytick={5.476, 5.48, 5.484, 5.488, 5.492, 5.496},
			xtick={1, 10, 100, 1000, 10000, 100000, 1000000, 10000000},
		xlabel= $L$, style={font=\normalsize},
			xmin = 1,
			xmax = 1e7,
			ymax = 5.496,
			ymin = 5.476,
			%ylabel = Estimated $\ln(Z)/N$,
			ylabel = $\ln(\hat{Z})/N$,
			yticklabel style = {font=\tiny,xshift=0.5ex},
            xticklabel style = {font=\tiny,yshift=0.5ex}			
			]
\pgfplotstableread{./5/aZS1.txt}\mydataone
\pgfplotstableread{./5/aZS2.txt}\mydatatwo
\pgfplotstableread{./5/aZS3.txt}\mydatathree
\pgfplotstableread{./5/aZS4.txt}\mydatafour
\pgfplotstableread{./5/aZS5.txt}\mydatafive
\pgfplotstableread{./5/aZU1.txt}\mydatasix
\pgfplotstableread{./5/aZU2.txt}\mydataseven
\pgfplotstableread{./5/aZU3.txt}\mydataeigth
\pgfplotstableread{./5/aZU4.txt}\mydatanine
\pgfplotstableread{./5/aZU5.txt}\mydataten

		\addplot [
		 color = black,
		 legend style = {at = {(10000, 5.478)}}
		]		
		 table[y = Z] from \mydataone;
		 
		 \addplot [
 		 color = blue,
 		 dashed,
 		 thin
		 ]
 		  table[y = Z] from \mydatasix;
 		 
 		 \addplot [
 		 color = black,
 		thin
 		 ]
 		  table[y = Z] from \mydatatwo;

		 \addplot [
 		 color = black,
 		thin
 		 ]
 		  table[y = Z] from \mydatathree;
 		  
 		 \addplot [
 		 color = black,
 		 thin
		 ]
 		  table[y = Z] from \mydatafour;
 		  
 		 \addplot [
 		 color = black
		 ]
 		  table[y = Z] from \mydatafive;

   		 \addplot [
 		 color = blue,
 		  		 dashed
		 ]
 		  table[y = Z] from \mydataseven;
 		  
   		 \addplot [
 		 color = blue,
  		 dashed
		 ]
 		  table[y = Z] from \mydataeigth;
 		  
   		 \addplot [
 		 color = blue,
  		 dashed
		 ]
 		  table[y = Z] from \mydatanine;
 		  
   		 \addplot [
 		 color = blue,
  		 dashed
		 ]
 		  table[y = Z] from \mydataten;
		\legend{Importance Sampling, Uniform Sampling};
		%\addlegendentry{Uniform Sampling};

%table [x index=0, y index=1]{\mytable};
\end{axis}
\end{tikzpicture}
%\end{subfigure}
%%%%%%%%%%%%%%%%%%%%%%%%%%%%%%%%%%%%%%%%%%%%%%%%
%\begin{subfigure}{0.48\textwidth}
%\begin{tikzpicture}
%\begin{axis}[
%			legend style={legend pos = south east,font=\tiny},
%			%width=\textwidth,
%			height = 40.0ex,
%			width = 46.0ex,
%			grid = major,
%			tick pos=left, 
%			xmode=log,
%		    y tick label style={
%        /pgf/number format/.cd,
%            fixed,
%            fixed zerofill,
%            precision=3,
%        /tikz/.cd
%    		}, 
%			ylabel = Estimated $\ln(Z)/N$,
%			xminorticks = false,			
%			ytick={4.971, 4.976, 4.981, 4.986, 
%			4.991, 4.996},
%			xtick={1, 10, 100, 1000, 10000, 100000, 1000000, 10000000, 1e8},
%		xlabel= Number of Samples, style={font=\normalsize},
%\end{axis}
%\end{tikzpicture}
%  \end{subfigure}
\caption{\label{fig:FerPotts1}%
Experimental results for a fixed realization of a Potts model with $q=3$, \mbox{$N=40\times 40$}, 
%The plots shows the 
%estimated $\ln(Z)/N$ vs.~number of samples $L$.
and coupling parameters $J_{e} \overset{\text{i.i.d.}}{\sim} \calU[2.5, 3.0]$ for all $e\in\EE$.
%Left: random couplings with $J_{b} \sim \calU[2.5, 3.0]$ for all $b\in\B$.
%Right: random couplings with $J_{b} \sim \calU[2.5, 3.0]$ for $b\in\B_\T$
The plot shows the estimated $\ln(Z)/N$ using importance sampling (solid black lines) 
and using uniform sampling (dashed blue lines) in the dual normal factor graph.
%using importance sampling (solid black lines) and uniform sampling 
%(dashed blue lines) in the dual NFG of a $40\times 40$ 3-state Potts model with 
%and $J_b \sim \calU[2.0, 2.5]$ for $b\in\B_\F$.
%$b \in \B_\F$ with \textbf{(left)} $J_b \sim \calU[2.5, 3.0]$; \textbf{(right)} 
%$J_b \sim \calU[2.0, 2.5]$.}
}
\end{figure}
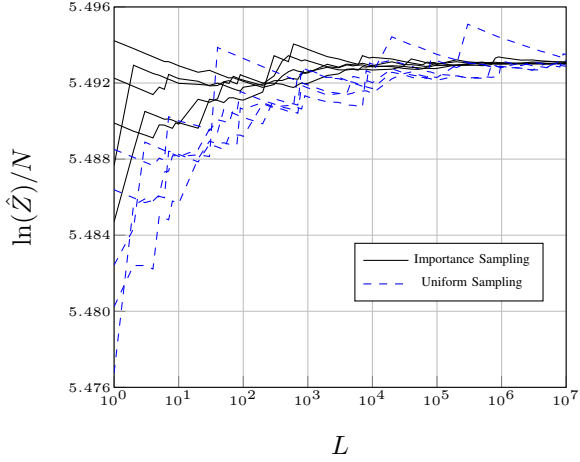

\Fig{fig:FerPotts1} shows results 
for a fixed realization of a Potts model with $q=3$ on a grid of size $N=40\times 40$ and with 
couplings $J_e  \overset{\text{i.i.d.}}{\sim} \calU[2.5, 3.0]$ for all $e\in\EE$.
%but only a single fixed realization of the couplings is used.
The plot shows $\ln(\hat{Z})/N$ vs.\ the number of samples $L$ %(without any averaging ???)
for five independent runs of the Monte Carlo algorithms. 
%(For actual applications, we recommend at least ten independent runs, 
%but \Fig{fig:FerPotts1} would become too crowded.)
It is obvious (and unsurprising) that importance sampling converges more quickly 
than uniform sampling. In this example, importance sampling yields the estimate
$\ln(\hat{Z})/N \approx 5.493$. % (with about ??? decimal digits).

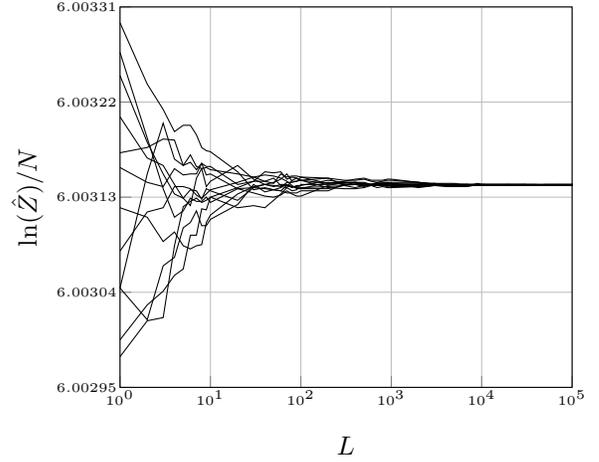
\begin{figure}[t]
\centering
%\begin{subfigure}{0.48\textwidth}
\begin{tikzpicture}
\begin{axis}[
			%legend pos=south east,
			legend style={legend pos = south east,font=\tiny},
			height = 42.0ex,
			width = 48.0ex,
			grid = major,
			tick pos=left, 
			xmode=log,
		    y tick label style={
        /pgf/number format/.cd,
            fixed,
            fixed zerofill,
            precision=5,
        /tikz/.cd
    		}, 
			xminorticks = false,			
			ytick={6.00295, 6.00304, 6.00313, 6.00322, 6.00331},
			xtick={1, 10, 100, 1000, 10000, 100000},
		xlabel= $L$, style={font=\normalsize},
			xmin = 1,
			xmax = 1e5,
			ymax = 6.00331,
			ymin = 6.00295,
			%ylabel = Estimated $\ln(Z)/N$,
			ylabel = $\ln(\hat{Z})/N$,
			yticklabel style = {font=\tiny,xshift=0.5ex},
            xticklabel style = {font=\tiny,yshift=0.5ex}			
			]
\pgfplotstableread{./6/Z1.txt}\mydataone
\pgfplotstableread{./6/Z2.txt}\mydatatwo
\pgfplotstableread{./6/Z3.txt}\mydatathree
\pgfplotstableread{./6/Z4.txt}\mydatafour
\pgfplotstableread{./6/Z5.txt}\mydatafive
\pgfplotstableread{./6/Z6.txt}\mydatasix
\pgfplotstableread{./6/Z7.txt}\mydataseven
\pgfplotstableread{./6/Z8.txt}\mydataeigth
\pgfplotstableread{./6/Z9.txt}\mydatanine
\pgfplotstableread{./6/Z10.txt}\mydataten
\pgfplotstableread{./6/Z11.txt}\mydataeleven
\pgfplotstableread{./6/Z12.txt}\mydatatw
		\addplot [
		 color = black
		]		
		 table[y = Z] from \mydataone;
		 
		 \addplot [
 		 color = black,
 		 thin
		 ]
 		  table[y = Z] from \mydatasix;
 		 
 		 \addplot [
 		 color = black
 		 ]
 		  table[y = Z] from \mydatatwo;

		 \addplot [
 		 color = black,
 		thin
 		 ]
 		  table[y = Z] from \mydatathree;
 		  
 		 \addplot [
 		 color = black
		 ]
 		  table[y = Z] from \mydatafour;
 		  
 		 \addplot [
 		 color = black
		 ]
 		  table[y = Z] from \mydatafive;

   		 \addplot [
 		 color = black
		 ]
 		  table[y = Z] from \mydataseven;
 		  
   		 \addplot [
 		 color = black
		 ]
 		  table[y = Z] from \mydataeigth;
 		  
   		 \addplot [
 		 color = black
		 ]
 		  table[y = Z] from \mydatanine;
 		  
   		 \addplot [
 		 color = black
		 ]
 		  table[y = Z] from \mydataten;

  		 \addplot [
 		 color = black
		 ]
 		  table[y = Z] from \mydataeleven;

  		 \addplot [
 		 color = black
		 ]
 		  table[y = Z] from \mydatatw;
		%\legend{Importance Sampling, Uniform Sampling};
		%\addlegendentry{Uniform Sampling};

%table [x index=0, y index=1]{\mytable};
\end{axis}
\end{tikzpicture}
\caption{\label{fig:IsingFerr50}%
Experimental results for a fixed realization of an Ising model 
with \mbox{$N=50\times 50$} and
%The plots shows the 
%estimated $\ln(Z)/N$ vs.~number of samples $L$.
couplings $J_{e} \overset{\text{i.i.d.}}{\sim} \calU[2.0, 3.5]$ for all $e\in\EE$.
The plot shows the estimated $\ln(Z)/N$ using importance sampling in the dual normal
factor graph.
%Left: random couplings with $J_{b} \sim \calU[2.5, 3.0]$ for all $b\in\B$.
%Right: random couplings with $J_{b} \sim \calU[2.5, 3.0]$ for $b\in\B_\T$
%using importance sampling (solid black lines) and uniform sampling 
%(dashed blue lines) in the dual NFG of a $40\times 40$ 3-state Potts model with 
%and $J_b \sim \calU[2.0, 2.5]$ for $b\in\B_\F$.
%$b \in \B_\F$ with \textbf{(left)} $J_b \sim \calU[2.5, 3.0]$; \textbf{(right)} 
%$J_b \sim \calU[2.0, 2.5]$.}
}
\end{figure}

\Fig{fig:IsingFerr50} shows results obtained from importance sampling
for a fixed realization of an Ising model of size $N=50\times 50$ and with 
couplings $J_{e} \overset{\text{i.i.d.}}{\sim} \calU[2.0, 3.5]$ for all $e\in\EE$. The estimated $\ln(Z)/N$ is 
about $6.00314$.

%%%%%%%%%%%%%%%%%%%%%%%%%%%%%%%%%%%%%%%%%%%%%%%%%%%%%%%%%%%%%%%%%

\section{The 2D Potts Model in an External Field}
\label{sec:PottsExt}

We now extend the proposed Monte Carlo methods 
to Potts models with an external field, 
%2D ferromagnetic $q$-state Potts model with periodic 
%boundary conditions, and in the presence 
%of a nonnegative external field. 
where the Hamiltonian (\ref{eqn:HamiltonianI}) (the energy of a configuration $\x$) 
is generalized to 
\begin{IEEEeqnarray}{c}
\label{eqn:HamiltonianP}
%\mathcal{H}(\x) = - \!\!\!\!\sum_{\text{$(k,\ell)\in \B$}} \!\!\!\!J_{k, \ell}\cdot[x_k = x_{\ell}]
%- \sum_{m = 1}^N \! H_m\cdot[x_m = 0] 
%\mathcal{H}(\x) = - \!\!\sum_{(k,\ell)\in \B} J_{k, \ell} \cdot \delta(x_k - x_{\ell})
%- \sum_{m = 1}^N H_m \delta(x_m) 
\mathcal{H}(\x) = - \!\!\sum_{(k,\ell)\in \EE} J_{k, \ell}\cdot\delta(x_k - x_{\ell})
- \sum_{k = 1}^N H_k\cdot\delta(x_k), 
\end{IEEEeqnarray}
where 
%the real coupling parameter $J_{k, \ell}$ controls the strength of
%the interaction between $(x_k, x_{\ell})$ and
the real parameters $H_k$ represent the external field. 
We restrict ourselves to the standard case where
%make the standard assumption that 
the external field affects 
the variable $x_k$ only if $x_k = 0$, cf.\ \cite[Chapt.~1]{NO:11}.

We will also assume 
\begin{IEEEeqnarray}{c} 
\label{eqn:PositiveHk}
H_k \geq 0
\end{IEEEeqnarray}
for all $k$. 
%(For ease of exposition, we have assumed that the external field affects 
%the variable $x_k$ only if $x_k = 0$,
%but our approach is easily generalized to more general external fields, 
%cf.\ \cite[Chapt.~1]{NO:11}.)
The partition function is 
\begin{IEEEeqnarray}{c}
\label{eqn:PartitionfunctionDefE2}
Z = \sum_{\x \in \calX^N} e^{-\mathcal{H}(\x)}.
\end{IEEEeqnarray}

Following our approach in Section~\ref{sec:NFGD}, we can construct the 
the primal normal factor graph of the model as shown in~\Fig{fig:2DGridExt}, where
the empty boxes represent~(\ref{eqn:IsingKernelDualMod2}) and
the small empty boxes represent factors given by
\begin{equation} \label{eqn:PottsH}
\tau_{k}(x_k) = \left\{ \begin{array}{ll}
       e^{H_{k}}, & \text{if $x_k = 0$} \\
       1, & \text{otherwise.}
      \end{array}\right.   
\end{equation}

\subsection{Dual Normal Factor Graph}

The corresponding dual normal factor graph is shown in \Fig{fig:2DGridDMDualExt}. 
The only change with respect to \Fig{fig:2DGridDM} 
is the additional factors $\lambda_k(\tilde z_k)$, $k=1,\ldots,N$,
which are the 1D Fourier transforms of the factors (\ref{eqn:PottsH}) given 
by\footnote{Here, in contrast to~(\ref{eqn:FourierEdgeFactor}), a local scale factor $1/q$ is included in the
definition of the 1D Fourier transform. For the 2D torus, this makes the scale factor $Z_\text{d}/Z$ equal to 
$q^{N}$ in Potts models with or without an external field. If we do not include the local scale factor $1/q$
in (\ref{eqn:FourierTransExternalFieldDual}), we need to distinguish between two cases:
$Z_\text{d} = q^{N}Z$ for the 2D torus without an external 
field, and $Z_\text{d} = q^{2N}Z$ for the 2D torus in the presence of an external field.}
\begin{IEEEeqnarray}{c}
      \label{eqn:FourierTransExternalFieldDual}
\lambda_k(\tilde z_k) 
  =\frac{1}{q}\sum_{x_k=0}^{q-1} \tau_k(x_k) e^{-\mathrm{i}2\pi x_k \tilde z_k /q}
\end{IEEEeqnarray}
Thus
\begin{IEEEeqnarray}{c}
\lambda_k(\tilde z_k) 
  = \left\{ \begin{array}{ll}
      \dfrac{e^{H_k} -1 + q}{q},  & \text{if $\tilde z_k = 0$} \\ [2.0ex]
      \dfrac{e^{H_k} - 1}{q},     & \text{otherwise.}
      \end{array} \right.
      \label{eqn:GammaDualVarEF}
\end{IEEEeqnarray}
Note that (\ref{eqn:GammaDualVarEF}) is nonnegative due to (\ref{eqn:PositiveHk}).

%%%%%%%%%%%%%%%%%%%%%%%%%%%%%%%%%%%%%%%%%%%%%%%%%
%\vspace{2mm}

Again, from the scale factor~(\ref{eqn:NDual}), for the 2D torus we have
\begin{IEEEeqnarray}{c}
\label{eqn:ZdZEF}
Z_\text{d} = q^{N}Z.
\end{IEEEeqnarray}

\subsection{Partitioning the Variables}
\label{sec:PartionEF}

For the Potts model in an external field, the variables in the dual normal factor graph consist of
$\tilde\Y = \{ \tilde Y_e: e\in \EE \}$ 
and $\tilde\ZZ = \{ \tilde Z_k: k\in \{ 1,\ldots,N \} \}$.
Again, we partition these variables into 
$(\tilde\Y,\tilde\ZZ)_\T$ and $(\tilde\Y,\tilde\ZZ)_\F$ 
such that, in any valid configuration, the variables in $(\tilde\Y,\tilde\ZZ)_\F$ are linearly 
independent and the variables in $(\tilde\Y,\tilde\ZZ)_\T$ 
are fully determined by $(\tilde\Y,\tilde\ZZ)_\F$ via a linear transformation. 
%such that the subgraph corresponding to $(\tilde\X,\tilde\Y)_\T$
%has no cycles and reaches every zero-sum node/factor. 
%Note that 
However, the variables $\tilde \ZZ$ are not independent. %~\cite{Mo:IZS2016}.
\begin{proposition} \label{prop:tildeXSum0}
In every valid configuration, it holds that
\begin{IEEEeqnarray}{c} 
\label{eqn:PropZeroSumYk}
\sum_{k=1}^N \tilde z_k = 0 
\end{IEEEeqnarray}
\eproofnegspace
\end{proposition}

\begin{proofof}{}
Because of the zero-sum constraints on the vertices of the dual normal factor graph,
each $\tilde z_k$ (in any valid configuration) can be written as 
the sum of the variables attached to the 
corresponding zero-sum indicator function. However,
each variable will appear exactly twice in~(\ref{eqn:PropZeroSumYk}),
%(with different $k$),
once as $y_e$ and once as $-y_e$. This completes the proof. 
\end{proofof}

In the absence of an external field, the number of valid configurations in the dual normal 
factor graph of the Potts model is $q^{|\F|}$, cf. Section~\ref{sec:SpanningTree}. Furthermore, according to 
Proposition~\ref{prop:tildeXSum0}, adding an external field increases the number of independent variables (free 
components) by $N-1$, and thus increases the number of valid configurations to $q^{|\F| + N -1}$. 
Therefore, for the 2D torus in an external field, the number of valid configurations in the dual normal factor graph 
is $q^{2N}$, whereas it is $q^N$ in the primal normal factor graph.

Both the uniform sampling algorithm 
and the importance sampling algorithm of Section~\ref{sec:IS} can be adapted to the present setting. 
We describe only the latter for two such partitionings. 
The first partitioning is suitable for models in the low-temperature regime, and the second one 
is designed for models 
that are in the presence of a strong
external field.
The performance of the proposed algorithms depends on the choice of the partitioning, as will be illustrated by our numerical experiments in Section~\ref{sec:NumExpExtFieldIS}.

%\subsection{Monte Carlo Algorithms}
\subsection{Importance Sampling for Models at Low Temperature}

An obvious choice for $(\tilde\Y,\tilde\ZZ)_\F$ with the required properties includes $\tilde\Y_\T$ (i.e., a spanning 
tree as in Section~\ref{sec:SpanningTree}) and 
$N-1$ components of $\tilde \ZZ$, i.e.,
\begin{IEEEeqnarray}{c}
\label{eqn:BasicPartitionE}
(\tilde\Y,\tilde\ZZ)_\F = (\tilde\Y_\F, \tilde \ZZ\!\setminus\! \tilde Z_1),
\end{IEEEeqnarray}
which implies
\begin{IEEEeqnarray}{c}
\label{eqn:BasicPartitionE1}
(\tilde\Y,\tilde\ZZ)_\T = (\tilde\Y_\T, \tilde Z_1).
\end{IEEEeqnarray}

%%%%%%%%%%%%%%%%%%%%%%%%%%%%%%%%%%%%%%%%%%%%%%%%%%%%%%%%%%%%%
\begin{figure}[t]
\setlength{\unitlength}{0.88mm}
\centering
\begin{picture}(89,78)(-6,-6)
%\put(-6,-6){\dashbox(89,80){}}
\small
\put(0,60){\framebox(4,4){$=$}}
\put(4,62){\line(1,0){8}}           %\put(8,63){\pos{bc}{$\X_{1,2}$}}
\put(12,60){\framebox(4,4){$+$}}
%%%%%%%%%%%%%%%%%%%
\put(15.95,61){$\circ$}
\put(17.38,62){\line(1,0){6.6}}
%%%%%%%%%%%%%%%%%%
\put(24,60){\framebox(4,4){$=$}}
\put(28,62){\line(1,0){8}}         %\put(32,63){\pos{bc}{$\tilde X_{2,3}$}}
\put(36,60){\framebox(4,4){$+$}}
%%%%%%%%%%%%%%%%%%
\put(39.95,61){$\circ$}
\put(41.38,62){\line(1,0){6.6}}
%%%%%%%%%%%%%%%%%
\put(48,60){\framebox(4,4){$=$}}
\put(52,62){\line(1,0){8}}      
\put(60,60){\framebox(4,4){$+$}}
%%%%%%%%%%%%%%%%%
\put(63.95,61){$\circ$}
\put(65.38,62){\line(1,0){6.6}}
%%%%%%%%%%%%%%%%
\put(72,60){\framebox(4,4){$=$}}
%%%%%%%%%%%%%%%%%%%
%%%%%%%%%%%%%%%%%%%
\put(0,50){\framebox(4,4){$+$}}
%%%%%%%%%%%%%
\put(1.2,48.45){$\circ$}
%%%%%%%%%%%%%
\put(24,50){\framebox(4,4){$+$}}
%%%%%%%%%%%%%
\put(25.2,48.45){$\circ$}
%%%%%%%%%%%%%
\put(48,50){\framebox(4,4){$+$}}
%%%%%%%%%%%%%
\put(49.2,48.45){$\circ$}
%%%%%%%%%%%%%
\put(72,50){\framebox(4,4){$+$}}
%%%%%%%%%%%%%
\put(73.2,48.45){$\circ$}
%%%%%%%%%%%%%%%%%%
%%%%%%%%%%%%%%%%%%
\put(0,40){\framebox(4,4){$=$}}
\put(4,42){\line(1,0){8}}
\put(12,40){\framebox(4,4){$+$}}
%%%%%%%%%%%%%%%%%
\put(15.95,41.2){$\circ$}
\put(17.4,42){\line(1,0){6.6}}
%%%%%%%%%%%%%%%%%
\put(24,40){\framebox(4,4){$=$}}
\put(28,42){\line(1,0){8}}
\put(36,40){\framebox(4,4){$+$}}
%%%%%%%%%%%%%%%%%
\put(39.95,41.2){$\circ$}
\put(41.4,42){\line(1,0){6.6}}
%%%%%%%%%%%%%%%%%
\put(48,40){\framebox(4,4){$=$}}
\put(52,42){\line(1,0){8}}
\put(60,40){\framebox(4,4){$+$}}
%%%%%%%%%%%%%%%%%
\put(63.98,41.2){$\circ$}
\put(65.38,42){\line(1,0){6.6}}
%%%%%%%%%%%%%%%%%
\put(72,40){\framebox(4,4){$=$}}
%%%%%%%%%%%%%%%%%%%%%
%%%%%%%%%%%%%%%%%%%%%
\put(0,30){\framebox(4,4){$+$}}
%%%%%%%%%%%%%
\put(1.2,28.45){$\circ$}
%%%%%%%%%%%%%
\put(24,30){\framebox(4,4){$+$}}
%%%%%%%%%%%%%
\put(25.2,28.45){$\circ$}
%%%%%%%%%%%%%
\put(48,30){\framebox(4,4){$+$}}
%%%%%%%%%%%%%
\put(49.2,28.45){$\circ$}
%%%%%%%%%%%%
\put(72,30){\framebox(4,4){$+$}}
%%%%%%%%%%%%%
\put(73.2,28.45){$\circ$}
%%%%%%%%%%%%%%%%%%%%
%%%%%%%%%%%%%%%%%%%%
\put(0,20){\framebox(4,4){$=$}}
\put(4,22){\line(1,0){8}}
\put(12,20){\framebox(4,4){$+$}}
%%%%%%%%%%%%%%%%
\put(15.95,21){$\circ$}
\put(17.38,22){\line(1,0){6.6}}
%%%%%%%%%%%%%%%%
\put(24,20){\framebox(4,4){$=$}}
\put(28,22){\line(1,0){8}}
\put(36,20){\framebox(4,4){$+$}}
%%%%%%%%%%%%%%%
\put(39.95,21){$\circ$}
\put(41.38,22){\line(1,0){6.6}}
%%%%%%%%%%%%%%%
\put(48,20){\framebox(4,4){$=$}}
\put(52,22){\line(1,0){8}}
\put(60,20){\framebox(4,4){$+$}}
%%%%%%%%%%%%%%%
\put(63.98,21){$\circ$}
\put(65.38,22){\line(1,0){6.6}}
%%%%%%%%%%%%%%%
\put(72,20){\framebox(4,4){$=$}}
%%%%%%%%%%%%%%%%%%%%%
%%%%%%%%%%%%%%%%%%%%%
\put(0,10){\framebox(4,4){$+$}}
%%%%%%%%%%%%%
\put(1.1,8.45){$\circ$}
%%%%%%%%%%%%%
\put(24,10){\framebox(4,4){$+$}}
%%%%%%%%%%%%%
\put(25.1,8.45){$\circ$}
%%%%%%%%%%%%%
\put(48,10){\framebox(4,4){$+$}}
%%%%%%%%%%%%%
\put(49.1,8.45){$\circ$}
%%%%%%%%%%%%%
\put(72,10){\framebox(4,4){$+$}}
%%%%%%%%%%%%%%
\put(73.1,8.45){$\circ$}
%%%%%%%%%%%%%%%%%%%%
\put(0,0){\framebox(4,4){$=$}}
\put(12,0){\framebox(4,4){$+$}}
%%%%%%%%%%%%%
\put(15.95,1.2){$\circ$}
%%%%%%%%%%%%%
\put(24,0){\framebox(4,4){$=$}}
\put(36,0){\framebox(4,4){$+$}}
%%%%%%%%%%%%%
\put(39.96,1.2){$\circ$}
%%%%%%%%%%%%%
\put(48,0){\framebox(4,4){$=$}}
\put(60,0){\framebox(4,4){$+$}}
%%%%%%%%%%%%%
\put(63.96,1.2){$\circ$}
%%%%%%%%%%%%%
\put(72,0){\framebox(4,4){$=$}}
%%%%%%%%%%%%%%%%%%%%%
%%%
\put(14,64){\line(0,1){2}}
\put(38,64){\line(0,1){2}}
\put(62,64){\line(0,1){2}}
\put(12,66){\framebox(4,4){}}   \put(10,67.2){\pos{cb}{$\kappa_{1}$}}% \put(14,71){\pos{cb}{$\kappa_{1}$}}
\put(36,66){\framebox(4,4){}}   %\put(38,71){\pos{cb}{$\gamma_{2,3}$}}
\put(60,66){\framebox(4,4){}}
\put(14,44){\line(0,1){2}}
\put(38,44){\line(0,1){2}}
\put(62,44){\line(0,1){2}}
\put(12,46){\framebox(4,4){$$}}
\put(36,46){\framebox(4,4){$$}}
\put(60,46){\framebox(4,4){$$}}
\put(14,24){\line(0,1){2}}
\put(38,24){\line(0,1){2}}
\put(62,24){\line(0,1){2}}
\put(12,26){\framebox(4,4){$$}}
\put(36,26){\framebox(4,4){$$}}
\put(60,26){\framebox(4,4){$$}}
\put(14,4){\line(0,1){2}}
\put(38,4){\line(0,1){2}}
\put(62,4){\line(0,1){2}}
\put(12,6){\framebox(4,4){$$}}
\put(36,6){\framebox(4,4){$$}}
\put(60,6){\framebox(4,4){$$}}
\put(0,52){\line(-1,0){2}}
\put(24,52){\line(-1,0){2}}
\put(48,52){\line(-1,0){2}}
\put(72,52){\line(-1,0){2}}
\put(-6,50){\framebox(4,4){$$}}
\put(18,50){\framebox(4,4){$$}}
\put(42,50){\framebox(4,4){$$}}
\put(66,50){\framebox(4,4){$$}}
\put(0,32){\line(-1,0){2}}
\put(24,32){\line(-1,0){2}}
\put(48,32){\line(-1,0){2}}
\put(72,32){\line(-1,0){2}}
\put(-6,30){\framebox(4,4){$$}}
\put(18,30){\framebox(4,4){$$}}
\put(42,30){\framebox(4,4){$$}}
\put(66,30){\framebox(4,4){$$}}
\put(0,12){\line(-1,0){2}}
\put(24,12){\line(-1,0){2}}
\put(48,12){\line(-1,0){2}}
\put(72,12){\line(-1,0){2}}
\put(-6,10){\framebox(4,4){$$}}
\put(18,10){\framebox(4,4){$$}}
\put(42,10){\framebox(4,4){$$}}
\put(66,10){\framebox(4,4){$$}}

\put(4,62){\line(1,0){8}}        

\put(28,62){\line(1,0){8}}       

\put(52,62){\line(1,0){8}}      

\put(4,42){\line(1,0){8}}

\put(28,42){\line(1,0){8}}

\put(52,42){\line(1,0){8}}

\put(4,22){\line(1,0){8}}

\put(28,22){\line(1,0){8}}

\put(52,22){\line(1,0){8}}

\put(14,64){\line(0,1){2}}
\put(38,64){\line(0,1){2}}
\put(62,64){\line(0,1){2}}
\put(14,44){\line(0,1){2}}
\put(38,44){\line(0,1){2}}
\put(62,44){\line(0,1){2}}
\put(14,24){\line(0,1){2}}
\put(38,24){\line(0,1){2}}
\put(62,24){\line(0,1){2}}

%%%%%%%%%%%%%%%%%%%%%%%%%%%%%%%%%
\put(2,54){\line(0,1){6}}
\put(2,48.7){\line(0,-1){4.6}}
\put(26,54){\line(0,1){6}}
\put(26,48.7){\line(0,-1){4.6}}
\put(50,54){\line(0,1){6}}
\put(50,48.7){\line(0,-1){4.6}}
\put(74,54){\line(0,1){6}}
\put(74,48.7){\line(0,-1){4.6}}
\put(2,34){\line(0,1){6}}
\put(2.0,28.7){\line(0,-1){4.6}}
\put(26,34){\line(0,1){6}}
\put(26.0,28.7){\line(0,-1){4.6}}
\put(50,34){\line(0,1){6}}
\put(50,28.7){\line(0,-1){4.6}}
\put(74,34){\line(0,1){6}}
\put(74,28.7){\line(0,-1){4.6}}
\put(2,14){\line(0,1){6}}
\put(2.0,8.7){\line(0,-1){4.6}}
\put(26,14){\line(0,1){6}}
\put(26.0,8.7){\line(0,-1){4.6}}
\put(50,14){\line(0,1){6}}
\put(50.0,8.7){\line(0,-1){4.6}}
\put(74,14){\line(0,1){6}}
\put(74,8.7){\line(0,-1){4.6}}
\put(4,2){\line(1,0){8}}
\put(17.38,2){\line(1,0){6.6}}
\put(28,2){\line(1,0){8}}
\put(41.38,2){\line(1,0){6.6}}
\put(52,2){\line(1,0){8}}
\put(65.38,2){\line(1,0){6.6}}

\put(4,60){\line(4,-3){4}}        %\put(7,58){\pos{tr}{$\tilde Y_1$}}
 \put(8,54){\framebox(3,3){}}     \put(12,55){\pos{cl}{$\tau_1$}}
\put(28,60){\line(4,-3){4}}       %\put(31,58){\pos{tr}{$\tilde Y_2$}}
 \put(32,54){\framebox(3,3){}}    %\put(36,55){\pos{cl}{$\lambda_2$}}
\put(52,60){\line(4,-3){4}}
 \put(56,54){\framebox(3,3){}}
 \put(76,60){\line(4,-3){4}}
 \put(80,54){\framebox(3,3){}}
\put(4,40){\line(4,-3){4}}
 \put(8,34){\framebox(3,3){}}
 \put(28,40){\line(4,-3){4}}
 \put(32,34){\framebox(3,3){}}
 \put(52,40){\line(4,-3){4}}
 \put(56,34){\framebox(3,3){}}
 \put(76,40){\line(4,-3){4}}
 \put(80,34){\framebox(3,3){}}
 \put(4,0){\line(4,-3){4}}
 \put(8,-6){\framebox(3,3){}}
 \put(28,0){\line(4,-3){4}}
\put(32,-6){\framebox(3,3){}}
 \put(52,0){\line(4,-3){4}}
 \put(56,-6){\framebox(3,3){}}
 \put(76,0){\line(4,-3){4}}
 \put(80,-6){\framebox(3,3){}}
\put(4,20){\line(4,-3){4}}
 \put(8,14){\framebox(3,3){}}
 \put(28,20){\line(4,-3){4}}
 \put(32,14){\framebox(3,3){}}
 \put(52,20){\line(4,-3){4}}
 \put(56,14){\framebox(3,3){}}
 \put(76,20){\line(4,-3){4}}
 \put(80,14){\framebox(3,3){}}
%%%%%%%%%%%%%%%%%%%%%
\end{picture}
\vspace{2ex}
\caption{\label{fig:2DGridExt}%
Primal normal factor graph of the 2D Potts model in an external field. 
The only change with respect to \Fig{fig:2DGridM}
is the additional factors $\tau_k$ given by (\ref{eqn:PottsH}). 
%The empty boxes represent the factors~(\ref{eqn:IsingA}), 
%the small empty boxes represent the factors~(\ref{eqn:PottsH}), 
%and the boxes labeled ``$=$'' are equality indicator functionss as in~(\ref{eqn:DefEqFact}).
The periodic boundary conditions are not shown.
%xxx where
%the unlabeled boxes represent factors~(\ref{eqn:IsingA}), the 
%unlabeled small boxes represent~(\ref{eqn:PottsH}), and 
%boxes containing ``$=$" symbols are equality indicator 
%factors given by~(\ref{eqn:DefEqFact}).
}
\end{figure}
%\vspace{\floatsep}

%We will assume 
%a partitioning of the variables as in (\ref{eqn:BasicPartitionE}).
The quantities $\Gamma_\F$ and $\Gamma_\T$ 
from (\ref{eqn:GammaF}) and (\ref{eqn:GammaT})
are then generalized to 
\begin{equation}
\Gamma_\F\big( (\tilde\y, \tilde\z)_\F \big)
 = \prod_{e\in\F} \gamma_e(\tilde y_e) \prod_{k=2}^N \lambda_k(\tilde z_k)
\end{equation}
and
\begin{equation}
\Gamma_\T\big( (\tilde\y, \tilde\z)_\T \big)
=  \lambda_1(\tilde z_1) \prod_{e\in \T} \gamma_e(\tilde y_e),
\end{equation}
respectively. 

The quantity $Z_\F$ from (\ref{eqn:ZFis}) is generalized to 
\begin{IEEEeqnarray}{rCl}
Z_\F & = & \sum_{(\tilde\y, \tilde\z)_\F} \Gamma_\F\big( (\tilde\y, \tilde\z)_\F \big) \\
 & = & \left( \prod_{e\in\F} \sum_{\xi=0}^{q-1} \gamma_e(\xi) \right)
       \left( \prod_{k=2}^N \sum_{\xi'=0}^{q-1} \lambda_k(\xi') \right) \\
 & = & \left( \prod_{e\in\F} q e^{J_e}\right) 
       \left( \prod_{k=2}^N e^{H_k} \right) \\
 & = & q^{|\F|} \exp\!\left(\, \sum_{e \in\F} J_e + \sum_{k=2}^N H_k \right).
\end{IEEEeqnarray}

The algorithm then goes as follows. 
\begin{enumerate}
\item
Generate $L$ independent samples 
$\smpl{(\tilde\x, \tilde\y)_\F}{1}, \ldots, \smpl{(\tilde\y, \tilde\z)_\F}{L}$ 
from the distribution
\begin{IEEEeqnarray}{c}
\label{eqn:AuxProbISE}
p_\F\big( (\tilde\y, \tilde\z)_\F \big) =
  \frac{\Gamma_\F\big( (\tilde\y, \tilde\z)_\F \big)}{Z_\F}
\end{IEEEeqnarray}
\item
For each sample $\smpl{(\tilde\y, \tilde\z)_\F}{\ell}$,
compute its unique extension $\smpl{(\tilde\y, \tilde\z)_\T}{\ell}$ 
to a valid configuration $\smpl{(\tilde\y, \tilde\z)}{\ell}$, including $\tilde z_1$, which
 can be computed as
\begin{IEEEeqnarray}{c}
\tilde z_1 = - \sum_{m=2}^N \tilde z_{m}
\end{IEEEeqnarray}
from~(\ref{eqn:PropZeroSumYk}).
\item
Compute the estimate
\begin{IEEEeqnarray}{c}
\label{eqn:hatZisE}
\hat Z^{\text{Imp}}_\text{d} = \frac{Z_\F}{L} \sum_{\ell=1}^L 
  \Gamma_\T\big( \smpl{(\tilde\y, \tilde\z)_\T}{\ell} \big).
\end{IEEEeqnarray}
\end{enumerate}

\begin{figure}[t!!!]
\setlength{\unitlength}{0.88mm}
\centering
\begin{picture}(89,78)(-6,-6)
%\put(-6,-6){\dashbox(89,80){}}
\small
\put(0,60){\framebox(4,4){$+$}}
\put(4,62){\line(1,0){8}}           %\put(8,63){\pos{bc}{$\tilde X_{1,2}$}}
\put(12,60){\framebox(4,4){$=$}}
%%%%%%%%%%%%%%%%%%%
\put(15.95,61){$\circ$}
\put(17.38,62){\line(1,0){6.6}}
%%%%%%%%%%%%%%%%%%
\put(24,60){\framebox(4,4){$+$}}
\put(28,62){\line(1,0){8}}         %\put(32,63){\pos{bc}{$\tilde X_{2,3}$}}
\put(36,60){\framebox(4,4){$=$}}
%%%%%%%%%%%%%%%%%%
\put(39.95,61){$\circ$}
\put(41.38,62){\line(1,0){6.6}}
%%%%%%%%%%%%%%%%%
\put(48,60){\framebox(4,4){$+$}}
\put(52,62){\line(1,0){8}}      
\put(60,60){\framebox(4,4){$=$}}
%%%%%%%%%%%%%%%%%
\put(63.95,61){$\circ$}
\put(65.38,62){\line(1,0){6.6}}
%%%%%%%%%%%%%%%%
\put(72,60){\framebox(4,4){$+$}}
%%%%%%%%%%%%%%%%%%%
%%%%%%%%%%%%%%%%%%%
\put(0,50){\framebox(4,4){$=$}}
%%%%%%%%%%%%%
\put(1.2,48.45){$\circ$}
%%%%%%%%%%%%%
\put(24,50){\framebox(4,4){$=$}}
%%%%%%%%%%%%%
\put(25.2,48.45){$\circ$}
%%%%%%%%%%%%%
\put(48,50){\framebox(4,4){$=$}}
%%%%%%%%%%%%%
\put(49.2,48.45){$\circ$}
%%%%%%%%%%%%%
\put(72,50){\framebox(4,4){$=$}}
%%%%%%%%%%%%%
\put(73.2,48.45){$\circ$}
%%%%%%%%%%%%%%%%%%
%%%%%%%%%%%%%%%%%%
\put(0,40){\framebox(4,4){$+$}}
\put(4,42){\line(1,0){8}}
\put(12,40){\framebox(4,4){$=$}}
%%%%%%%%%%%%%%%%%
\put(15.95,41.2){$\circ$}
\put(17.4,42){\line(1,0){6.6}}
%%%%%%%%%%%%%%%%%
\put(24,40){\framebox(4,4){$+$}}
\put(28,42){\line(1,0){8}}
\put(36,40){\framebox(4,4){$=$}}
%%%%%%%%%%%%%%%%%
\put(39.95,41.2){$\circ$}
\put(41.4,42){\line(1,0){6.6}}
%%%%%%%%%%%%%%%%%
\put(48,40){\framebox(4,4){$+$}}
\put(52,42){\line(1,0){8}}
\put(60,40){\framebox(4,4){$=$}}
%%%%%%%%%%%%%%%%%
\put(63.98,41.2){$\circ$}
\put(65.38,42){\line(1,0){6.6}}
%%%%%%%%%%%%%%%%%
\put(72,40){\framebox(4,4){$+$}}
%%%%%%%%%%%%%%%%%%%%%
%%%%%%%%%%%%%%%%%%%%%
\put(0,30){\framebox(4,4){$=$}}
%%%%%%%%%%%%%
\put(1.2,28.45){$\circ$}
%%%%%%%%%%%%%
\put(24,30){\framebox(4,4){$=$}}
%%%%%%%%%%%%%
\put(25.2,28.45){$\circ$}
%%%%%%%%%%%%%
\put(48,30){\framebox(4,4){$=$}}
%%%%%%%%%%%%%
\put(49.2,28.45){$\circ$}
%%%%%%%%%%%%
\put(72,30){\framebox(4,4){$=$}}
%%%%%%%%%%%%%
\put(73.2,28.45){$\circ$}
%%%%%%%%%%%%%%%%%%%%
%%%%%%%%%%%%%%%%%%%%
\put(0,20){\framebox(4,4){$+$}}
\put(4,22){\line(1,0){8}}
\put(12,20){\framebox(4,4){$=$}}
%%%%%%%%%%%%%%%%
\put(15.95,21){$\circ$}
\put(17.38,22){\line(1,0){6.6}}
%%%%%%%%%%%%%%%%
\put(24,20){\framebox(4,4){$+$}}
\put(28,22){\line(1,0){8}}
\put(36,20){\framebox(4,4){$=$}}
%%%%%%%%%%%%%%%
\put(39.95,21){$\circ$}
\put(41.38,22){\line(1,0){6.6}}
%%%%%%%%%%%%%%%
\put(48,20){\framebox(4,4){$+$}}
\put(52,22){\line(1,0){8}}
\put(60,20){\framebox(4,4){$=$}}
%%%%%%%%%%%%%%%
\put(63.98,21){$\circ$}
\put(65.38,22){\line(1,0){6.6}}
%%%%%%%%%%%%%%%
\put(72,20){\framebox(4,4){$+$}}
%%%%%%%%%%%%%%%%%%%%%
%%%%%%%%%%%%%%%%%%%%%
\put(0,10){\framebox(4,4){$=$}}
%%%%%%%%%%%%%
\put(1.1,8.45){$\circ$}
%%%%%%%%%%%%%
\put(24,10){\framebox(4,4){$=$}}
%%%%%%%%%%%%%
\put(25.1,8.45){$\circ$}
%%%%%%%%%%%%%
\put(48,10){\framebox(4,4){$=$}}
%%%%%%%%%%%%%
\put(49.1,8.45){$\circ$}
%%%%%%%%%%%%%
\put(72,10){\framebox(4,4){$=$}}
%%%%%%%%%%%%%%
\put(73.1,8.45){$\circ$}
%%%%%%%%%%%%%%%%%%%%
\put(0,0){\framebox(4,4){$+$}}
\put(12,0){\framebox(4,4){$=$}}
%%%%%%%%%%%%%
\put(15.95,1.2){$\circ$}
%%%%%%%%%%%%%
\put(24,0){\framebox(4,4){$+$}}
\put(36,0){\framebox(4,4){$=$}}
%%%%%%%%%%%%%
\put(39.96,1.2){$\circ$}
%%%%%%%%%%%%%
\put(48,0){\framebox(4,4){$+$}}
\put(60,0){\framebox(4,4){$=$}}
%%%%%%%%%%%%%
\put(63.96,1.2){$\circ$}
%%%%%%%%%%%%%
\put(72,0){\framebox(4,4){$+$}}
%%%%%%%%%%%%%%%%%%%%%
%%%
\put(14,64){\line(0,1){2}}
\put(38,64){\line(0,1){2}}
\put(62,64){\line(0,1){2}}
\put(12,66){\framebox(4,4){}}    \put(10,67.2){\pos{cb}{$\gamma_{1}$}}%\put(14,71){\pos{cb}{$\gamma_{1}$}}
\put(36,66){\framebox(4,4){}}   %\put(38,71){\pos{cb}{$\gamma_{2,3}$}}
\put(60,66){\framebox(4,4){}}
\put(14,44){\line(0,1){2}}
\put(38,44){\line(0,1){2}}
\put(62,44){\line(0,1){2}}
\put(12,46){\framebox(4,4){$$}}
\put(36,46){\framebox(4,4){$$}}
\put(60,46){\framebox(4,4){$$}}
\put(14,24){\line(0,1){2}}
\put(38,24){\line(0,1){2}}
\put(62,24){\line(0,1){2}}
\put(12,26){\framebox(4,4){$$}}
\put(36,26){\framebox(4,4){$$}}
\put(60,26){\framebox(4,4){$$}}
\put(14,4){\line(0,1){2}}
\put(38,4){\line(0,1){2}}
\put(62,4){\line(0,1){2}}
\put(12,6){\framebox(4,4){$$}}
\put(36,6){\framebox(4,4){$$}}
\put(60,6){\framebox(4,4){$$}}
\put(0,52){\line(-1,0){2}}
\put(24,52){\line(-1,0){2}}
\put(48,52){\line(-1,0){2}}
\put(72,52){\line(-1,0){2}}
\put(-6,50){\framebox(4,4){$$}}
\put(18,50){\framebox(4,4){$$}}
\put(42,50){\framebox(4,4){$$}}
\put(66,50){\framebox(4,4){$$}}
\put(0,32){\line(-1,0){2}}
\put(24,32){\line(-1,0){2}}
\put(48,32){\line(-1,0){2}}
\put(72,32){\line(-1,0){2}}
\put(-6,30){\framebox(4,4){$$}}
\put(18,30){\framebox(4,4){$$}}
\put(42,30){\framebox(4,4){$$}}
\put(66,30){\framebox(4,4){$$}}
\put(0,12){\line(-1,0){2}}
\put(24,12){\line(-1,0){2}}
\put(48,12){\line(-1,0){2}}
\put(72,12){\line(-1,0){2}}
\put(-6,10){\framebox(4,4){$$}}
\put(18,10){\framebox(4,4){$$}}
\put(42,10){\framebox(4,4){$$}}
\put(66,10){\framebox(4,4){$$}}

\put(4,62){\line(1,0){8}}        

\put(28,62){\line(1,0){8}}       

\put(52,62){\line(1,0){8}}      

\put(4,42){\line(1,0){8}}

\put(28,42){\line(1,0){8}}

\put(52,42){\line(1,0){8}}

\put(4,22){\line(1,0){8}}

\put(28,22){\line(1,0){8}}

\put(52,22){\line(1,0){8}}

\put(14,64){\line(0,1){2}}
\put(38,64){\line(0,1){2}}
\put(62,64){\line(0,1){2}}
\put(14,44){\line(0,1){2}}
\put(38,44){\line(0,1){2}}
\put(62,44){\line(0,1){2}}
\put(14,24){\line(0,1){2}}
\put(38,24){\line(0,1){2}}
\put(62,24){\line(0,1){2}}

%%%%%%%%%%%%%%%%%%%%%%%%%%%%%%%%%
\put(2,54){\line(0,1){6}}
\put(2,48.7){\line(0,-1){4.6}}
\put(26,54){\line(0,1){6}}
\put(26,48.7){\line(0,-1){4.6}}
\put(50,54){\line(0,1){6}}
\put(50,48.7){\line(0,-1){4.6}}
\put(74,54){\line(0,1){6}}
\put(74,48.7){\line(0,-1){4.6}}
\put(2,34){\line(0,1){6}}
\put(2.0,28.7){\line(0,-1){4.6}}
\put(26,34){\line(0,1){6}}
\put(26.0,28.7){\line(0,-1){4.6}}
\put(50,34){\line(0,1){6}}
\put(50,28.7){\line(0,-1){4.6}}
\put(74,34){\line(0,1){6}}
\put(74,28.7){\line(0,-1){4.6}}
\put(2,14){\line(0,1){6}}
\put(2.0,8.7){\line(0,-1){4.6}}
\put(26,14){\line(0,1){6}}
\put(26.0,8.7){\line(0,-1){4.6}}
\put(50,14){\line(0,1){6}}
\put(50.0,8.7){\line(0,-1){4.6}}
\put(74,14){\line(0,1){6}}
\put(74,8.7){\line(0,-1){4.6}}
\put(4,2){\line(1,0){8}}
\put(17.38,2){\line(1,0){6.6}}
\put(28,2){\line(1,0){8}}
\put(41.38,2){\line(1,0){6.6}}
\put(52,2){\line(1,0){8}}
\put(65.38,2){\line(1,0){6.6}}

\put(4,60){\line(4,-3){4}}        %\put(7,58){\pos{tr}{$\tilde Y_1$}}
 \put(8,54){\framebox(3,3){}}     \put(12,55){\pos{cl}{$\lambda_1$}}
\put(28,60){\line(4,-3){4}}       %\put(31,58){\pos{tr}{$\tilde Y_2$}}
 \put(32,54){\framebox(3,3){}}    %\put(36,55){\pos{cl}{$\lambda_2$}}
\put(52,60){\line(4,-3){4}}
 \put(56,54){\framebox(3,3){}}
 \put(76,60){\line(4,-3){4}}
 \put(80,54){\framebox(3,3){}}
\put(4,40){\line(4,-3){4}}
 \put(8,34){\framebox(3,3){}}
 \put(28,40){\line(4,-3){4}}
 \put(32,34){\framebox(3,3){}}
 \put(52,40){\line(4,-3){4}}
 \put(56,34){\framebox(3,3){}}
 \put(76,40){\line(4,-3){4}}
 \put(80,34){\framebox(3,3){}}
 \put(4,0){\line(4,-3){4}}
 \put(8,-6){\framebox(3,3){}}
 \put(28,0){\line(4,-3){4}}
\put(32,-6){\framebox(3,3){}}
 \put(52,0){\line(4,-3){4}}
 \put(56,-6){\framebox(3,3){}}
 \put(76,0){\line(4,-3){4}}
 \put(80,-6){\framebox(3,3){}}
\put(4,20){\line(4,-3){4}}
 \put(8,14){\framebox(3,3){}}
 \put(28,20){\line(4,-3){4}}
 \put(32,14){\framebox(3,3){}}
 \put(52,20){\line(4,-3){4}}
 \put(56,14){\framebox(3,3){}}
 \put(76,20){\line(4,-3){4}}
 \put(80,14){\framebox(3,3){}}
\end{picture}
\vspace{2ex}
\caption{\label{fig:2DGridDMDualExt}%
Dual normal factor graph of the 2D Potts model in an external field. 
The only change with respect to \Fig{fig:2DGridDM}
is the additional factors $\lambda_k$ given by (\ref{eqn:GammaDualVarEF}). 
The periodic boundary conditions are not shown.
%for $k=1,\ldots,N$. 
%Dual NFG of the 2D Potts model, where 
%the unlabeled boxes represent~(\ref{eqn:IsingKernelDual}), 
%unlabeled small boxes represent~(\ref{eqn:IsingKernelDualExternalF}), 
%boxes containing ``$=$" 
%symbols are given by~(\ref{eqn:DefEqFact}),
%boxes containing ``$+$" symbols are given by~(\ref{eqn:DefModqsumFact}), and the 
%small circles denoted by ``$\circ$" represent 
%sign inverters.
}
\end{figure}

%%%%%%%%%%%%%%%%%%%%%%%%%%%%%%%%%%%%%%%%%
\begin{figure*}[t!!]
\centering
\begin{tikzpicture}[scale = 0.81]
\begin{axis}[
			title = {$H = 0.05$},
			height = 39.0ex,
			grid = major,
			tick pos=left, 
			ymode=log,
			xminorticks = false,
		    yminorticks = false,	
		    y tick label style={
        /pgf/number format/.cd,
            fixed,
        /tikz/.cd
    		}, 				
			ytick={1e-7, 1e-6, 1e-5, 1e-4, 1e-3, 1e-2, 1e-1, 1},
			xtick={0.1, 0.5, 0.9, 1.3, 1.7, 2.1, 2.5},
		        xlabel= \phantom{$J$} ={font=\normalsize},
			xmin = 0.1,
			xmax = 2.5,
			ymin = 0.0000001,
			ymax = 1,
			ylabel = Relative Error,
			yticklabel style = {font=\tiny,yshift=0.5ex},
            xticklabel style = {font=\tiny,xshift=0.0ex}			
			]
\pgfplotstableread{./7/BP.txt}\mydataone
\pgfplotstableread{./7/GBPLoop4.txt}\mydatatwo
\pgfplotstableread{./7/Tree.txt}\mydatathree
\pgfplotstableread{./7/Unif.txt}\mydatafour
\pgfplotstableread{./7/Imp.txt}\mydatafive

		\addplot [
		 color = black,
		 mark = o,
		]		
		 table[y = Z] from \mydataone;
		 
 		 \addplot [
 		 color = black,
 		 mark = triangle,
 		 ]
 		  table[y = Z] from \mydatatwo;	 
 
  		 \addplot [
 		 color = black,
 		 mark = star,
 		 ]
 		  table[y = Z] from \mydatathree;	
 		  
 		 \addplot [
 		 color = blue,
 		 mark = x,
 		 ]
 		  table[y = Z] from \mydatafour;	
 		  
 		 \addplot [
 		 color = blue,
 		 mark = diamond,
 		 ]
 		  table[y = Z] from \mydatafive;

 		 %\legend{BP, GBPLoop, TreeEP, Uni. Sam., Imp. Sam.};	  	

\end{axis}
\end{tikzpicture}
\hfill
\begin{tikzpicture}[scale = 0.81]
\begin{axis}[
			title = {$H = 0.1$},
			height = 39.0ex,
			grid = major,
			tick pos=left, 
			ymode=log,
			xminorticks = false,	
		    yminorticks = false,
		    ymajorticks = false,
                        ymajorgrids= true,		
		    y tick label style={
                    /pgf/number format/.cd,
                    fixed,
                    /tikz/.cd
    		    }, 	
                        yticklabels=\empty,						
			ytick= {1e-7, 1e-6, 1e-5, 1e-4, 1e-3, 1e-2, 1e-1, 1},
			xtick={0.1, 0.5, 0.9, 1.3, 1.7, 2.1, 2.5},
		xlabel= $J$ ={font=\normalsize},
			xmin = 0.1,
			xmax = 2.5,
			ymin = 0.0000001,
			ymax = 1,
			%ylabel = Relative Error,
			%yticklabel style = {font=\tiny,yshift=0.5ex},
            xticklabel style = {font=\tiny,xshift=0.0ex}			
			]
\pgfplotstableread{./8/BP.txt}\mydataone
\pgfplotstableread{./8/GBPLoop4.txt}\mydatatwo
\pgfplotstableread{./8/Tree.txt}\mydatathree
\pgfplotstableread{./8/Unif.txt}\mydatafour
\pgfplotstableread{./8/Imp.txt}\mydatafive

		\addplot [
		 color = black,
		 mark = o,
		]		
		 table[y = Z] from \mydataone;
		 
 		 \addplot [
 		 color = black,
 		 mark = triangle,
 		 ]
 		  table[y = Z] from \mydatatwo;	 
 
  		 \addplot [
 		 color = black,
 		 mark = star,
 		 ]
 		  table[y = Z] from \mydatathree;	
 		  
 		 \addplot [
 		 color = blue,
 		 mark = x,
 		 ]
 		  table[y = Z] from \mydatafour;	
 		  
 		 \addplot [
 		 color = blue,
 		 mark = diamond,
 		 ]
 		  table[y = Z] from \mydatafive;

 		 %\legend{BP, GBP, TreeEP, Uni, Imp};	  
\end{axis}
\end{tikzpicture}
\hfill
\begin{tikzpicture}[scale = 0.81]
\begin{axis}[
			title = {$H = 0.2$},
			legend style={at = {(0.975,0.72)} ,font=\tiny},		
			height = 39.0ex,
			grid = major,
			tick pos=left, 
			ymode=log,
			xminorticks = false,	
		    yminorticks = false,
		    ymajorticks = false,
                        ymajorgrids= true,		
		    y tick label style={
        /pgf/number format/.cd,
            fixed,
        /tikz/.cd
    		}, 	
                        yticklabels=\empty,			
			ytick = {1e-7, 1e-6, 1e-5, 1e-4, 1e-3, 1e-2, 1e-1, 1},
			xtick={0.1, 0.5, 0.9, 1.3, 1.7, 2.1, 2.5},
		        xlabel= \phantom{$J$} ={font=\normalsize},
			xmin = 0.1,
			xmax = 2.5,
			ymin = 0.0000001,
			ymax = 1,
			%ylabel = Relative Error,
			%yticklabel style = {font=\tiny,yshift=0.5ex},
            xticklabel style = {font=\tiny,xshift=0.0ex}			
			]
\pgfplotstableread{./9/BP.txt}\mydataone
\pgfplotstableread{./9/GBPLoop4.txt}\mydatatwo
\pgfplotstableread{./9/Tree.txt}\mydatathree
\pgfplotstableread{./9/Unif.txt}\mydatafour
\pgfplotstableread{./9/Imp.txt}\mydatafive

		\addplot [
		 color = black,
		 mark = o,
		]		
		 table[y = Z] from \mydataone;
		 
 		 \addplot [
 		 color = black,
 		 mark = triangle,
 		 ]
 		  table[y = Z] from \mydatatwo;	 
 
  		 \addplot [
 		 color = black,
 		 mark = star,
 		 ]
 		  table[y = Z] from \mydatathree;	
 		  
 		 \addplot [
 		 color = blue,
 		 mark = x,
 		 ]
 		  table[y = Z] from \mydatafour;	
 		  
 		 \addplot [
 		 color = blue,
 		 mark = diamond,
 		 ]
 		  table[y = Z] from \mydatafive;

 		 \legend{BP, GBP, TreeEP, Uni, Imp~(\ref{eqn:hatZisE})};	  
\end{axis}
\end{tikzpicture}
%  \end{subfigure}
%%%%%%%%%%%%%%%%%%%%%%%%%%%%%%%%%%%%%%%%%%%%%%%%
%\hspace{8ex}
%\begin{subfigure}{0.24\textwidth}
%\begin{tikzpicture}
%
%\end{tikzpicture}
%
%  \end{subfigure}
%%%%%%%%%%%%%%%%%%%%%%%%%%%%%%%%%%%%%%%%%%%%%%%%%%%%%%%%%%%%%%%%%%%%%%%%%%%%%%%%%%%%%%%%%%%%%%%%%%%%%%%
%\hspace{8ex}
%\begin{subfigure}{0.24\textwidth}
%  \end{subfigure}
\caption{\label{fig:FerPottsExternal}%
Comparison with deterministic algorithms (BP, GBP, and TreeEP): experimental results 
for a Potts model with $q=3$, $N=8\times 8$, constant coupling parameter $J$, and in a 
constant external field $H$.
The plots show the relative error~(\ref{eqn:RelLogError}) 
as a function of $J$.
Left: $H=0.05$; middle: $H=0.1$; right: $H=0.2$.
%as a function of the coupling parameter $J$ 
%for an $8\times 8$ \mbox{3-state} Potts model with periodic boundary conditions, 
%with constant couplings, and in a constant external field $H$, with~\textbf{(left)} $H = 0.05$, 
%\textbf{(middle)} $H = 0.1$, and \textbf{(right)} $H = 0.2$.
}
\end{figure*}
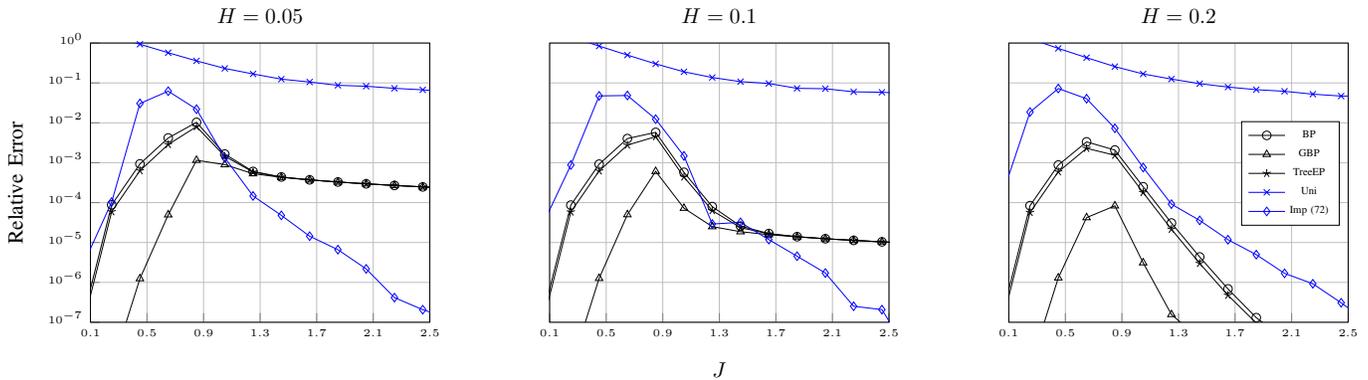
%%%%%%%%%%%%%%%%%%%%%%%%%%%%%%%%%%%%%%%%%%%

The estimate (\ref{eqn:hatZisE}) is easily verified to be unbiased, 
cf.~(\ref{eqn:hatZisUnbiasedProved}).
Creating the samples $\smpl{(\tilde\y, \tilde\z)_\F}{\ell}$ in Step~1 
is straightforward since the distribution in~(\ref{eqn:AuxProbISE}) decomposes into a product:
%is easy since (\ref{eqn:AuxProbISE}) is a product distribution: 
first sample the components $\smpl{\tilde y_e}{\ell}$, $e\in\F$, of each sample
%of $\smpl{\tilde\x_\F}{\ell}$ 
independently according to (\ref{eqn:SampleProbIS}), 
then sample the components $\smpl{\tilde z_2}{\ell},\ldots,\smpl{\tilde z_N}{\ell}$ 
of each sample independently according to
\begin{multline} 
\label{eqn:SampleProbISYE}
P\big(\smpl{\tilde{z}_k}{\ell} = \xi\big) \\ =
 {%
 \renewcommand{\arraystretch}{1.5}
 \left\{\begin{array}{ll}
   \dfrac{1+(q-1)e^{-{H_k}}}{q}, & \text{if $\xi=0$} \\ [1.0ex]
   \dfrac{1-e^{-H_k}}{q},  & \text{for $\xi = 1,2,\ldots, q-1$.}
 \end{array}\right.
 }
\end{multline}
With this choice of partitioning, it can be verified that~(\ref{eqn:VarISChi2div})
vanishes when $\text{$e^{J_e} \gg q$ for $e\in \T$}$. 

One can design a slightly different algorithm with rejections by drawing all
the components $\smpl{\tilde z_1}{\ell}, \smpl{\tilde z_2}{\ell},\ldots,\smpl{\tilde z_N}{\ell}$ 
according to~(\ref{eqn:SampleProbISYE}); but accept only the samples that satisfy~(\ref{eqn:PropZeroSumYk}).
The corresponding partitioning in the dual normal factor graph is 
\begin{IEEEeqnarray}{c}
\label{eqn:SecondPartitionE}
(\tilde\Y,\tilde\ZZ)_\F = (\tilde\Y_\F,\tilde\ZZ).
\end{IEEEeqnarray}

Accordingly
\begin{IEEEeqnarray}{c}
\Gamma_\F\big( (\tilde\y, \tilde\z)_\F \big)
 = \prod_{e\in \F} \gamma_e(\tilde y_e) \prod_{k=1}^N \lambda_k(\tilde z_k)
\end{IEEEeqnarray}
and
\begin{IEEEeqnarray}{c}
\Gamma_\T\big( (\tilde\y, \tilde\z)_\T \big)
 = \prod_{e\in\T} \gamma_e(\tilde y_e).
\end{IEEEeqnarray}
For more details, see~\cite{Mo:IZS2016}.

\subsection{Importance Sampling for Models in a Strong External Field}

%Following~\cite{MeMo:2014a}, 
We assume another partitioning of $(\tilde\Y,\tilde\ZZ)$ given by
\begin{IEEEeqnarray}{c} 
(\tilde \Y, \tilde \ZZ)_\F = \tilde \Y, \label{eqn:BasicPartitionE2}
\end{IEEEeqnarray}
which implies
\begin{IEEEeqnarray}{c} 
(\tilde \Y, \tilde \ZZ)_\T = \tilde \ZZ. \label{eqn:BasicPartitionE3}
\end{IEEEeqnarray}

This partitioning generalizes $\Gamma_\F$ and $\Gamma_\T$ 
%from (\ref{eqn:GammaF}) and (\ref{eqn:GammaT}) 
to 
\begin{IEEEeqnarray}{c}
\Gamma_\F\big(\tilde\y \big)
 = \prod_{e\in\EE} \gamma_e(\tilde y_e) 
\end{IEEEeqnarray}
and
\begin{IEEEeqnarray}{c}
\Gamma_\T\big(\tilde\z\big)
= \prod_{k=1}^N \lambda_k(\tilde z_k).
\end{IEEEeqnarray}

The partition function $Z_\F$ 
%from (\ref{eqn:ZFis}) 
is then generalized to 
\begin{IEEEeqnarray}{rCl}
Z_\F &= & \sum_{\tilde\y} \Gamma_\F\big(\tilde\y \big) \\
 & = & q^{|\EE|} \exp\!\Big( \sum_{e\in\EE} J_e \Big).
\end{IEEEeqnarray}

The importance sampling algorithm goes as follows. 
\begin{enumerate}
\item
Generate $L$ independent samples 
$\smpl{\tilde\y}{1}, \ldots, \smpl{\tilde\y}{L}$ 
according to
\begin{IEEEeqnarray}{c}
\label{eqn:AuxProbISEF}
p_\F\big(\tilde\y\big) = 
  \frac{\Gamma_\F(\tilde\y)}{Z_\F}.
\end{IEEEeqnarray}
\item
From each sample $\smpl{\tilde\y}{\ell}$,
compute $\smpl{\tilde\z}{\ell}$.
\item
Compute the unbiased estimate
\begin{IEEEeqnarray}{c}
\label{eqn:hatZisEExternal}
\hat Z^{\text{Imp}}_\text{d} = \frac{Z_\F}{L} \sum_{\ell=1}^L 
  \Gamma_\T\big( \smpl{\tilde\z}{\ell} \big).
\end{IEEEeqnarray}
\end{enumerate}

In this setting, it can be verified that~(\ref{eqn:VarISChi2div})
vanishes in the limit of the strong field (i.e., for $e^{H}\gg q$).
For more details, see~\cite{MeMo:2014a}.
%both $p_\text{d}$ and $p_\F$ become uniform over the valid configurations
%and both (\ref{eqn:VarISChi2div}) and (\ref{eqn:VarUnifChi2div}) vanish. 
The partitioning 
in~(\ref{eqn:BasicPartitionE2}) and (\ref{eqn:BasicPartitionE3}) will also be used in 
Section~\ref{sec:HightTSExt} to establish connections between the dual normal factor 
graph representation 
of the 2D Ising model in an external field and the high-temperature series expansions of the partition function.

\subsection{Numerical Experiments}
\label{sec:NumExpExtFieldIS}

Experimental results for a Potts model with $N=8\times 8$, $q=3$, and periodic boundary conditions
are shown in Figs.~\ref{fig:FerPottsExternal} and~\ref{fig:FerPottsExternalSH}, for constant coupling 
parameter $J_e=J$ 
and for constant external field $H_k=H$.
The partition function is computed exactly 
%(essentially by brute force), which allows to compare the accuracy 
via the junction tree algorithm, which allows to compare the accuracy 
of different algorithms. For the Monte Carlo algorithms, 
the estimates are averaged over 50 trials, each with $L=10^8$ samples. 
The accuracy of the estimates 
is then compared with deterministic algorithms: 
%(implemented in~\cite{Mooij:2010})
BP, GBP, and TreeEP as in Section~\ref{sec:Num1}.
The figures show 
the relative error~(\ref{eqn:RelLogError}) as a function of $J$.
%we compare the proposed methods 

\Fig{fig:FerPottsExternal} shows results for $H=0.05$ (left), $H=0.1$ (middle), and $H=0.2$ (right). 
We find that GBP works well, especially for $H =0.2$;
but the importance sampling algorithm~(\ref{eqn:hatZisE}) is more accurate 
at low temperatures (i.e., for large $J$) and in weak external fields. 

For $H = 1.0$, experimental results are shown in~\Fig{fig:FerPottsExternalSH}, where we consider both importance sampling algorithms proposed in~(\ref{eqn:hatZisE}) and~(\ref{eqn:hatZisEExternal}). As expected, we observe that the importance sampler in~(\ref{eqn:hatZisEExternal}) outperforms~(\ref{eqn:hatZisE}) for stronger external fields (stronger compared to the coupling parameter). However, GBP performs extremely 
well in this setting, indeed its relative error is below $10^{-7}$ in the whole range.

%We find that GBP works very well in this setting, 
%but the proposed importance sampling algorithm is more accurate 

The accuracy of the importance sampling algorithms can be improved, of course, 
by increasing the number of samples.

\section{The 2D Ising Model in an External Field, \\ the high-temperature series expansion,\\ 
and the subgraphs-world process}
\label{sec:HightTSExt}

In this section, we show the equivalence between the valid configurations in the dual normal factor 
graph and the configurations in Jerrum and Sinclair subgraphs-world process~\cite{JS:93} for a 
ferromagnetic Ising model on a 2D torus.
Following~\cite{JS:93}, we restrict our focus to Ising models in a
constant external field $H$.
 
For this model, the energy 
of a configuration $\x \in \calX^N$ is given by the Hamiltonian~\cite[Chapter 3]{Yeo:92}
\begin{multline}
\label{eqn:HamiltonianIsing}
\mathcal{H}(\x) = -\sum_{\text{$(k,\ell)\in \EE$}}\!\!\!J_{k, \ell}\cdot\big(2\delta(x_k - x_{\ell}) - 1\big) \\
- H\sum_{k = 1}^N \big(1-2\delta(x_k)\big).
\end{multline}
The primal normal factor graph of the model is shown 
in~\Fig{fig:2DGridExt}, where the empty boxes represent~(\ref{eqn:IsingPrimal}) and 
the small empty boxes represent the factors 
\begin{equation} 
\label{eqn:PottsHIsingExt}
\tau(x_{k}) = \left\{ \begin{array}{ll}
      e^{-H}, & \text{if $\tilde{x}_{k} = 0$} \\
      e^H, & \text{if $\tilde{x}_{k} = 1.$}
  \end{array} \right.
\end{equation}

\Fig{fig:2DGridDMDualExt} shows 
the dual normal factor graph of the 2D Ising model in an external field, where the empty boxes represent~(\ref{eqn:IsingDualAdding}) and 
the small empty boxes represent the factors $\lambda(\tilde z_k)$, which are the 1D Fourier transforms of~(\ref{eqn:PottsHIsingExt}), and are given by
%$\kappa_k(x_k)$:
\begin{IEEEeqnarray}{rCl}
\lambda(\tilde z_k) 
  & = & \left\{ \begin{array}{ll}
      \cosh(H),  & \text{if $\tilde z_k = 0$} \\% [2.0ex]
      {-}\sinh(H),     & \text{if $\tilde z_k = 1.$}
      \end{array} \right.
      \label{eqn:GammaDualVarEFIsing}
\end{IEEEeqnarray}
Again, for the Ising model, the ``$\circ$'' symbols are immaterial and can be removed from normal factor graphs.

Since the partition function is invariant under 
the change of sign of the external field~\cite[Chapter 1]{Baxter07}, we will assume $H \le 0$.  
%\begin{equation} \label{eqn:PositiveHk}
%H \le 0
%\end{equation}
Therefore, factors~(\ref{eqn:GammaDualVarEFIsing}) are nonnegative.
The invariance of the partition function under the change of sign of $H$ is  
implied by Proposition~\ref{prop:tildeXSum0}, as in any valid configuration in the dual normal factor graph, 
it holds that
\begin{IEEEeqnarray}{c} 
\label{eqn:PropZeroSumYkIsing}
\sum_{k=1}^N \tilde z_k = 0,
\end{IEEEeqnarray}
i.e., the Hamming weight of $\tilde \ZZ$ is always even, where 
the Hamming weight of a configuration is the number of nonzero components of 
that configuration~\cite{RJM:77}. Indeed, $\prod_{k=1}^N \lambda(\tilde z_k)$ takes on the same positive value 
regardless of the sign of $H$.

%%%%%%%%%%%%%%%%%%%%%%%%%%%%%%%%%%%%%%%%%%%%%%%%%%%%%%%%%%%%%%%%
\begin{figure}[t!]
\centering
\begin{tikzpicture}
\begin{axis}[
			legend style={at = {(0.272,0.93)} ,font=\tiny},		
			height = 42.0ex,
			width = 48.0ex,
			grid = major,
			tick pos=left, 
			ymode=log,
			xminorticks = false,	
		    yminorticks = false,	
		    y tick label style={
        /pgf/number format/.cd,
            fixed,
        /tikz/.cd
    		}, 				
			ytick={1e-8,1e-7, 1e-6, 1e-5, 1e-4, 1e-3, 1e-2, 1e-1},
			xtick={0.1, 0.3, 0.5, 0.7, 0.9, 1.1},
		xlabel= $J$ ={font=\normalsize},
			xmin = 0.1,
			xmax = 1.1,
			ymin = 0.00000001,
			ymax = 0.1,
			ylabel = Relative Error = {font=\tiny},
			yticklabel style = {font=\tiny,yshift=0.5ex},
            xticklabel style = {font=\tiny,xshift=0.0ex}			
			]
\pgfplotstableread{./10/BP.txt}\mydataone
\pgfplotstableread{./10/GBP.txt}\mydatatwo
\pgfplotstableread{./10/TreeEP.txt}\mydatathree
\pgfplotstableread{./10/Imp1.txt}\mydatafour
\pgfplotstableread{./10/Imp2.txt}\mydatafive
		\addplot [
		 color = black,
		 mark = o,
		]		
		 table[y = Z] from \mydataone;
		 
 		 \addplot [
 		 color = black,
 		 mark = triangle,
 		 ]
 		  table[y = Z] from \mydatatwo;	 
 
  		 \addplot [
 		 color = black,
 		 mark = star,
 		 ]
 		  table[y = Z] from \mydatathree;	
 		  
 		 \addplot [
 		 color = blue,
 		 mark = diamond,
 		 ]
 		  table[y = Z] from \mydatafive;
 		  
 		  \addplot [
 		 color = blue,
 		 mark = x,
 		 ]
 		  table[y = Z] from \mydatafour;	

 		 \legend{BP, GBP, TreeEP, Imp~(\ref{eqn:hatZisE}), Imp~(\ref{eqn:hatZisEExternal})};	  	

%		\legend{BP, GBPLoop, TREEEP, Unif. Sampling, Imp. 
%       Sampling};
		%\addlegendentry{Uniform Sampling};

%table [x index=0, y index=1]{\mytable};
\end{axis}
\end{tikzpicture}
  %%%%%%%%%%%%%%%%%%%%%%%%%%%%%%%%%%%%%%%%%%%%%%%%%%%%%%%%%
\caption{\label{fig:FerPottsExternalSH}%
Comparison with deterministic algorithms (BP, GBP, and TreeEP): experimental results 
for a Potts model with \mbox{$q=3$}, $N=8\times 8$, constant coupling parameter $J$, and in a 
constant external field $H = 1.0$.
The plots show the relative error~(\ref{eqn:RelLogError}) 
as a function of $J$.
}
\end{figure}

In~\cite{JS:93}, the authors propose a Markov chain (called the 
subgraphs-world process), which is defined on the set of edges $\calW \subseteq \EE$ of the interaction 
graph of the model (as in~\Fig{fig:2DGridDMDualExt}). 
The scheme of Jerrum and Sinclair then uses the following expansion of the partition function in 
powers of $\tanh(H)$ and $\tanh(J)$
\begin{IEEEeqnarray}{c} 
\label{eqn:HTExpandExt}
%\kappa_{k, \ell}(x_k, x_{\ell}) \eqdef e^{J_{k, \ell}\cdot\delta(x_k - x_{\ell})}.
Z = A\sum_{\calW \subseteq \EE}\tanh(H)^{|\text{odd$(\calW)$}|}\prod_{(k,\ell) \in \calW} \tanh(J_e),
\end{IEEEeqnarray}
where $\text{odd$(\calW)$}$ denotes the set of all odd-degree vertices in the subgraph of $\EE$
induced by $\calW$, and
\begin{IEEEeqnarray}{c}
\label{eqn:eqA}
A = \big(2\cosh(H)\big)^{N}\prod_{(k,\ell) \in \calW} \cosh(J_e).
\end{IEEEeqnarray}
The sum in~(\ref{eqn:HTExpandExt}) is known as the high-temperature series 
expansion in statistical physics~\cite{Newell:53},~\cite[p.~94]{Yeo:92}.

In the dual normal factor graph, we adopt the partitioning of $(\tilde\Y,\tilde\ZZ)$ proposed 
in~(\ref{eqn:BasicPartitionE2}) 
and~(\ref{eqn:BasicPartitionE3}). We 
can thus freely choose the variables $\tilde \Y = \{\tilde Y_e: e\in\EE \}$, and therefrom 
compute the variables $\tilde\ZZ = \big\{\tilde Z_k: k\in \{ 1,\ldots,N \}\big\}$. 

The partition function $Z_\text{d}$ can then be written as
\begin{IEEEeqnarray}{rCl} \label{eqn:ZWE1}
Z_\text{d} & = & \sum_{\text{valid $(\tilde\y, \tilde \z)$}} \prod_{k = 1}^N\lambda(\tilde z_k)\prod_{e\in\EE} \gamma_e(\tilde{y}_e)  \label{eqn:ZWE1}\\
 & = & 2^{|\EE|}\cosh(H)^{N}\prod_{e \in \EE}\cosh(J_e)\cdot \nonumber\\ 
 &&\sum_{\text{valid $(\tilde\y, \tilde \z)$}}
 \prod_{k = 1}^N \tanh(|H|)^{\tilde z_k}\prod_{e\in\EE}\tanh(J_e)^{\tilde y_e}.  \label{eqn:ZWE7} 
 \end{IEEEeqnarray}
In a 2D torus $|\EE| = 2N$, thus
\begin{equation}
Z_\text{d} = 2^{N}A\sum_{\text{valid $(\tilde\y, \tilde \z)$}} 
\tanh(|H|)^{\sum_{k = 1}^N \tilde z_k}\prod_{e\in\EE}\tanh(J_e)^{\tilde y_e},   \label{eqn:ZWE2}
\end{equation}
where $A$ is as in~(\ref{eqn:eqA}).

Accordingly, we 
define % the bijection 
$\calS \subseteq \EE$ as 
\begin{IEEEeqnarray}{c} 
\label{eqn:WF}
\calS(\tilde \y) \eqdef \{e : \tilde Y_e = 1\}.
\end{IEEEeqnarray}

Here, as $\tilde \y$ runs over the configurations in the dual normal factor graph, $\calS(\tilde \y)$ 
runs over all the $2^{|\EE|}$ subsets of $\EE$.
Let us consider the subgraph of $\EE$ induced by $\calS$, in which $\tilde z_k = 1$ if it is connected to 
a zero-sum indicator function with odd degree, and  $\tilde z_k = 0$ otherwise. Therefore,
$\sum_{k = 1}^N\tilde z_k$ counts the number of odd-degree vertices in $\calS$. 
Thus, from~(\ref{eqn:ZWE2}) we obtain
\begin{IEEEeqnarray}{c} 
Z_\text{d} = 2^{N}A\sum_{\calS \subseteq \EE}\tanh(|H|)^{|\text{odd($\calS$)}|}\prod_{e\in \calS(\tilde \y)}\tanh(J_e).  \label{eqn:ZWE4}%\\
%&=& 2^{2N}\cosh(J)^{2N}\cosh(H)^{N}\sum_{\text{valid $\tilde\x$}}\tanh(J)^{W(\tilde \x)}, \label{eqn:ZWE2}
\end{IEEEeqnarray}

After applying the scale factor~(\ref{eqn:ZdZEF}) in~(\ref{eqn:ZWE4}), we will 
obtain the high-temperature series expansion of the partition function given by~(\ref{eqn:HTExpandExt}). We conclude that the configurations in the subgraphs-world process coincide
with the valid configurations in the dual normal factor graph of the Ising model in an external field. 
(The number of vertices with odd degree in the subgraph $\calS$ is always even, which is also implied by 
Proposition~\ref{prop:tildeXSum0}.)

The scheme of Jerrum and Sinclair works on a Markov chain whose states are configurations of 
the subgraphs-world and whose stationary distribution is given by $p_\text{d}$, cf.~Section~\ref{sec:IS}. Transitions
occur between states that differ in a single edge according to the Metropolis rule.
Remarkably, the mixing time of the proposed Markov chain is only polynomial in the size of the model 
at \emph{all} temperatures. Indeed, a rigorous analysis shows that the expected running time of the generator for the subgraphs-world 
configurations is upper bounded by $O\big(|E|^2N^8(\log \epsilon^{-1} + |E|)\big)$, where $\epsilon$ is the confidence parameter. 
The scheme is randomized, i.e., it provides approximations to the 
partition function, which fall within arbitrary small error bounds with 
high probability~\cite[Section 4]{JS:93}. 

Our proposed unbiased Monte Carlo methods in the dual normal factor graph draw independent 
samples according to an auxiliary distribution. The partition function is then estimated by averaging according to the 
importance sampling weights of the independent samples, cf.~(\ref{eqn:EstIS}). 
Monte Carlo methods of this paper work particularly well in the low-temperature regime.

The main focus of this paper is on the (nonbinary) Potts model, where Monte Carlo methods in the dual normal factor graph 
outperform the state-of-the-art methods
at low temperatures (with no external field or in a weak external field). 
However, approximating the partition function of the ferromagnetic 
Potts model is already as hard as approximating the number of independent sets in 
a bipartite graph, which is among the presumably intractable (the $\#$BIS-hardness) 
problems. For more details see~\cite{goldberg2012,Gold:12,Galan:16}.

\section{Conclusion}

We reviewed representations of the Ising and Potts models of statistical physics in terms of normal factor graphs, and further explored the idea 
that Monte Carlo algorithms 
in the dual normal factor graph can yield 
good estimates of the partition function at low temperatures. 
%that MC methods can use the dual factor graph for good convergence at low temperature. 
Specifically, we proposed and investigated such algorithms 
for the ferromagnetic 
%$q$-state 
Potts model, 
and we observed good convergence for strong couplings (i.e., at low temperatures).
%we proposed such
%methods for estimating the partition function of the
%2D ferromagnetic $q$-state Potts model. We investigated 
%a uniform sampling and an importance sampling scheme, both
%of which were shown to work very well in 
%models with strong couplings (i.e., at low temperature).
In our numerical experiments, for the 2D ferromagnetic 
%$q$-state 
Potts models, the importance sampling algorithms %of ~\cite{Mo:IZS2016}
in the dual normal factor graph 
yield more accurate estimates 
at low temperatures (and with no, or weak, external field) 
than 
%clearly outperforms the
the state-of-the-art deterministic and Monte Carlo methods. % (BP, GBP, and TreeEP). 
%in 
%models with strong couplings and in models with a mixture 
%of weak and
%strong couplings.
%We presented an importance sampling algorithm
%in the dual NFG of the 2D ferromagnetic Potts model to estimate 
%the partition function. The algorithm works by 
%simulating a subset of the variables 
%according to an auxiliary distribution, 
%followed by %Vicen doing
%doing computations on the remaining ones. In contrast to MC methods 
%in the primal domain, convergence of the proposed 
%algorithm improves when the coupling parameters, on only 
%a subset of the edges (which forms a spanning tree) 
%become stronger. In terms of accuracy, the algorithm
%outperforms the
%state of the art methods by more than an order of magnitude. 
We expect such Monte Carlo methods in the dual normal factor graph 
to work well also for three-dimensional grids 
and in graphical models with more general topologies. Using deterministic algorithms (e.g., GBP and TreeEP) 
in the dual graph is certainly possible,
%and likely to work very well at low temperatures, 
but it has not been tried yet.

We also showed 
the equivalence between the valid configurations in the dual normal factor graph and the terms that appear in the 
high-temperature series expansion of the partition function of the ferromagnetic Ising model in an 
external field, and discussed connections with the subgraphs-world process in
the randomized approximate scheme of Jerrum and Sincalir.

Finally, it should be mentioned that the factors in the dual normal factor graph 
can, in general, be negative or even complex-valued, which could be a serious 
challenge for Monte Carlo methods.
%Monte Carlo algorithms in the dual normal factor graph have a critical limitation. 
%%However, Monte Carlo algorithms in the dual factor graph have two 
%%limitations that should not be ignored. 
%, which is 
Such issues were avoided in this paper 
by considering only ferromagnetic models,
but they must be faced in any attempt to deal with antiferromagnetic models,
spin glasses, and computational problems in quantum information processing~\cite{MoLo:ITW2012, CV:ISIT2017, CV:2017, LoVo:IT2017}.

\appendices

%%%%%%%%%%%%%%%%%%%%%%%%%%%%%%%%%%%%%%%%%

\section{Comparing the Variance of Monte Carlo methods\\ in the Primal and in the Dual 
Normal Factor Graphs of the 2D Ising Model}
\label{appsec:VarIsing}
%\label{appsec:DualDetails}

We compare the variance of the uniform sampling and importance sampling estimators in the primal and in 
the dual domains for estimating the partition function of the Ising model on a 2D torus,
with constant coupling parameter $J$, without an external field, and in the thermodynamic limit (i.e., as $N \to \infty$). 
The choice of the model and the parameters is due to the fact that the
partition function is analytically available from 
Onsager's solution in this case, see~\cite{Onsager:44},\cite[Chapter 7]{Baxter07}. 

For this model, the critical coupling (i.e., the phase transition) is located at
\begin{equation} \label{eqn:CriticalJ}
J_\text{c} = \frac{1}{2}\ln(1+\sqrt{2}) \approx 0.4407
\end{equation}
and, at criticality, the derivative of $\ln Z$ with respect to $J$ (i.e., the internal energy of the model) 
is given by
\begin{IEEEeqnarray}{c} 
\label{eqn:CriticalEnergy}
 \lim_{N \to \infty}\frac{1}{N}\frac{\partial \ln Z(J_\text{c})}{\partial J_\text{c}} = \sqrt{2},
\end{IEEEeqnarray}
see~\cite{Mccoy:14}. 

In the primal domain, the analytical solution of the partition function allows us to calculate the exact value of 
the variance of the uniform sampling estimator as a function of $J$. In the dual domain,  we provide upper 
and lower bounds on the variance of the estimators. The derived bounds are not necessarily tight
for all values of $J$, however, they are good enough to illustrate the opposite behavior of
the estimators in the primal and in the dual domains.   

We recall from~(\ref{eqn:CardBF}) and~(\ref{eqn:CardBFPeriodic}) that for the 2D torus
$|\T| = N-1$ and $|\F| = N+1$.

\subsection{Uniform Sampling}
\label{ref:VarIsingUnif}

Following our analysis in Section~\ref{sec:Analysis}, the variance of the 
uniform sampling estimator in the primal domain~(\ref{eqn:UnifPrimal}) 
can be expressed as
\begin{IEEEeqnarray}{c}
\label{eqn:VarUnifChi2divP}
\Var[\hat Z^{\text{Uni}}] \frac{L}{Z^2} = \chi^2\big( p_\text{B}, p_\text{u} \big),
\end{IEEEeqnarray}
where $p_{\text{B}}(\x)$ is the Boltzmann distribution given by~(\ref{eqn:Prob}) and $p_\text{u}(\x)$ is the uniform 
distribution over all the configurations.

It follows that
\begin{IEEEeqnarray}{rCl}
1+\Var[\hat Z^{\text{Uni}}] \frac{L}{Z(J)^2} & = & \sum_{\x}
        \frac{p_{\text{B}}(\x)^2}{p_\text{u}(\x)} \\
& = & \frac{2^N}{Z(J)^2} \sum_{\x} f(\x)^2 \label{eqn:VarPrimalSubs3}\\
& = & 2^N\frac{Z(2J)}{Z(J)^2}, \label{eqn:VarPrimalSubs4}
\end{IEEEeqnarray}
where $Z(J)$ denotes the partition function evaluated at $J$, and the last step is due to the following identity
\begin{equation} \label{eqn:PartFunctionPower}
Z(2J) = \sum_{\x} f(\x)^2.
\end{equation}

Thus, in the thermodynamic limit we obtain
\begin{multline} 
\label{eqn:VarUnifChi2divPLimit}
\lim_{N \to \infty} \frac{1}{N} \ln \Big(1+\Var[\hat Z^{\text{Uni}}] \frac{L}{Z(J)^2}\Big) = \\
\ln(2) + \lim_{N \to \infty}\frac{\ln Z(2J)}{N} - \lim_{N \to \infty}\frac{2\ln Z(J)}{N}.
\end{multline}

\begin{figure}[t!!]
\centering
\begin{tikzpicture}
\begin{axis}[
			legend style={at = {(0.975,0.81)} ,font=\tiny},		
			height = 42.0ex,
			width = 48.0ex,
			grid = major,
			tick pos=left, 
			%ymode=log,
			xminorticks = false,	
		    yminorticks = false,	
		    y tick label style={
        /pgf/number format/.cd,
            fixed,
        /tikz/.cd
    		}, 				
			ytick={0,  0.2, 0.4, 0.6, 0.8},
			xtick={0.0, 0.5, 1.0, 1.5, 2.0, 2.5, 3.0},
		xlabel= $J$ ={font=\normalsize},
			xmin = 0.0,
			xmax = 2.5,
			ymin = 0.0,
			ymax = 0.8,
			ylabel = $\lim_{N \to \infty}\frac{1}{N} \ln\big(1+\Var \hat Z\cdot \frac{L}{Z^2}\big)$ = {font=\normalsize},
			yticklabel style = {font=\tiny,yshift=0.5ex},
            xticklabel style = {font=\tiny,xshift=0.0ex}			
			]
\pgfplotstableread{./11/VarUP.txt}\mydataone
\pgfplotstableread{./11/VarUDLower1.txt}\mydatathree
\pgfplotstableread{./11/VarUDUpper.txt}\mydatatwo
\pgfplotstableread{./11/VarUImp.txt}\mydatafive

		\addplot [
		 line width = 0.24mm,
		 smooth,
		 color = black,
		]		
		 table[y = Z] from \mydataone;
		 
 		 \addplot [
		 line width = 0.27mm,
		 smooth,
 		 color = blue,
		 densely dotted,
 		 ]
 		  table[y = Z] from \mydatatwo;	 

		 \addplot [
		 line width = 0.25mm,
		 smooth,
 		 color = blue,
	     dashed,
 		 ]
 		  table[y = Z] from \mydatathree;	 

		 \addplot [
		 line width = 0.27mm,
		 smooth,
 		 color = red,
	     dash dot,
 		 ]
 		  table[y = Z] from \mydatafive;	 

%		 \addplot [
%		 line width = 0.25mm,
%		 smooth,
% 		 color = red,
%	     %dashed,
% 		 %mark = x,
% 		 ]
% 		  table[y = Z] from \mydataseven;	 
%
%		 \addplot [
%		 line width = 0.25mm,
%		 smooth,
% 		 color = red,
%		 dashed,
% 		 %mark = x,
% 		 ]
% 		  table[y = Z] from \mydataeight;	 

% 

 		 \legend{Uni primal exact result~(\ref{eqn:VarUnifChi2divPLimit}), Uni dual upper bound~(\ref{eqn:VarUnifChi2divDUpper}), Uni dual lower bound~(\ref{eqn:VarUnifChi2divDLower}), 
Imp dual upper bound~(\ref{eqn:VarUnifChi2divDUpperIS})};	  	

%		\legend{BP, GBPLoop, TREEEP, Unif. Sampling, Imp. 
%       Sampling};
		%\addlegendentry{Uniform Sampling};

%table [x index=0, y index=1]{\mytable};
\end{axis}
\end{tikzpicture}
%%%%%%%%%%%%%%%%%%%%%%%%%%%%%%%%%%%%%%%%%%%%%%%%%%%%%%%%%
\caption{\label{fig:Var}%
Comparing the variance of Monte Carlo methods in the primal and in the 
dual domains
%as a function of the coupling parameter $J$ 
for the Ising model on a 2D torus, with constant coupling $J$, and in the thermodynamic limit. 
The solid black line shows~(\ref{eqn:VarUnifChi2divPLimit}); for the uniform sampling algorithm 
%the solid blue line shows the upper bound in~(\ref{eqn:VarUnifChi2divDUpper2}), 
in the dual domain the dotted blue line shows the upper 
bound in~(\ref{eqn:VarUnifChi2divDUpper}) and the dashed blue line shows the lower 
bound in~(\ref{eqn:VarUnifChi2divDLower}); for the importance sampling algorithm in the
dual domain the dashed-dotted red line shows the upper bound in~(\ref{eqn:VarUnifChi2divDUpperIS}).}
\end{figure}
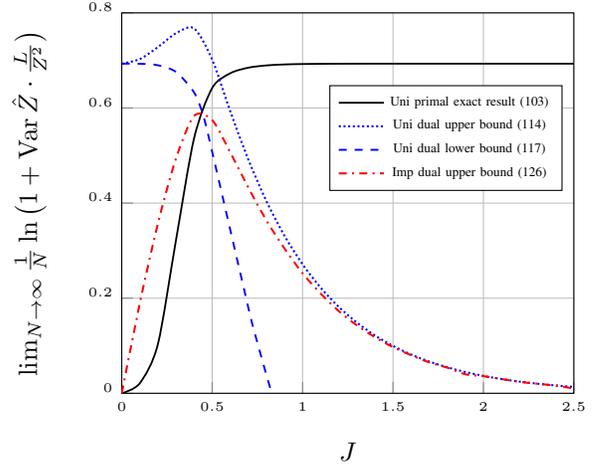

We use the closed-form solution of the partition function to evaluate~(\ref{eqn:VarUnifChi2divPLimit}) numerically as a
function of $J$, which is plotted by the solid black line in~\Fig{fig:Var}. As expected, we observe 
that uniform sampling in the primal domain can provide good estimates of the partition 
function when $J$ is small (i.e., at high temperature), while it is an inefficient estimator 
for larger values of $J$ (i.e., at low temperature).

%We recall that the factors in the dual normal factor graph of the Ising model are 
%given by~(\ref{eqn:IsingDualAdd}).
From~(\ref{eqn:VarUnifChi2div}), we expand the variance of the uniform sampling algorithm in the 
dual domain~(\ref{eqn:EstU}) as
\begin{IEEEeqnarray}{rCl}
1+\Var[\hat Z^{\text{Uni}}_\text{d}] \frac{L}{Z_\text{d}(J)^2} & = & \sum_{\text{valid $\tilde \y$}} 
        \frac{ p_\text{d}(\tilde\y)^2}{p_\text{u}(\tilde \y)} \\
& = & \frac{2^{|\F|}}{Z_\text{d}(J)^2} \sum_{\text{valid $\tilde \y$}} \Gamma(\tilde\y)^2 \label{eqn:VarDualSubs2}\\
& = & \frac{2^{N+1}}{Z_\text{d}(J)^2} R, \label{eqn:VarDualSubs3}
\end{IEEEeqnarray}
where $R \eqdef \sum_{\text{valid $\tilde \y$}} \Gamma(\tilde\y)^2$.
%and in the last step we set $|\B_\F| = N+1$.

From~(\ref{eqn:NDual}), we have $Z_\text{d} = 2^NZ$ . Thus
\begin{IEEEeqnarray}{c}
1+\Var[\hat Z^{\text{Uni}}_\text{d}] \frac{L}{Z_\text{d}(J)^2} = 
\frac{2^{-N+1}}{Z(J)^2}R.  \label{eqn:VarDualSubs}
\end{IEEEeqnarray}

%Therefore, in thermodynamic limits
%\begin{equation} \label{eqn:VarUnifChi2divDLimit}
%\lim_{N \to \infty} \ln \Big(1+\Var[\hat Z^{\text{Uni}}_\text{d}] \frac{L}{Z_\text{d}(J)^2}\Big) = 
%-\ln(2) + \lim_{N \to \infty}\frac{S_\text{d}}{N} - \lim_{N \to \infty}\frac{2\ln Z(J)}{N}
%\end{equation}
In the sequel, we will derive upper and lower bounds on $R$, which 
%We apply the obvious inequality 
%\begin{equation} \label{eqn:SimpleDUpper}
%S_\text{d} \le Z_\text{d}(J)^2
%\end{equation}
%in~(\ref{eqn:VarDualSubs3}) to obtain
%\begin{equation} \label{eqn:VarUnifChi2divDUpper2}
%\lim_{N \to \infty} \frac{1}{N}\ln 
%\Big(1+\Var[\hat Z^{\text{Uni}}_\text{d}] \frac{L}{Z_\text{d}(J)^2}\Big) \le \ln(2),
%\end{equation}
%which is plotted by the solid blue line in~\Fig{fig:Var}.
is the partition function of a dual normal factor 
graph (as shown in~\Fig{fig:2DGridDM}) with factors given by
\begin{IEEEeqnarray}{c}
\label{eqn:IsingDualAddPower}
\rho(\tilde{y}_e) = \left\{ \begin{array}{ll}
      4\cosh(J)^2, & \text{if $\tilde{y}_e = 0$} \\
      4\sinh(J)^2, & \text{if $\tilde{y}_e = 1$}.
  \end{array} \right.
\end{IEEEeqnarray}

%%%%%%%%%%%%%%%%%%%%%%%%%%%%%%%%%%%%%%%%%%%%%%%%%%%%%%%%%%%%%%%%
%%%%%%%%%%%%%%%%%%%%%%%%%%%%%%%%%%%%%%%%%%%%%%%%%%%%%%%%%%

Thus
\begin{IEEEeqnarray}{rCl}  
%Z_\text{d} = \sum_{\text{valid $\tilde\x$}} \Gamma_\F(\tilde\x_\F)\, \Gamma_\T(\tilde\x_\T).
R &=& \sum_{\text{valid $\tilde\y$}} \prod_{e\in \EE} \rho(\tilde y_e) \\
& \le & \big(4\cosh(J)^2\big)^{|\T|}\sum_{\tilde\y_\F} \prod_{e\in\F} \rho(\tilde y_e) \\
& = & \big(2\cosh(J)\big)^{2(N-1)}R_\F. \label{eqn:DualPartitionSumGammaPower}
\end{IEEEeqnarray}
%\begin{equation}
%p'_\F(\tilde \x_\F) \eqdef \frac{\Gamma'_\F(\tilde\x_\F)}{S_\F} \label{eqn:ISPropProbPower}
%\end{equation}

Here, $R_\F$ is the partition function of a subgraph of the dual normal factor graph induced by $\F$, which can be
computed exactly as
\begin{IEEEeqnarray}{rCl} 
R_\F & = & \big(\rho(0) + \rho(1)\big)^{|\F|} \\
        &=& \big(4\cosh(2J)\big)^{N+1}. \label{eqn:ZFisPower}
\end{IEEEeqnarray}

Combining~(\ref{eqn:VarDualSubs}),~(\ref{eqn:DualPartitionSumGammaPower}), and~(\ref{eqn:ZFisPower}) yields the following upper bound
\begin{multline} 
\label{eqn:VarUnifChi2divDUpper}
\lim_{N \to \infty} \frac{1}{N}\ln \Big(1+\Var[\hat Z^{\text{Uni}}_\text{d}] \frac{L}{Z_\text{d}(J)^2}\Big) \le  
3\ln(2) \\ + \ln\big(\cosh(2J)\cdot\cosh(J)^2\big) - \lim_{N \to \infty}\frac{2\ln Z(J)}{N},
\end{multline}
which is plotted by the dotted blue line in~\Fig{fig:Var}.

To obtain the lower bound, we note that
\begin{IEEEeqnarray}{rCl}  
%Z_\text{d} = \sum_{\text{valid $\tilde\x$}} \Gamma_\F(\tilde\x_\F)\, \Gamma_\T(\tilde\x_\T).
R &=& \sum_{\text{valid $\tilde\y$}} \prod_{e\in \EE} \rho(\tilde y_e) \\
& \ge & \big(4\cosh(J)^2\big)^{2N}. \label{eqn:DualPartitionSumGammaPowerLower}
\end{IEEEeqnarray}

Combining~(\ref{eqn:VarDualSubs}) and~(\ref{eqn:DualPartitionSumGammaPowerLower}) gives the following lower bound
\begin{multline} 
\label{eqn:VarUnifChi2divDLower}
\lim_{N \to \infty} \frac{1}{N}\ln \Big(1+\Var[\hat Z^{\text{Uni}}_\text{d}] \frac{L}{Z_\text{d}(J)^2}\Big) \ge 
3\ln(2) \\ + 4\ln\big(\cosh(J)\big) - \lim_{N \to \infty}\frac{2\ln Z(J)}{N},
\end{multline}
which is plotted by the dashed blue line in~\Fig{fig:Var}. 
%The lower 
%bound in~(\ref{eqn:VarUnifChi2divDLower}) is tighter than the lower bound 
%presented in~\cite[Section~VII]{Mo:2017Arxiv}.

From~\Fig{fig:Var}, we observe that uniform sampling in the 
dual domain is inefficient for small values of $J$, however,  compared to uniform sampling 
in the primal domain,
it can provide more reliable estimates of the partition function when $J$ is large. 
(Recall from Section~\ref{sec:Analysis} that~(\ref{eqn:VarUnifChi2div}) vanishes as $J \to \infty$, i.e, in the 
low-temperature limit.) 

Both estimators seem to be inefficient in the mid-temperature regime and 
near criticality~(\ref{eqn:CriticalJ}).

\subsection{Importance Sampling}

From~(\ref{eqn:VarISChi2div}), the variance of the importance sampling algorithm 
%in the dual domain~(\ref{eqn:EstIS}) 
can be expressed as
\begin{IEEEeqnarray}{rCl}
1+\Var[\hat Z^{\text{Imp}}_\text{d}] \frac{L}{Z_\text{d}(J)^2}
 & = & \sum_{\text{valid $\tilde \y$}} 
        \frac{ p_\text{d}(\tilde\y)^2}{p_\F(\tilde \y)} \\
 & = & \frac{Z^2_\F}{Z_\text{d}(J)^2} \sum_{\text{valid $\tilde \y$}} p_\F(\tilde \y)
        \Gamma_\T(\tilde\y_\T)^2. \IEEEeqnarraynumspace \label{eqn:VarImpUpperIntermediate}
%& = & \frac{Z_\F}{Z_\text{d}(J)^2}S_\text{d} - 1, 
\end{IEEEeqnarray}
From~(\ref{eqn:NDual}), we have $Z_\text{d} = 2^NZ$. Moreover,
\begin{IEEEeqnarray}{rCl}
Z_\F & = &\sum_{\tilde \y_\F} \Gamma_\F(\tilde \y_\F) \\
  & = & 2^{|\F|} e^{J|\F|} \\
       & = & 2^{N+1} e^{J(N+1)}, \label{eqn:ZFisIsing}
\end{IEEEeqnarray} 
cf.~(\ref{eqn:ZFis}). From~(\ref{eqn:IsingDualAdding}) and (\ref{eqn:GammaT}), we obtain
\begin{IEEEeqnarray}{rCl}  
%Z_\text{d} = \sum_{\text{valid $\tilde\x$}} \Gamma_\F(\tilde\x_\F)\, \Gamma_\T(\tilde\x_\T).
\sum_{\text{valid $\tilde \y$}} p_\F(\tilde \y)
        \Gamma_\T(\tilde\y_\T)^2 & \le & \big(4\cosh(J)^2\big)^{|\T|} \\
& = & \big(2\cosh(J)\big)^{2(N-1)}. \label{eqn:DualPartitionSumGammaPowerIS}
\end{IEEEeqnarray}

Thus
\begin{IEEEeqnarray}{c}
1+\Var[\hat Z^{\text{Imp}}_\text{d}] \frac{L}{Z_\text{d}(J)^2} =
\frac{2^{2N}e^{2J(N+1)}}{Z(J)^2}\cosh(J)^{2(N-1)},  \IEEEeqnarraynumspace\label{eqn:VarDualSubsIs}
\end{IEEEeqnarray} 
which, in the thermodynamic limit $N \to \infty$, gives the following upper bound
\begin{multline} \label{eqn:VarUnifChi2divDUpperIS}
\lim_{N \to \infty} \frac{1}{N}\ln \Big(1+\Var[\hat Z^{\text{Imp}}_\text{d}] \frac{L}{Z_\text{d}(J)^2}\Big) \le 
2\ln(2) \\ + 2J + 2\ln\big(\cosh(J)\big) - \lim_{N \to \infty}\frac{2\ln Z(J)}{N}.
\end{multline}
%which is plotted by the dashed-dotted red line in~\Fig{fig:Var}.

Let us denote the upper bound in~(\ref{eqn:VarUnifChi2divDUpperIS}) by $U(J)$, which is plotted by 
the dashed-dotted red line in~\Fig{fig:Var}. The 
derivative of $U(J)$ with respect to the coupling 
parameter $J$ is 

\begin{IEEEeqnarray}{c} 
\label{eqn:VarUnifChi2divISCriticality}
\frac{\partial U(J)}{\partial J} = 2 + 2\tanh(J) - \lim_{N \to \infty}\frac{2}{N}\frac{\partial \ln Z(J)}{\partial J}.
\end{IEEEeqnarray}

From~(\ref{eqn:CriticalEnergy}), it is straightforward to verify that (\ref{eqn:VarUnifChi2divISCriticality}) is zero at the critical 
coupling $J_\text{c}$ given by~(\ref{eqn:CriticalJ}). From~\Fig{fig:Var} we observe that 
the upper bound $U(J)$ grows to attain its maximum at $J_\text{c}$, but then decays 
for $J > J_\text{c}$.
(Again, recall from Section~\ref{sec:Analysis} 
that~(\ref{eqn:VarISChi2div}) vanishes as $J \to \infty$.)

\section*{Acknowledgments}

The authors are grateful to Hans-Andrea Loeliger for his comments and discussions related to the 
topics of this paper.
The authors would like to thank David Forney, Pascal Vontobel, Justin Dauwels, Alistair Sinclair,
and Stefan Moser for their helpful 
comments on an earlier draft of this manuscript. 
The authors wish to thank
the associate editor, Yongyi Mao, for his constructive suggestions during the review process.
Part of this work was 
done during the first author's stay at the Institut Henri 
Poincar\'e -- Centre Emile Borel, France, at the Information Theory and Coding Group, University of Pompeu 
Fabra, Spain, and at 
the Department of Statistics and Actuarial Science, University of Waterloo, Canada. The first 
author thanks these institutions for hospitality and support. This work is supported in part by the Spanish Ministry of Economy and Competitiveness under the María de Maeztu Units of Excellence Programme (MDM-2015-0502) and the Ramon y Cajal program RYC-2015-18878 (AEI/MINEICO/FSE,UE).

\newcommand{\IT}{IEEE Trans.\ Inf.\ Theory}
\newcommand{\CASI}{IEEE Trans.\ Circuits \& Systems~I}
\newcommand{\COM}{IEEE Trans.\ Comm.}
\newcommand{\COMLet}{IEEE Commun.\ Lett.}
\newcommand{\COMMag}{IEEE Communications Mag.}
\newcommand{\ETT}{Europ.\ Trans.\ Telecomm.}
\newcommand{\SPMag}{IEEE Signal Proc.\ Mag.}
\newcommand{\ProcIEEE}{Proceedings of the IEEE}

\end{document}